\documentclass[10pt,twocolumn,english]{elsarticle}
\usepackage{mathptmx}
\usepackage[T1]{fontenc}
\usepackage[latin9]{inputenc}
\usepackage{geometry}
\usepackage{mathrsfs}
\usepackage{amsmath}
\usepackage{amsthm}
\usepackage{amssymb}
\usepackage{graphicx}
\usepackage{esint}

\makeatletter

\newcommand{\noun}[1]{\textsc{#1}}

\theoremstyle{plain}
\newtheorem{thm}{\protect\theoremname}
\theoremstyle{definition}
\newtheorem{defn}[thm]{\protect\definitionname}
\theoremstyle{remark}
\newtheorem{rem}[thm]{\protect\remarkname}
\ifx\proof\undefined
\newenvironment{proof}[1][\protect\proofname]{\par
\normalfont\topsep6\p@\@plus6\p@\relax
\trivlist
\itemindent\parindent
\item[\hskip\labelsep\scshape #1]\ignorespaces
}{%
\endtrivlist\@endpefalse
}
\providecommand{\proofname}{Proof}
\fi

\@ifundefined{date}{}{\date{}}
\journal{xxxx}

\makeatother

\usepackage{babel}
\providecommand{\definitionname}{Definition}
\providecommand{\remarkname}{Remark}
\providecommand{\theoremname}{Theorem}

\begin{document}

\begin{frontmatter}{}

\title{\noindent Prediction Errors Drive UCS Revaluation and not Classical
Conditioning: Evidence and Neurophysiological Consequences}

\author[rvt]{Luca Puviani}

\ead{luca.puviani@unimore.it}

\author[focal]{Sidita Rama}

\author[rvt]{Giorgio Vitetta}

\address[rvt]{{\small{}Department of Engineering ``Enzo Ferrari'' , University
of Modena and Reggio Emilia, Via Vivarelli 10, int.1-41125 Modena
, Italy}}

\address[focal]{{\small{}MD at Local Health Unit of Modena, Via S.Giovanni del Cantone
23, 41121 Modena , Italy}}
\begin{abstract}
\noindent Nowadays, the experimental study of emotional learning is
commonly based on classical conditioning paradigms and models, which
have been thoroughly investigated in the last century. On the contrary,
limited attention has been paid to the revaluation of an unconditioned
stimulus (UCS), which, as experimentally observed by various researchers
in the last four decades, occurs out of classical conditioning. For
this reason, no analytical or quantitative theory has been developed
for this phenomenon and its dynamics. Unluckily, models based on classical
conditioning are unable to explain or predict important psychophysiological
phenomena, such as the failure of the extinction of emotional responses
in certain circumstances. In this manuscript an analytical representation
of UCS revaluation learning is developed; this allows us to identify
the conditions determining the \textquotedblleft inextinguishability\textquotedblright{}
(or resistant-to-extinction) property of emotional responses and reactions
(such as those observed in evaluative conditioning, in the nonreinforcement
presentation of a conditioned inhibitor, in post-traumatic stress
disorders and in panic attacks). Furthermore, an analysis of the causal
relation existing between classical conditioning and UCS revaluation
is provided. Starting from this result, a theory of implicit emotional
learning and a novel interpretation of classical conditioning are
derived. Moreover, we discuss how the proposed theory can lead to
the development of new methodologies for the detection and the treatment
of undesired or pathological emotional responses, and can inspire
animal models for resistant-to-extinction responses and reactions.\end{abstract}
\begin{keyword}
Amygdala\sep classical conditioning\sep conditioned inhibitor\sep
emotional learning\sep evaluative conditioning\sep misattribution\sep
prediction error\sep PTSD\sep resistant-to-extinction\sep UCS revaluation\sep
unconscious emotion. 
\end{keyword}

\end{frontmatter}{}

\clearpage{}

\newpage{}

\part*{Introduction\label{sec:Introduction} }

Emotions are critical for the environmental adaptation of individuals
and for their survival. In fact, emotions prepare any organism to
act in some specific ways without the need of experiencing a physical
effect that originates from a specific source of stimulation; for
instance, the perception of a snake evokes an innate reaction \cite{Flykt2007,Ohman1993,Ohman1993b}
before experiencing (or re-experiencing) a physical elicitation (e.g.,
a snake attack). This is due to the fact that the representation of
some stimuli and the associated emotional reactions are innate \cite{Flykt2007,Ohman1993,Ohman1993b},
being shaped by evolution.

The emotional system plays also an important role in many psychiatric
and psychological diseases, in decision making \cite{Damasio1996}
and in the clinical field (e.g., in drug treatment; \cite{Benedetti2008,Enck2008,Petrovic2002,Watson2009,Weimer2015}).
Moreover, experimental evidence suggests that subliminal emotional
stimulation (through emotional pictures) can improve sport or endurance
performance \cite{Blanchfield2014,Radel2009}, and that emotions influence
pain perception \cite{Roy2009,Wagener2009,Wiech2009}. This leads
to the conclusion that any influence at emotional level may substantially
affect human behavior and perception. 

Nowadays, the experimental study of emotions and of their related
phenomena is usually based on \emph{classical conditioning paradigms}
\cite{Pavlov1927}. Classical conditioning occurs when a \emph{conditioned
stimulus} (CS) is paired with an \emph{unconditioned stimulus} (UCS).
Generally speaking, a CS is a neutral/innocuous stimulus (e.g., a
sound or a neutral visual cue), whereas the associated UCS is a source
of stimulation (e.g., an electric shock or food). By repeated CS-UCS
pairings, a CS can come to elicit a \emph{conditioned response} (CR),
which is often similar to the \emph{unconditioned response} (UCR),
i.e. to the response directly elicited by the paired UCS \cite{Fanselow2005,Kim2006,Pavlov1927};
however, the repetitive presentation of the CS without that of the
paired UCS leads to CR \emph{extinction}, since the CS itself will
not be longer able to signal (or predict) the UCS.

Classical conditioning theory provides an explanation of the emotional
learning mechanism involved in repeated CS-UCS pairings. However,
this is not the only emotional learning mechanism, as originally understood
by \cite{Rescorla1974}. In fact, Rescorla noticed the difference
between the learning mechanisms involving two independent emotional
memories, one concerning CS-UCS pairing, the other one the UCS outcome
evaluation (and re-evaluation). This viewpoint has been shared by
other researchers (e.g., see \cite{Davey1989,Gottfried2004,Hosoba2001,Schultz2013}).
However, as far as we know, until now no mathematical (or quantitative)
model has been developed for the second type of emotional learning,
i.e. for \emph{UCS revaluation} (in other words, for the learning
and the re-valuation of an UCS outcome). For this reason, no quantitative
description can be given for many psychophysiological effects and
phenomena driven by the emotional system. In fact, this requires the
knowledge of a model able to porperly describe the dynamics of the
interactions between the emotional system and a primary stimulus (i.e.,
an UCS). 

In this manuscript we tackle the problems of developing a novel theory
and a new model for implicit emotional learning; we also lay foundations
for the development of new methods for the modulation of emotional
reactive responses. Our solutions to the above mentioned problems
are based on the analysis and the interpretation of various experimental
results acquired in different disciplines. In fact, we propose a unifying
framework that provides a comprehensive and coherent description of
the emotional learning system. 

The article begins by illustrating a multidimensional representation
of the emotional response to a given stimulus. The proposed representation,
which is supported by experimental evidence showing the involvement
of distinct and specific neuronal populations in emotional responses
\cite{Schultz2000,Schultz2006} (and is also supported by other experimental
results based on pharmacological conditioning \cite{Amanzio1999,Guo2010,Haour2005}),
accounts for the different types of emotional response and, in particular,
for \emph{active responses} (which are actively sustained by external
physical mechanisms, like pharmacological or mechanical stimulation),
\emph{reactive responses} (which are not sustained by any physical
mechanism, but ``self-instantiated'' by the emotional system) and
\emph{passive residual responses} (due to the excitatory residuals
from previous emotional elicitations; \cite{Zillmann1971,Zillmann1972}).
Moreover, in our representation plays an important role the \emph{source
attribution and misattribution} phenomena \cite{Anderson1989,Bryant2003a,Cotton1981,Jones2009,Uleman1987}
and the modeling of \emph{contrast effects} \cite{flaherty1982}.
We then develop a novel model describing the emotional response of
an organism elicited by a source of stimulation (i.e., an UCS). The
derivation of our model relies on the assumption that, similarly as
various mathematical models describing classical conditioning (e.g.,
Rescorla-Wagner model \cite{Miller1995,Rescorla1972} or \emph{temporal
difference }(TD)\emph{ models} \cite{Doherty2003,Schultz1997,Sutton1988,Sutton1990}),
or probabilistic (Bayesian) ``perception'' and ``action'' learning
models (i.e., the \emph{predictive coding }(PC) \cite{Friston2008}
and the \emph{active inference models} \cite{friston2009,friston2010}),
coding of emotional and behavioral responses involves the computation
of a specific \emph{error-signal}. In our work this error signal is
defined as the difference between the response expected from the considered
source of stimulation and the response actually perceived by the elicited
organism. This definition is equivalent to that adopted in TD and
PC models, and relies on experimental observations acquired in functional
imaging studies \cite{Berns2001,Doherty2003,Garrison2013}, or directly
measured in dopaminergic circuits (e.g., in the \emph{ventral tegmental
area}, VTA) or in other fear-related circuits \cite{Bray2007,Delgado2008b,Li2014,McNally2011,Schultz2000,schultz2000b,Schultz2006,Steinberg2013,Waelti2001}.
The relationship between the mechanisms of classical conditioning
learning and that of UCS revaluation learning is taken into consideration
next. In particular, we analyse the different mechanisms involved
in the encoding of a CS and of an UCS, and show how the two resulting
neural representations lead to specific different properties. Furthermore,
we show that a) what is observed during classical conditioning is
actually an interrelated effect of both associative and UCS revaluation
learning mechanisms, b) only the latter is driven by the emotional
system through the computation of emotional error signals (this important
result is supported by experimental evidence obtained from optogenetic
manipulations \cite{Gore2015,Redondo2014}). After showing how the
Rescorla-Wagner equation for classical conditioning can be analitically
derived from the proposed model for implicit emotional learning under
specific conditions, we propose a more complete model for Pavlovian
conditioning; in doing so, the stochastic Hebbian plasticity rule
\cite{Amit1994,Fusi2002,Fusi2007,Hebb1949,Soltani2006,Soltani2010}
is exploited. The proposed model is able to predict specific phenomena
which cannot be explained by the currently available classical conditioning
models (such as\emph{ the dependence of asymptotic responding on CS
intensity and US intensity}; \cite{Miller1995,Young1976}). Furthermore,
we discuss how the \emph{conditioned inhibition }\cite{Rescorla1969}
and some related phenomena (such as the failure of the extinction
of conditioned inhibition through nonreinforced presentations of the
inhibitor; \cite{devito1987,Harris2014}), which cannot be described
in terms of classical conditioning \cite{Harris2014,Miller1995},
can be quantitatively predicted by the proposed theory. Then, we analyse
the case in which the emotional system is elicited by a continuous
time-varying source of stimulation (e.g., a time-varying acoustic
stimulation, such as music; \cite{Koelsch2014}), and we show how
our model can be extended to provide some mathematical and neurophysiological
indications. Our model is then exploited to quantitatively describe
some (system-level) mechanisms leading to a resistance-to-extinction
of an emotional response (i.e., inextinguishability over successive
trials or over time) under certain conditions, and to illustrate some
mechanisms through which an emotional reactive response associated
with a primary stimulus can be artificially strengthened/mitigated.
This provides new insights on various psychiatric pathologies, like
panic attacks \cite{Meuret2006}, \emph{post traumatic stress disorder}
(PTSD; e.g., see \cite{book_traumatic2012,Parsons2013,Perusini2016}),
and psychological phenomena like \emph{evaluative conditioning} (EC)
\cite{Baeyens1992,Baeyens2005,Dawson2007a,DeHouwer2001,Gast2012,Gawronski2014,Hardwick2000,Hofmann2010,Hutter2013,Jones2009,Sweldens2010,Sweldens2014}
(which are known to be resistant to extinction \cite{Baeyens2005,Jones2009}).

Finally, we summarise our main findings and discuss the potential
implications of our results on future research on emotion processing.

\part*{Methods}

\section{Response representation\label{sec:Response-representation}}

This Section is organized as follows. In Section \ref{sub:Stimulus-induced-response:}
the emotional and non-emotional components of the \emph{central nervous
system} (CNS) response to a given stimulus are analysed and vector
representations for such a response are developed. Then, in Section
\ref{sub:Active-and-reactive} the active and reactive components
of an emotional response are introduced. Finally, the specific features
of the reactive response are discussed in Section \ref{sub:On-the-specificity-of-error}.

\subsection{Components and vector representation of a brain response\label{sub:Stimulus-induced-response:}}

Generally speaking, in a mammalian CNS the \emph{response} (i.e.,
the effect) elicited by a given stimulus consists of distinct \emph{components},
each involving a multitude of neuronal and receptor families; this
response (or effect) can be determined by hormonal, mechanical, acoustic,
pharmacological and/or other effects originating from the peripherical
systems or from the CNS itself. In the following each component of
a CNS response is associated with \emph{a specific neuronal }(e.g.,
dopaminergic, serotonergic, opiodergic, etc.)\emph{ population} \emph{within
a given brain region} (e.g., the population of dopaminergic neurons
in the \emph{ventral tegmental area}, VTA). For this reason, if $N$
distinct components can be identified in the CNS, the considered response
can be represented as a row vector $\mathbf{y}=[y_{1},y_{2},...,y_{N}]$
belonging to a $N$-dimensional space, called CNS \emph{space} in
the following; moreover, this vector can be expressed as the linear
combination of $N$ \emph{versors} (i.e., unit norm vectors) $\left\{ \mathbf{v}_{i};i=1,2,...,N\right\} $;
this set of versors, being associated with different neuronal populations,
form a complete basis $\mathcal{B}$ for the considered space. Then,
we have that

\begin{equation}
\mathbf{y}=\sum_{i=1}^{N}y_{i}\mathbf{v}_{i},\label{eq:rappresentaz_base}
\end{equation}
where $y_{i}$ is a real quantity representing the product between
the mean number of elicited neurons and their mean firing rates for
the $i$-th neuronal population (with $i=1,2,...,N$); consequently,
$y_{i}$ takes on a positive (negative) value if the response produces
an increase of (a decrease or inhibition of) the activity for the
$i$-th population, and is equal to zero whenever the response does
not involve any adjustment for the baseline activity of the population.
It is also important to point out that, generally speaking, the versors
forming the basis $\mathcal{B}$ are not orthogonal, since a specific
component can be (linearly or non-linearly, and directly or indirectly)
related with other components. This fact is exemplified by the dopaminergic
nigrostriatal population (associated with motor functions) and the
dopaminergic neurons within the mesolimbic system (associated with
motivation and reward functions); in fact, these populations are not
simply differentiated from an anatomic viewpoint and significant functional
interactions between them have been observed (e.g., see \cite{Wise2009}
and references therein). Furthermore, given a specific region, a neuronal
population can influence another population (e.g., the noradrenergic
neuronal population can interact with the dopamine neurons \cite{Paladini2004}).
From a neurochemical perspective, the motivation for the interdependencies
between different components is represented by the fact that a given
neurotransmitter may simultaneously interact with all the various
isoforms of its receptor on neurons that also are under the influence
of multiple other afferent pathways and their transmitters \cite{Goodman2006}.

The components forming the response elicited by a given stimulus can
be classified on the basis of different criteria. A fundamental criterion
consists in distinguishing \emph{emotional }components from \emph{non-emotional
}ones. In practice, the \emph{emotional} component of a given response
is due to all the emotional and motivational neuronal systems and
influences the mesocortical limbic structures contributing to approach
and avoidance behaviors \cite{Parsons2015}. On the contrary, the
\emph{non-emotional} components originate from all the neuronal systems
not belonging to the emotional/motivational systems (e.g., to the
nigrostriatal dopaminergic system or the sensorimotor system). Note
that, generally speaking, emotional and non-emotional components can
be interdependent; for instance, they can be causally related (e.g.,
\emph{physical} and \emph{emotional} pains are correlated) or indirectly
related (e.g., emotional components can influence the immune response
\cite{Barak2006,Benedetti2008,cacioppo2007,Leary1990,Steinman2004}).
This differentiation between emotional and non-emotional components
can be included in the vector model sketched above by identifying
an emotional sub-space within the CNS space, as exemplified by Fig.
1. 

\begin{figure}
\noindent \centering{}\includegraphics{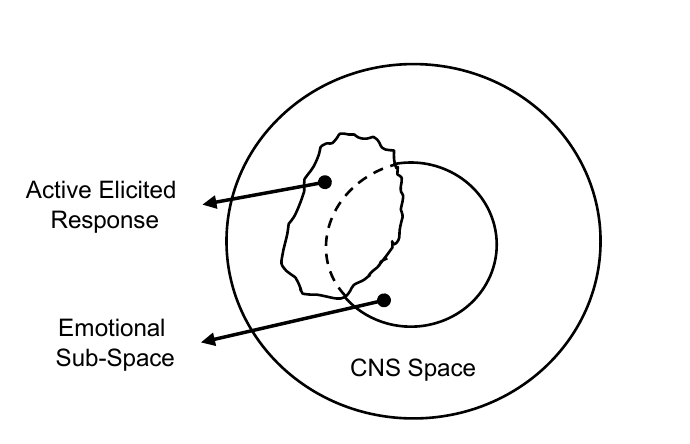}\caption{Bidimensional representation of the CNS space and the emotional sub-space.
A generic elicited response in the CNS space is also sketched. }
\end{figure}

Furthermore, the emotional (non-emotional) response components are
collected in the vector $\mathbf{y}_{em}$ ($\mathbf{y}_{ne}$) and
overall CNS can be represented as the concatenation of the two aforementioned
vectors, i.e. as

\begin{equation}
\mathbf{y}=[\mathbf{y}_{em},\mathbf{y}_{ne}].\label{eq:em-ne-components}
\end{equation}

Finally, it is worth noting that our response representation is suited
to properly model various experimental results evidencing that the
mammalian CNS is able to discriminate different emotional rewarding
components (e.g., different dopaminergic populations related to appetitive
or rewarding stimulations). In particular, some results illustrated
in \cite{Schultz2000,Schultz2006} have lead to the conclusion that
neurons can discriminate between cocaine and liquid rewards (or between
cocaine and heroin), possibly even better than between natural rewards,
and that in rats some neurons can be activated by lever pressing regardless
of drug injection, whereas other neurons by pressing a lever of a
specific drug or natural rewards only. This shows that every external
stimulus, such as food, a natural liquid reward or a rewarding drug,
is able to elicit in the rat CNS some specific emotional components
related to specific neuronal populations.

\subsection{Active and reactive emotional responses\label{sub:Active-and-reactive}}

In this manuscript the following definitions are adopted in relation
to a source of emotional stimulation.
\begin{defn}
\emph{\label{def:Source-of-emotional}Source of emotional stimulation}
\end{defn}
An \emph{emotional source of stimulation }is defined as any source
or primary stimulus able to elicit a response involving an emotional
or motivational component. \medskip{}

Note that the term ``primary stimulus'' is employed here to exclude
any secondary stimulus (i.e., CS \cite{Pavlov1927,Rescorla1972})
conditioned to a source stimulus through classical conditioning; nevertheless,
even a CS previously paired with a primary stimulus (i.e., an UCS)
is able to elicit an emotional response. In the following we assume
that a given UCS is always able to elicit an emotional component (e.g.,
an electric shock device or food), so that both the terms UCS and
emotional source of stimulation could be used indiscriminately. 
\begin{defn}
\emph{Active stimulation and active response\label{def:Active-stimulation}}
\end{defn}
An \emph{active emotional response} is defined as any emotional component
elicited by a primary stimulus through direct physical (e.g. mechanical,
chemical, pharmacological, etc.) mechanisms. Furthermore, the stimulus
elicitation is defined as \emph{active stimulation.} 

\medskip{}
It is worth mentioning that the adjective ``active'' refers to the
fact that the considered response is substained through an active
and physical action of the considered external source of stimulation. 

Real world examples of a stimulus eliciting an active response are
provided by a drug able to stimulate an emotional or motivational
component (through pharmacological mechanisms), a pain stimulus (originating,
for instance, from an electric shock delivery), or food.

Generally speaking, the active response can be represented as a non-linear
and time-varying function of the physical stimulation generating it
and of the internal physiological states (note that the time-variance
of this function accounts, for instance, for \emph{habituation} effect
or for \emph{receptor upregulation and downregulation} \cite{Goodman2006}).
Moreover, the active response may be influenced by various physical
features of the source stimulus, such as its frequency, rate of change
and magnitude of the elicitation.
\begin{defn}
\emph{Reactive stimulation and reactive response}
\end{defn}
A \emph{reactive emotional response} (or non-active response) is defined
as any emotional component ``self-induced'' by the perception of
a stimulus; such a response is not actively sustained by any external
physical (e.g. mechanical, chemical, pharmacological, etc.) mechanism.
Furthermore, the stimulus elicitation is defined as \emph{reactive
stimulation.} 

\medskip{}

It is worth noting that the adjective ``reactive'' refers to the
fact that the considered response is reactively eicited as the stimulus
is perceived and it is not substained through any direct physical
mechanism.

A real word example of a stimulus eliciting a reactive response is
provided by a CS previously paired with a primary stimulus; in this
case the elicited reactive response is called \emph{coditioned reflex}
\cite{Pavlov1927}. The non-active response can be elicited also through
the mere perception of a primary source of stimulation, such as a
threatening stimulus (e.g. a spider or a snake) or food. Such a source
of stimulation could be innate (e.g., represented by a biological/phylogenetic
fear-related threat, such as a spider, a snake or an angry face \cite{Flykt2007,Ohman1993,Ohman1994})
or learned (e.g., an ontogenetic source like a gun \cite{Flykt2007}). 

It is important to point out that, from an evolutionary perspective,
emotional reactive responses are fundamental for the survival of individuals
since they can elicit behavioral responses without the need for individuals
to phisically re-experiencing a given source elicitation (e.g., a
snake bite), or they act to turn individual attention to positive
valenced sources, such as food. For this reason, reactive responses
are learned through the emotional system in order to induce a proper
reaction whenever individuals face sources of stimulation or other
cues signalling a primary stimulus. As it will be discussed in more
detail in Section \ref{sub:amygdala-role}, reactive responses are
elicited through the amygdala, since, in the absence of an intact
amygdala, no reactive response or conditioned reflex can occur \cite{Bechara1995}.
It is also worth mentioning that a primary stimulus is able to elicit
both an active emotional response and a reactive emotional response;
on the contrary a secondary stimulus (i.e., a CS) can elicit a reactive
response only. 

Given the definitions and the considerations illustrated above, the
emotional vector $\mathbf{y}_{em}$ can be expressed as

\begin{equation}
\mathbf{y}_{em}=\mathbf{y}_{aem}+\mathbf{y}_{rem}\label{eq:2}
\end{equation}
where $\mathbf{y}_{aem}$ ($\mathbf{y}_{rem}$) represents the active
(reactive) portion of $\mathbf{y}_{em}$.

In the literature various results are available about the latency
and decay of emotional responses. First of all, the latency of an
emotional response is usually deemed negligible; however, its decay
is a time-consuming process for all nontrivial or pronunced emotional
states \cite{Bryant2003a,Hull1943,Scott2007,Zillmann1971}. In analysing
the decay of a response, it should be always kept into account that
any emotional response can be decomposed in two main \emph{functional
factors}, namely the \emph{dominantly neurally controlled factor}
and the \emph{dominantly humorally controlled factor}. The former
factor has a faster decay after the end of the stimulation, whereas
the latter one is relatively slower \cite{Zillmann1971} and can elicit
neural responses. For this reason, the different factors of both reactive
and active responses extinguish after their elicitation at different
speeds. For this reason, it can be always assumed that at least a
portion of both active and reactive responses extinguishes quite slowly
(e.g., this portion is related to hormones and neuromodulators decay).
Moreover, at a given instant a generic active or reactive response
can be in its \emph{active state} (e.g., instantiated, or during an
active elicitation) or in its \emph{passive state }(i.e., during the
decay interval following the end of an active elicitation).

The most important ideas illustrated in this Section can be summarized
in Fig. 2, which shows the different types of the emotional response
that can be elicited by a generic source of stimulation and the mathematical
notation adopted to denote them.

\begin{figure*}
\noindent \begin{centering}
\includegraphics{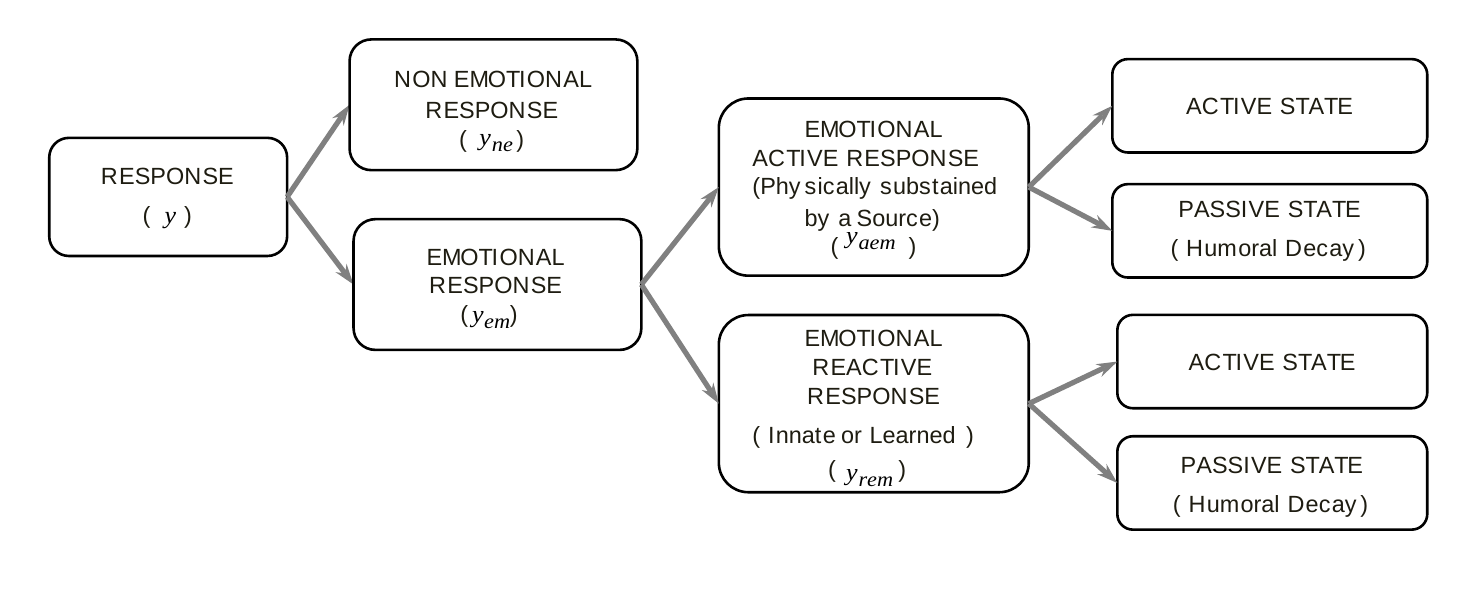}
\par\end{centering}

\caption{{\footnotesize{}Representation of the different types of CNS response
elicited by a generic source of stimulation.}}
\end{figure*}

\subsection{On the nature of reactive emotional responses\label{sub:On-the-specificity-of-error} }

As already stated in the previous Section, a generic CS can elicit
a reactive emotional response only, unless it also represents a primary
source of stimulation. For this reason, the observation of the responses
elicited by different CSs can unveil the specific properties of reactive
responses. Various results are already available about this issue
in the literature. In particular, it is well known that the \emph{unconscious/implicit}
placebo/nocebo effect can be interpreted as a conditioning process
in which a primary stimulus (e.g., a drug) become paired with a CS
(e.g., the substance administration or other cues) \cite{Benedetti2008,Colloca2008a,Colloca2008,Colloca2011,Colloca2010,Jensen2012,Jensen2014,Lui2010,Montgomery1997,Nolan2012,Watson2009,Williams2004}.
After some effective pairings, a successive presentation of the CS
(e.g., the administration of an inerte substance), in the absence
of any active physical/pharmachological stimulation, is able to elicit
a purely reactive response which mimics the previous active response
due to the primary stimulus. This claim is confirmed by various brain
imaging data, evidencing that placebos can mimic the effect of active
drugs and activate the same brain areas; this occurs for placebo-dopamine
in Parkinson's disease, for placebo-analgesics or antidepressants,
and for placebo-caffeine in healthy subjects (see \cite{Haour2005}
and references therein). Moreover, in \cite{Amanzio1999} it is shown
that \emph{pharmacological conditioning}, like conditioning with opioids,
produces placebo analgesia mediated via opioid receptors administering
saline infusions, and that, if conditioning is performed with nonopioid
drugs, other nonopioid mechanisms are involved, so that conditioning
activates the same specific neuronal populations as the primary stimulus.
Similar results and conclusions have been obtained in mice experiments
\cite{Guo2010}. This line of reasoning is also supported by a further
experiment \cite{Ito2000} showing that an increase in dopamine release
in the \emph{ventral striatum}, measured through microdialysis, are
observed not only when rats self administer cocaine (representing
the UCS in the considered experiment), but also when they are solely
presented with a tone (representing the CS) that has been previously
paired with cocaine administration. Furthermore, experimental verification
of the influence of nonconscious conditioned stimuli on placebo/nocebo
effects \cite{Jensen2012,Jensen2014} show that a reactive stimulus
is able to interfere with a given active stimulation (e.g., an active
drug or a painful stimulation), by increasing or decreasing the effect
of the active response. This suggests that common active and reactive
response components can be additive or competing and, hence, both
contribute to the determination of the overall elicited response.
The last observation is also supported by further experimental results
\cite{Roy2009,Wagener2009,Wiech2009} which show that emotional reactive
stimulations (e.g., the perception of emotional pictures or other
reactive stimuli) modulate pain perception. For these reasons, if
the active and the reactive elicited responses involve some common
components, these, in turn, add up in an algebraic sense (see Eqs.
(\ref{eq:2}) and (\ref{eq:rappresentaz_base})). A further support
of this conclusion come from the experimental evidences described
in \cite{Bryant2003a,Hull1943,Zillmann1971,Zillmann1972} which show
that, under some conditions, an emotional excitation can be transferred
to a successive independent source of stimulation, energizing it,
because of the residual excitation due to the incomplete decay of
the previous emotional elicitation. This phenomenon is called \emph{excitation
transfer} \cite{Zillmann1971,Zillmann1972}.

Additional results illustrated in \cite{Singer2004} reveal the differentiation
between emotional and non-emotional components, and reactive and active
responses in nociceptive stimulation (in this case the pain network
involves both affective or emotional components, and sensory components).
In particular, functional imaging experiments have evidenced that
both physically receiving a pain stimulus (i.e., an \emph{active}
stimulation) or observing a signal indicating that a loved person
- present in the same room - is receiving a similar pain stimulus
(i.e., a \emph{reactive} emotional pain-related stimulus mediated
through empathy) activates the bilateral \emph{anterior insula} (AI),
the rostral \emph{anterior cingulate cortex} (ACC), the brainstem
and and the cerebellum. These results have also shown that, on the
contrary, activity in the posterior insula/secondary somatosensory
cortex, the \emph{sensorimotor cortex} (SI/MI), and the caudal ACC
was specific to physically receiving pain (i.e., that non-emotional
components were exclusively due to the active response) \cite{Singer2004}.

All the results mentioned above evidence that a CS (or a generic reactive
stimulus) can elicit specific neural populations and that these populations
reflect or mimic the same systems elicited by the primary stimulus
paired with the conditioned CS; in other words, as far as the emotional
components are concerned, a \emph{reactive} response mimics the original
\emph{active} response. Moreover, active and reactive responses which
involve the same emotional components add up in algebraic sense. This
observation leads us to the following two important remarks: 
\begin{rem}
\label{rem:emotional_mimicking}
\end{rem}
A \emph{reactive} response can mimic, at least, the emotional sub-space
of a previously associated elicited \emph{active} response. 

 Nonetheless, we argue that a reactive response can certainly determine
also some non-emotional components provided that these are causally
related to the emotional ones (e.g., the emotional system can influence
the immune response \cite{Barak2006,Benedetti2008,cacioppo2007,Leary1990,Steinman2004},
or the dopaminergic mesolimbic system can interact with the nigrostriatal
dopaminergic system \cite{DeLaFuente2001,DelaFuente2002,Wise2009}).
Furthermore, we argue that even the \emph{humoral immune response
sub-space }(in particular the components of the CNS such as the hypothalamic-pituitary-adrenal
axis, HPA, or the sympathetic nervous system, SNS; \cite{cacioppo2007}),\emph{
}other than the emotional sub-space, can be mimicked from an active
response elicitation; this last assertion is supported by experimental
results related to conditioned immune response and pharmacological
conditioning with immunosuppressive drugs \cite{Benedetti2008,cacioppo2007,Vits2011}.
Note also that Remark \ref{rem:emotional_mimicking} is fundamental
on the description of the emotional dynamics (see Section \ref{sec:Quantitative-Analysis-of}).
\begin{rem}
\label{rem:Ii+Ie}
\end{rem}
If a \emph{reactive} and an \emph{active} responses involve the same
components, the \emph{reactive} response can strenghten/weaken (energize/inhibit)\emph{
}the\emph{ }emotional components of the primary \emph{active} \emph{response}
since they are qualitatively indistinguishable. Hence, Eq. (\ref{eq:2})
can be applied.

As it will become clearer in Section \ref{sec:Quantitative-Analysis-of},
this remark will play a fundamental role in the description of emotional
dynamics.

\medskip{}

One can argue that, in some cases, a reactive response learned from
its active counterpart does not mimic the original emotional response.
For instance, this occurs in \emph{aversive conditioning}, when an
electric shock is used as unconditioned stimulus; in particular, it
has been observed that heart rate decreases during the presentation
of a CS and, on the contrary, it increases when an active electric
shock is given \cite{Chance2008}. For this reason, at first glance
reactive and active responses seem to be qualitatively different as
they lead to distinct behaviors. Actually, a different interpretation
of these results can be formulated. In fact, it should not be forgotten
that the active response to a pain stimulus consists of two components
(one emotional, the other one non-emotional) and the non-emotional
component could result in a behavior substantially different than
that produced by the elicitation of the emotional component \emph{alone}.
Moreover, generally speaking, in the complex mammalian CNS, different
level of a specific neurotransmitter within a given brain region (determined
by the elicitation of a specific emotional component) could lead to
different (even opposite) behavioral observations \cite{Goodman2006}
(this last assertion will be discussed in more detail in Section \ref{sub:On-the-learning}).

\section{Error-driven learning\label{sec:Error-driven-learning}}

In this Section the basic mechanisms involved in emotional learning
are reviewed; in particular we focus on the role played by the amygdala
(and other important systems) in emotional learning and on the mechanisms
for the generation of the error signals on which such a learning is
based.

\subsection{On the role of the amygdala in emotional learning \label{sub:amygdala-role}}

In complex vertebrates the amygdala represents the core center in
the formation and storage of emotional events and in the elicitation
of emotional responses. In particular, it is well known that the amygdala
plays a fundamental role in encoding \emph{emotional memories}, in
\emph{fear responses} and in \emph{skin conductance response }(SCR)\emph{
classical conditioning}. In fact, various results are available in
the literature about the involvement of the amygdala in the \emph{acquisition
and encoding of relevant emotional memories} \cite{Cardinal2002,Davis1992,Namburi2015,Richardson2004,Uwano1995},
and about the fact that these mechanisms are based on its synaptic
plasticity \cite{Maren2005,Namburi2015,Sigurdsson2007}. In addition,
in a growing body of literature \cite{Amano2011,Choi2010,Glascher2003,paton2006,Sangha2013,Schoenbaum1999}
it is shown that amygdala is necessary for fear responses, and no
reactive fear responses are instantiated in the absence of an intact
amygdala \cite{Choi2010}. Finally, the analysis of patients with
a damaged brain has evidenced the importance of amygdala in SCR classical
conditioning. This emerges, for instance, from the results illustrated
in \cite{Bechara1995}, where the case of a patient with selective
bilateral destruction of his amygdala is analysed; in fact, this patient
shows an unconditioned SCR to a UCS, but no SCR conditioning to the
paired CS, although he is well aware of the CS-UCS relation.

All the results illustrated above lead to the conclusion that the
amygdala is necessary for the elicitation of an emotional reactive
response; note, however, that, if the amygdala is damaged, an active
elicited response (e.g., an unconditioned painful stimulus) can be
still elicited. Further results available in the literature evidence
that the amygdala mediates the emotional reaction, and that directly
and indirectly elicits emotional and motivational areas of human brain
\cite{Tovote2015,Gore2015,Janak2015}. In particular, on the basis
of information coming from the sub-cortical (i.e. thalamus) and cortical
pathways (from which perception and the representation of the features
of any source originates) the amygdala arouses both the cortex and
the emotional and motivational brain regions directly and indirectly
through different systems \cite{Sah2003,LeDoux2000}, that is the\emph{
nucleus accumbens} (NAcc), the prefrontal cortex, the midbrain, the
hypothalamus, the autonomic nervous system, the endocrine system and
others). In the following analysis we briefly refer to this group
of systems as \emph{system chain}. In fact, the amygdala sends projections
to a variety of systems, and it consists of several interacting subnuclei
that may provide specific individual contribution to the overall emotional
computation \cite{Killcross1997,LeDoux2000,Sah2003} (in particular,
different subnuclei of the amygdala can process and elicit specific
emotional components \cite{Balleine2006,Blundell2001,Corbit2005,Gore2015,Killcross1997}).
Research activities in this field have also evidenced that that the
representation of any UCS is stored within the \emph{basolateral amygdala}
(BLA) \cite{Gore2015,Redondo2014}. Note also that the amygdala mediates
both appetitive (i.e. rewarding) and aversive stimuli \cite{Amano2011,Gore2015,Muramoto1993,paton2006,Sangha2013,Schoenbaum1999,Shabel2009};
in the former case the BLA neurons project onto the NAcc, whereas
in the latter one onto the \emph{centromedial amygdala} (CeM) \cite{Namburi2015}.

\subsection{On the generation of error signals in emotional learning\label{sub:Error-signals-in_emotional_learning}}

Neurons in several brain structures appear to code specific signals,
that are called \emph{error signals} and, generally speaking, represent
the difference between a really experienced response and its expected
counterpart \cite{schultz2000b,Schultz2006}. In the literature a
number of results are available about the role played by specific
neuronal populations in coding error signals and the nature of such
signals. In particular the error signals are coded in relation to
rewards, punishments, external stimuli, and behavioral reactions \cite{Delgado2008b,schultz2000b,Waelti2001}.
In some cases, dopamine neurons, norepinephrine neurons, and nucleus
basalis neurons broadcast prediction errors as teaching signals to
large postsynaptic structures; in other cases, error signals are coded
by selected neurons in the cerebellum, superior colliculus, frontal
eye fields, parietal cortex, striatum, and visual system, where they
influence specific subgroups of neurons. In general, prediction errors
can be used in postsynaptic structures for the immediate selection
of behavior or for synaptic changes underlying emotional and behavioral
learning \cite{Garrison2013,schultz2000b}. Evidences of coding of
error signals during learning have been found in various neuroimaging
studies \cite{Berns2001,Doherty2003,Garrison2013}.

More specifically, as evidenced by a growing body of literature \cite{Bourdy2012,Delgado2008b,Schultz1998,Schultz2000,Schultz2006,Waelti2001},
in \emph{emotional learning,} populations of dopaminergic neurons
encode the error signal evaluating the difference between what is
expected (i.e., the expected reward) and what is really occuring;
furthermore, this error signal is exploited to correct and modulate
the individual's emotional and behavioral response. The error signal
computed in these dopaminergic regions can be positive or negative
and can drive appetitive or aversive emotional reactions \cite{Delgado2008b}. 

On the basis of all the above illustrated results, it can be stated
that error signals driving emotional responses are evaluated in different
brain regions, depending on the nature of the involved emotional components;
however, in emotional learning (and in the computation of the associated
error signal), a fundamental role is played by the \emph{orbitofrontal
cortex} (OFC) \cite{Doherty2007}, which represents a key structure
in coding and maintaining the representations of a stimulus response
(i.e., the representation of the expected outcome associated with
a stimulus) \cite{Doherty2004,Doherty2007}. In fact, various experimental
results have evidenced that the OFC generates information about expected
outcomes (e.g., see \cite{Takahashi2009} and references therein),
which are demeed critical in the computation of prediction errors;
these results are consistent with the relation between the reward-related
activity in OFC and VTA dopamine neurons \cite{Takahashi2009}. Experimental
results have also evidenced that, when OFC and midbrain data are juxtaposed,
anticipatory activity observed in the OFC is inversely related to
dopaminergic error signaling downstream \cite{stalnaker2015}. This
suggests that the error signals in other brain areas might depend
partly on OFC input for properly calculating the errors \cite{Schoenbaum2009,stalnaker2015}.
This idea has been partially confirmed for error signals in midbrain
dopamine neurons (at least in rats; see \cite{Takahashi2011}). The
OFC region has been also shown to respond to both appetitive and aversive
outcomes \cite{Morrison2011}, to integrate multiple sources of information
regarding outcome signals, to code outcomes and supervise the amygdala
(which, as already explained in the previous Section, is the key element
for both positive and negative-valence emotions) \cite{EdmundT.Rolls2008,Kennerley2011,wallis2007},
and to integrate cognitive information \cite{Plassmann2008} (also
coming from the dorsolateral prefrontal cortex \cite{Li2011}) in
its evaluation and coding of emotional outcomes \cite{Doherty2007}.
Finally, it is worth pointing out that OFC is not necessary for Pavlovian
conditioning, since this can be driven by prior experience; however,
it is certainly needed for modifying the response if the predicted
outcome is revaluated (i.e., UCS inflation and devaluation) \cite{stalnaker2015,Gallagher1999}.

\subsection{On error-based emotional learning\label{sub:On-error-based-emotional}}

On the basis of the experimental results summarised in the previous
Section, it can be assumed that, during emotional learning, an \emph{error
signal} is computed in different brain regions (e.g., the VTA), depending
on the nature of the elicited emotional response. In this process
the OFC plays an important role since it interacts with these brain
regions and codes the expected response to a given stimulus (i.e.,
the expected UCR). Moreover, the error signal which is broadcasted
to other brain regions including the OFC, may undergo processing and
fusion with other information (even at a higher cognitive level) and
is certainly sent to the amygdala, since this part of the brain is
necessary for the elicitation of an emotional reactive response, for
conditioning and for the storing process of emotional stimuli.

It is also important to point out that the extended network for computing
and coding the emotional error-signals (which we call \emph{error
computing distributed network}) can be represented as a distributed
system consisting of different sub-systems and, in particular, the
OFC, the \emph{prefrontal cortex} (PFC), the VTA, the midbrain, the
striatum and others; all these interact in an iterative fashion. However,
the study of these multiple neural interactions is behind the scope
of our analysis. For this reason, in the following we assume that,
whenever a stimulus elicits an emotional response, an error signal
is computed and transmitted to the amygdala for updating the corresponding
reactive emotional response; note that this error-driven mechanism
ensures adaptivity in emotional learning. As it will be shown in the
following sections, this mechanism can be quantitatively described
and analysed by representing the emotional learning system as a dynamic
system, characterised by memory elements (due to the amygdala and
to the OFC), fed by a time-variyng error signal, and generating the
emotional responses (i.e., the vector $\mathbf{y}_{em}$ in Eq. (\ref{eq:2})).

It is useful to mention that the ideas of relating brain learning
to an \emph{error-prediction} signal and of modelling this process
as a \emph{dynamic adaptive control system} have been already developed
in various theories. The most relevant example of this approach is
provided by Rescorla and Wagner in their theory about \emph{classical
conditioning}. In fact, in this case the human brain is assumed to
learn from a \emph{prediction error}, defined as the discrepancy between
a reference value and what is actually perceived by the considered
subject \cite{Miller1995,Rescorla1972}. In particular, learning occurs
through a mechanism that updates the expectations about the outcome
in proportion to a prediction error, so that, across trials, the expected
outcome converges to the actual outcome \cite{Rescorla1972}. A variant
of the Rescorla-Wagner theory is represented by the so called \emph{temporal
difference} (TD) \emph{learning} \cite{Doherty2003,Schultz1997,Sutton1988,Sutton1990},
which accounts for the time evolution of the response within each
trial. The goal of TD learning is providing a prediction, for each
instant $t$ in the trial during which a CS is presented, of the total
future reward to be gained in that trial from time $t$ to the end
of the trial itself \cite{Doherty2003}. 

A more recent theory (known as \emph{predictive coding} theory \cite{Friston2008})
formalizes the notion of the Bayesian brain, in which neural representations
in the higher levels of cortical hierarchies generate predictions
of representations in lower levels. These \emph{top-down }predictions
are compared with representations at the lower level to compute a
\emph{prediction error}. The resulting error-signal is passed back
up the hierarchy to update higher representations; this recursive
exchange of signals lead to the minimization (ideally the suppression)
of the prediction error at each and every level to provide a hierarchical
explanation for sensory inputs that enter at the lowest (sensory)
level. In the Bayesian jargon neuronal activity encodes beliefs or
probability distributions over states in the world that causes sensations
\cite{Friston2008}. In predictive coding theory the notion of precision
(or confidence, which is the inverse of the variance) of the error
signals is also formalised, and the mechanism through which the brain
has to estimate and encode the precision associated with the prediction
errors is explained. The prediction errors are then weighted with
their precision before being assimilated at a high hierarchical level.
Generally speaking, predictive coding assumes that organisms minimize
an upper bound on the entropy of sensory signals (the free energy),
which, under certain simplifying (Gaussian) assumptions is equivalent
to the prediction error. The minimization of the error-signal (at
the different hierarchical levels of a neural network) is generally
computed through a generalized gradient descent \cite{friston2010}.
Hence, predictive coding theory leads to model the brain activity
by means of a \emph{neural network} with multiple neuronal layers
(from the top high hierarchical level, to the bottom sensory level)
which is governed by the backpropagation algorithm \cite{bishop2006};
this algorithm leads to the minimization of the error signal, adjusting,
iteratively, the weights of the prediction errors at different network
layers (such weights correspond to the precisions). In \emph{machine
learning} and engineering fields artificial neural networks have been
extensively applied to a large number of learning problems \cite{bishop2006}
(for classification, regression or parameters estimation). Hence through
a neural network architecture, in principle, any nonlinear function
or distribution can be learned. Predictive coding theory and the above
mentioned concepts have been succesfully applied to perception \cite{Friston2008}
and motion (action) learning (the so called \emph{active inference}
theory; \cite{friston2009,friston2010}) to communication learning
\cite{Friston2015}; moreover, a predictive coding version of interoception
(i.e., a model for the autonomic, metabolic, immunological regulations
so, generally speaking, for homeostasis and allostasis, and, potentially,
even for emotional regulations) has been proposed in \cite{Barrelt2015,Pezzulo2015,Seth2013}.
Nonetheless, we will show (see Section \ref{sub:diff_interoception})
that the interoceptive predictive coding model, which could hold for
interoceptive visceromotor functions (i.e., homeostasis and allostasis),
cannot be directly applied to implicit emotional learning, since it
would lead to the instability of the emotional system.

\subsection{A new model for error-based emotional learning\label{sub:General-hypothesis}}

In the following an error-based mathematical model for describing
the dynamics of implicit emotional learning is developed. The main
features of the proposed model, and its similarities and differences
with the models mentioned at the end of Section \ref{sub:On-error-based-emotional}
are summarised below.

\emph{A model for primary stimulus learning} - The proposed model
describes the implicit learning of a UCS outcome (i.e., the UCS acquisition,
inflation and devaluation: the UCS revaluation) and, consequently,
does not immediately refer, unlike the classical conditioning models
(like the Rescorla-Wagner and TD models), to the learning of the associative
connection between a UCS and a paired CS \cite{Rescorla1974}. Further
details about this are provided in Section \ref{sub:On-the-learning},
where the relationship between these two learning mechanisms is analysed
in detail. 

\emph{Implicit emotional evaluation} - In the development of our model
it is assumed that no cognitive or suggestion processes occur during
the interactions between a subject and the considered UCS; for this
reason, our attention focuses on only \emph{implicit} (or automatic)
emotional evaluation (in other words, information such as verbal suggestions,
cognitive expectations, beliefs and so on, are avoided). Nevertheless,
the source of stimulation has to brought to the attention of the subject.
In particular it has to be evident that the UCS is responsible for
the response elicitation, so that the outcome will be correctly attribuited
to that source. This assumption exludes the situation in which the
outcome is misattribuited to another insignificant stimulus \cite{Bryant2003a,Cotton1981,Jones2009,Uleman1987},
or even not attributed (like in the case of hidden drug administration).
Furthermore, the stimulus remains unchanged in the sequence of trials.
Under such assumptions, without any loss of generality, it can be
assumed that the predicted response (i.e., the expected outcome) in
a given trial, for the given stimulus, will coincide with the response
experienced during the last source-subject interaction. Such an assumption
simplifies the following mathematical computations and does not lead
to any loss of the generality, since, generally speaking, one might
even assume that the predicted outcome converges during trials (or
during time) to the real experienced outcome through a learning mechanism.
In the latter case the convergence of the emotional response towards
a steady state will simply be slower.

\emph{Multidimensional nature of the error-signal and evaluation of
a single component} - As already mentioned in Section \ref{sub:On-the-specificity-of-error},
a mathematical representation based on multidimensional vectors should
be adopted to properly describe the error signal involved in emotional
learning and defined as the difference between the actually perceived
response and its expected counterpart; this is due to the fact that
multiple (and specific) emotional components could be elicited by
a primary stimulus. However, in the following we focus on the dynamics
of a \emph{single component} to ease the reading. This choice, however,
does not entail any loss of generality, since our model can be applied
to any component.

\emph{Indistinguishability between the emotional component of an active
response and the corresponding reactive learned response} - As illustrated
in Section \ref{sub:On-the-specificity-of-error}, a learned reactive
response mimics the emotional components of the active response elicited
by a primary stimulus. For this reason, we assume that the reactive
and the active responses add up, so generating the emotional component
$\mathbf{y}_{em}$ (see Eq. (\ref{eq:em-ne-components}) and Remark
\ref{rem:Ii+Ie}). 

\emph{The essential neural network and the related neural functions}
- The considered error signal is computed by a distribuited and dynamically
interconnected neural network whose master element is supposed to
be the OFC (see Sections \ref{sub:Error-signals-in_emotional_learning}
and \ref{sub:On-error-based-emotional}). This error-signal is transmitted
by the network to the amygdala, which, in turn, updates the emotional
reactive response associated with the representation of the primary
stimulus. The processing of the error signal inside the amygdala is
represented by the (unknown) \emph{amygdala function}, denoted $F_{A}\left(\cdot\right)$
in the following. 

As already discussed in Section \ref{sub:amygdala-role}, whenever
a primary stimulus is perceived, the amygdala elicits the reactive
response previously coded for that stimulus. In particular, the amygdala
is be able to arouse the cortex, and the emotional and motivational
brain regions directly and indirectly; this process involves the \emph{system
chain} described in Section \ref{sub:On-error-based-emotional} and
operating between the amygdala output and the emotional brain systems.
In the proposed model, the biological functionality of this\emph{
system chain} is described by the (unknown) \emph{system chain function
$F_{Ch}\left(\cdot\right)$}. Note that this function does not lend
itself to a simple description, since it is influenced by iterative
mechanisms involving neuronal and hormonal systems, and various brain
regions. In practice, $F_{Ch}(\cdot)$ represents the processing which
turns all the amygdala output signals into emotional and motivational
responses. These ideas are summarised in Fig. 3, which shows the functional
representation proposed for the implicit emotional learning system.

\begin{figure*}
\noindent \begin{centering}
\includegraphics{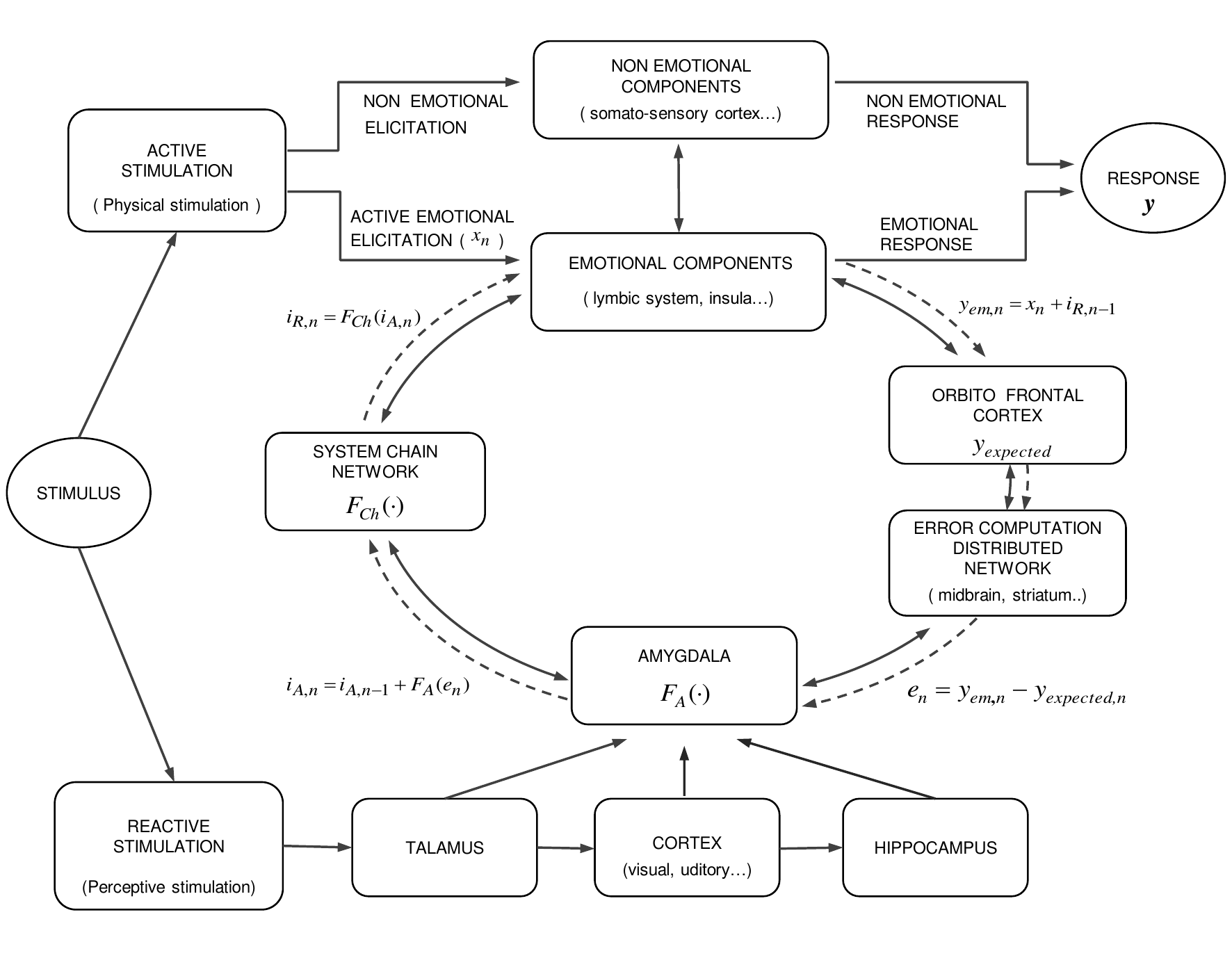}
\par\end{centering}

\caption{Discrete-time representation of the mechanisms on which the implicit
emotional learning system is based. The functional connections between
the involved actors and the processing task accomplished by each block
are shown. In particular, a stimulus could exert an active stimulation,
a reactive stimulation or both; in case of a reactive stimulation,
the stimulus is perceived and, successively, the amygdala elicits
the associated reactive response. The amygdala output is being processed
by the system chain network, which involve all the systems between
the amygdala output and the emotional/motivational brain regions (e.g.,
the nucleus accumbens, the sympathetic system, the hypothalamus and
others). The orbitofrontal cortex updates the expected response associated
to a given stimulus and drives the error computation network for the
computation of the emotional error signals. The error signals are
sent to the amygdala which updates the reactive responses associated
to the given stimulus. The processes involved in the discrete-time
emotional system are discuss in detail in the following sections.}
\end{figure*}

\emph{Functional approximations} - Generally speaking, the functions
$F_{A}\left(\cdot\right)$ and $F_{Ch}\left(\cdot\right)$ may change
over time, and are expected to be influenced by the internal physiological
states at the time the stimulus is encountered \cite{Toates1986,Cabanac1992,Dickinson1994,Balleine1998},
external circumstances and various features of the eliciting stimulus.
Moreover, it is expected that the amygdala function depends also on
the elicited sensory modality (e.g., auditory, visual, olfactory).
Finally, the behavior of these functions change from component to
component. However, in the following it is always assumed, for simplicity,
that these functions are static; consequently, the influence of all
the above mentioned factors and possible habituation effects are not
taken into consideration. Moreover, their dependence on the considered
emotional component is not highlighted by the adopted mathematical
notation since, as already mentioned above, we focus on a single emotional
component.

\section{A novel discrete-time dynamic model for the implicit acquisition
(and inflation) and devaluation (and extinction) of an emotional source\label{sec:Quantitative-Analysis-of}}

In this Section the discrete-time dynamic models for \emph{implicit
emotional learning} during the \emph{source acquisition} (and \emph{inflation})
and \emph{devaluation }\cite{Rescorla1974,Schultz2013,Gottfried2004}
(and extinction) are developed. In particular, the source acquisition
represents the process through which an emotional source of stimulation
(i.e., UCS) is detected and coded within the brain; the inflation
(devaluation), instead, represents the process through which an UCS
outcome (i.e., an UCR) increases (decreases) with respect to the previously
coded UCR. At the beginning of our study some additional hypotheses
needed in the development of our model (and complementing the hypotheses
illustrated in Section \ref{sub:General-hypothesis}) are listed and
properly motivated, and some details about the adopted mathematical
notation are provided.

\subsection{Additional hypothesis and mathematical notation\label{sub:specific_hyp_discrete}}

Unless explicitly stated, the following hypotheses hold in the development
of our model.

H.1 - \emph{Discrete trials} - Multiple trials in the interaction
between a source and a subject are considered; the trial duration
$\triangle T$ is assumed to be relatively small and, in particular,
negligible with respect to the \emph{inter-trial interval }(ITI) $T$.
For this reason, each trial can be ideally associated with a specific
point on the time axis and the corresponding emotional response can
be deemed constant, so that a \emph{discrete time scale }\cite{Bohner2001}
can be adopted in the representation of the considered phenomena.

H.2 - \emph{Residual response from previous trials} - The time constant
$\tau$ associated with the decay of the response elicited during
each trial is deemed negligible respect to the inter-trial interval
$T$; consequently, when a new trial takes place, the emotional response
due to the previous trials has already vanished.

H.3 - \emph{Novelty of the} \emph{source of stimulation} - The stimulus
eliciting the emotional response is assumed to be neutral before the
start of the trials (e.g., it is not a phylogenetic innate source
of stimulation). 

H.4 - \emph{Stimulus perception} - The perceived stimulus is the same
in each trial, so as the associated contextual information and boundary
conditions. This assumption states that, if a stimulus elicits a subject
during the first trial in a specific context (e.g., place, timing,
and specific boundary conditions), it has to be considered that the
\emph{stimulus perception} in the following trials involves exactly
the same contextual and boundary conditions. In the absence of such
an assumption the reactive response elicited by the stimulus perception
might be modulated by the different contextual information and boundary
conditions. For instance, the perception of a threatening stimulus
(e.g., a snake) at a short distance and without barriers should elicit
a reactive response stronger than that due to the same stimulus perceived
at a larger distance or in the presence of a separation barrier (i.e.,
with different boundary conditions and context).

H.5 - \emph{Response evolution} - During the process of source acquisition,
the emotional response increases monotonically over successive trials;
this assumption is motivated by the definition of acquisition and
consolidation of a source stimulus (hence, the extinction or de-valuation
processes are not taken into consideration in this case). On the contrary,
during the process of source devaluation or extinction, a monotonical
decrease of the emotional response is observed.

H.6 - \emph{Stability of the emotional system} - The emotional response
does not diverge (i.e., does not tend to infinity) as the number of
trials increases. 

In our analysis the following mathematical notation is adopted. The
emotional response and the active emotional component characterizing
the $n$-th trial are denoted $y_{n}$ and $x_{n}$, respectively
(note that these quantities correspond to the terms $\mathbf{y}_{em}$
and $\mathbf{y}_{aem}$, respectively, in Eq. (\ref{eq:2})), where
$n$ is the trial index (with \emph{$n=1,2,...$}). In principle,
the dependence of $y_{n}$ and $x_{n}$ (and that of the amygdala
function $F_{A}\left(\cdot\right)$ and of the corresponding system
chain function\emph{ $F_{Ch}\left(\cdot\right)$}; see the last part
of Section \ref{sub:General-hypothesis}) on the emotional space component
they refer to should be indicated; in the following, however, such
a dependence is omitted to ease the reading. Note also that H.3 entails
that
\begin{equation}
y_{0}=0,\label{eq:zero_instant}
\end{equation}
since the first trial corresponds to $n=1$, whereas H.5 leads to
the inequality
\begin{equation}
y_{n}\geq y_{n-1}\label{eq:ineq1}
\end{equation}
for the acquisition (and inflation) process and to
\begin{equation}
y_{n}\leq y_{n-1}\label{eq:ineq2}
\end{equation}
for the devaluation process (with $n\geq2$ in both (\ref{eq:ineq1})
and (\ref{eq:ineq2})). Finally, H.6 can be formulated as
\begin{equation}
\underset{n\rightarrow\infty}{\lim}|y_{n}|<\infty\label{eq:emotional_stability}
\end{equation}

\subsection{Quantitative analysis of source acquisition and inflation\label{sub:Quantitative-analysis-d-a}}

In the following analysis the considered active source of stimulation
(see Definitions \ref{def:Source-of-emotional} and \ref{def:Active-stimulation})
could be represented, for instance, by an electric shock device or
by a drug administration. We argue that the last mentioned case represents
the acquisition of an implicit or unconscious placebo response \cite{Benedetti2003,Benedetti2008,Colloca2008,Colloca2010,Siegel1975,Williams2004}. 

In the first trial (i.e., for $n=1$) the source stimulates the subject
for the first time, so eliciting the emotional response

\begin{equation}
y_{1}=x_{1}.\label{eq:state1}
\end{equation}
The \emph{error }(or \emph{difference}) \emph{signal }(defined the
as difference between the expected response and the really perceived
response; see (\ref{eq:zero_instant}) and (\ref{eq:state1}))

\begin{equation}
e_{1}\triangleq y_{1}-y_{0}=x_{1}\label{eq:error1}
\end{equation}
is evaluated in parallel. Then, this signal is transmitted from the
error evaluation network to the amygdala, which computes and stores
(within the amygdala itself) the amygdala reactive response 

\begin{equation}
i_{A,1}=F_{A}(e_{1})\label{eq:function_amygdala}
\end{equation}
this quantity is associated with the representation of the eliciting
stimulus (i.e., the acquired source of stimulation, UCS). Such a stimulus
representation is also stored within the amygdala and, in particular
in the BLA \cite{Redondo2014,Gore2015} (other related contextual
information are stored in the hippocampus \cite{Redondo2014}).

As already mentioned previously (see subsection \ref{sub:amygdala-role}),
after the \emph{acquisition process} described above, whenever the
subject perceives the source of stimulation, his/her amygdala elicits
the previously coded (and stored) reactive response $i_{A,1}$ (the
stimulus perception process has to be interpreted in the sense given
in H.4). At the level of emotional and motivational brain areas the
direct and indirect effects of $i_{A,1}$ result in the reactive response

\begin{equation}
i_{R,1}=F_{Ch}(i_{A,1}),\label{eq:Ii1}
\end{equation}
which depends on the full \emph{system chain} (i.e., on all the systems
elicited by the amygdala and their interactions). Note that $i_{R,1}$
(\ref{eq:Ii1}) is elicited independently of the fact that the source
stimulates actively (i.e., physically) the subject, because of its
reactive nature.

In the second trial, if the subject, after perceiving the stimulus
and having triggered the emotional reactive response associated with
it, is also physically stimulated, the emotional response is updated
on the basis of the active response $x_{2}$ elicited by the source
and the elicited reactive response $i_{R,1}$ (determined by the amygdala
in the previous trial, during which the emotional response has been
\emph{learned}). Then, we have

\begin{equation}
y_{2}=x_{2}+i_{R,1};\label{eq:y2-d-a}
\end{equation}
here $x_{2}$, which represents the response originating from the
physical interaction between the subject and the source stimulus in
the second trial, cannot be smaller than $x_{1}$, since the \emph{source
acquisition/inflation} is being considered (see H.5). After the response
$y_{2}$ has been elicited, the new error signal (see (\ref{eq:error1}))

\begin{equation}
e_{2}\triangleq y_{2}-y_{1}\label{eq:error2-d-a}
\end{equation}
encoding the difference between what was expected ($y_{1}$) and what
is experienced ($y_{2}$) is sent to the amygdala; this, in turn,
updates the stored emotional reactive response accordingly (note that
the stimulus remains unchanged over all the trials, so that in the
absence of cognitive or suggestion processes the expected value $y_{1}$
is exclusively due to the value implicitly stored at the end of the
previous source-subject interaction; see H.4). Moreover, the updated
reactive response, which is computed and stored within the amygdala,
is given by the sum of the previous reactive component with the increase
due to the last error signal $e_{2}$. For this reason, it is given
by

\begin{equation}
i_{A,2}=i_{A,1}+F_{A}(e_{2}).\label{eq:Iam2-d-a}
\end{equation}
The amygdala reaction $i_{A,2}$, in turn, reflects the emotional
response in the emotional/motivational brain areas, which is given
by

\begin{equation}
i_{R,2}=F_{Ch}\left(i_{A,2}\right).\label{eq:Ii2-d-a}
\end{equation}
Similarly, in the third trial, the response

\begin{equation}
y_{3}=x_{3}+i_{R,2}\label{eq:y3-d-a}
\end{equation}
is elicited. The last equation can be easily rewritten (see Eqs. (\ref{eq:function_amygdala}),
(\ref{eq:error2-d-a}), (\ref{eq:Iam2-d-a}) and (\ref{eq:Ii2-d-a}))
\begin{equation}
\begin{array}{c}
y_{3}=x_{3}+F_{Ch}\left(i_{A,2}\right)\\
=x_{3}+F_{Ch}\left(i_{A,1}+F_{A}(e_{2})\right)\\
=x_{3}+F_{Ch}\left(F_{A}(e_{1})+F_{A}(e_{2})\right)\\
=x_{3}+F_{Ch}\left(F_{A}(y_{1}-y_{0})+F_{A}(y_{2}-y_{1})\right).
\end{array}\label{eq:y3}
\end{equation}
Following the line of reasoning illustrated above leads easily to
the general expression 

\begin{equation}
y_{n}=x_{n}+F_{Ch}\left(\sum_{k=1}^{n-1}F_{A}\left(e_{k}\right)\right)\label{eq:yn-d-a}
\end{equation}
holding for $n\geq2$. It is important to point out that the sum appearing
in the right hand side of (\ref{eq:yn-d-a}) represents the storing
process within the amygdala (i.e., the emotional learning process
involving all the past experience); however, no quantitative result
can be inferred from (\ref{eq:yn-d-a}) in the absence of some information
about the structure of the functions $F_{A}\left(\cdot\right)$ and
$F_{Ch}\left(\cdot\right)$. As far as the last point is concerned,
in the following it is assumed that $F_{A}(0)=0$ and $F_{Ch}(0)=0$,
and that both $F_{A}\left(arg\right)$ and $F_{Ch}\left(arg\right)$
can be properly approximated as linear functions for small values
of $\left|arg\right|$. Consequently, the first order Taylor approximations

\begin{equation}
F_{A}(arg)\simeq\gamma\cdot arg\label{eq:approx-famygdala}
\end{equation}

\begin{equation}
F_{Ch}(arg)\backsimeq\Gamma\cdot arg\label{eq:approx-fchain}
\end{equation}

can be adopted, where $\gamma$ and $\Gamma$ denote the first derivatives
of $F_{A}\left(arg\right)$ and $F_{Ch}\left(arg\right),$ respectively,
evaluated at $arg=0$ ($\gamma$ and $\Gamma$ denote the reactivity
of the amygdala and the system chain, respectively, and take on real
values and, similarly as the parameter $\alpha$ defined below, depend
on the selected emotional component). Then, if the parameter

\begin{equation}
\alpha\triangleq\gamma\cdot\Gamma\label{eq:alfa}
\end{equation}
is defined, Eq. (\ref{eq:yn-d-a}) can be approximated as

\begin{equation}
\begin{array}{c}
y_{n}=x_{n}+\alpha\cdot\sum_{k=1}^{n-1}e_{k}=\\
x_{n}+\alpha\cdot\sum_{k=1}^{n-1}(y_{k}-y_{k-1}),
\end{array}
\end{equation}
which represents a first order approximation model. The last equation
can be reformulated, after some manipulations, as (see also (\ref{eq:zero_instant}))

\begin{equation}
\begin{array}{c}
y_{n}=x_{n}+\alpha\cdot\left(y_{n-1}-y_{0}\right)\\
=x_{n}+\alpha\cdot y_{n-1}
\end{array}\label{eq:final_d_a}
\end{equation}
which shows the recursive nature of the \emph{emotional dynamic learning
system}, described by a \emph{first order non-homogeneous model} \cite{May1976};
note that emotional stability (see H.6) requires the modulus of the
parameter $\alpha$ to be strictly less than unity \cite{May1976}
(i.e., $\left|\alpha\right|<1$).

From the last result and our previous assumptions it easily inferred
that a complete dynamic model describing emotional learning in an
approximate fashion is expressed by (\ref{eq:final_d_a}),

\begin{equation}
y_{n+1}>y_{n}\label{eq:second_eq}
\end{equation}
for any $n\in\mathbb{N}$ (with $y_{0}=0$) and

\begin{equation}
y_{n}=y_{n-1}\label{eq:third eq}
\end{equation}
when the error signal $e_{n}$ is equal to zero. 

Let us now analyse the implications of Eqs. (\ref{eq:final_d_a}-\ref{eq:third eq}),
we assume that, without any loss of generality, the active elicited
response is constant at every trial (e.g., the amplitude of the stimulus
is constant, so as the elicited active response), so that

\begin{equation}
x_{n}=X\label{eq:recursive-1}
\end{equation}
for $n\geq1$, where $X$ denotes a constant. Then, (\ref{eq:final_d_a})
turns into

\begin{equation}
y_{n}=X+\alpha\cdot y_{n-1},\label{eq:recursive-2}
\end{equation}
which can be interpreted in terms of eq. (\ref{eq:2}), since $X$
represents the active response, whereas $\alpha\cdot y_{n-1}$ the
associated reactive response ($i_{R}$). It is easy to show that,
in this case, the emotional response in the $n$-trial is given by

\begin{equation}
y_{n}=X\cdot\left(\frac{\alpha^{n}}{\alpha-1}-\frac{1}{\alpha-1}\right)\label{eq:solution-d-a-1}
\end{equation}
so that it approaches the asymptotic value

\begin{equation}
y^{*}=X\cdot\frac{1}{1-\alpha}\label{eq:fixed-point-d-a}
\end{equation}
as $n$ increases (in practice, after a few source-subject interactions,
$y_{n}$ closely approaches $y^{*}$). Then, substituting (\ref{eq:fixed-point-d-a})
in the right-hand side of (\ref{eq:recursive-2}) yields

\begin{equation}
y^{*}=X+\frac{\alpha}{1-\alpha}X\label{eq:solution-decomposition-d-a}
\end{equation}
which, once again, shows that the emotional response consists of an
active component ($X$) and a reactive component

\begin{equation}
i_{R}^{*}=\frac{\alpha}{1-\alpha}X\label{eq:asymptotic reactive}
\end{equation}
On the basis of the last results, it is not difficult to show that,
if $\alpha>0.5$, the emotional reactive response ($i_{R}$) becomes
greater than its active counterpart as the number of trial increases.
This phenomenon can occur, for instance, in a limited number of repeated
trials if the reactivity of the amygdala, at least for a specific
emotional component, is relatively strong (i.e., $\gamma$ takes on
a large value). In this case the acquired emotional reaction could
become very intense after some trials, even in the presence of a modest
active response due to a source physical stimulation. It is worth
mentioning that the amygdala reactivity could be increased, for instance,
by stress hormones (through direct and indirect mechanisms) \cite{Kim2002}.

Finally, it is worth mentioning that, if the source of stimulation
is a \emph{phylogenetic source} (i.e., a prepared biological and evolutionary
fear relevant stimulus coded in the mammalian amygdala since birth
\cite{Esteves1994,Flykt2007,Ohman1993,Ohman1994}), an emotional reaction
$Y_{0}$ can be natively coded and stored within the amygdala \cite{Ohman1994}.
For this reason, as soon as this source is perceived, an emotional
reactive response $Y_{0}$ could be elicited even in the absence of
previous source-subject interactions. In this case the dynamic model
for the increase (inflation) of the emotional response 

\begin{equation}
y_{n}=X+Y_{0}+\alpha\cdot\left(y_{n-1}-Y_{0}\right)\label{eq:phylogenetic-d-a}
\end{equation}
can be easily derived from (\ref{eq:final_d_a}).

\subsubsection{Theorem: On the necessity of both the \emph{reactive response} and
the \emph{expected (predicted) outcome} for the stability of the emotional
system\label{sub:Theorem:-On-the}}

It is important to point out that the reactive response determined
by the amygdala ($i_{R}$) and the expected outcome (i.e. to $y_{expected}$;
this, on the basis of our assumptions, can be considered equal to
the last occured outcome, i.e. to $y_{n-1}$) are both required in
order to ensure that the elicited response does not diverge if the
number of successive trials increases (in other words, the response
does not becomes arbitrarily large as the number of UCS stimulation
trials increases; see H.6).
\begin{proof}
\textbf{\emph{Proof by contradiction (reductio ad absurdum)\label{Proof-by-contradiction}}}
\end{proof}
\medskip{}

\emph{Hypothesis} 1: the reactive response associated with a stimulus
representation (i.e., with an UCS representation) coincides with the
expected (predicted) outcome, and the expected or predicted outcome
converges to the experienced outcome.

\emph{Hypotheses} H.1 - H.6 and the \emph{Remarks} \ref{rem:emotional_mimicking}-\ref{rem:Ii+Ie}
holds.

\medskip{}

Hypothesis 1 asserts that a unique reactive signal predicting the
UCS outcome exists, and that this signal coincides with the reactive
response elicited when the UCS is perceived; furthermore, the predicting
signal converges (by learning) to the actual experienced elicitation.
The last assumption has been formulated to include a more general
scenario than that considered in our initial assumptions, in which
the expected outcome coincides with the last experienced outcome.
From a mathematical viewpoint, the expected outcome can be computed
using any supervised learning method (or, alternatively, TD methods
\cite{Sutton1988}) in which the predicted outcome is evaluated on
the basis of the past $m$ predictions (i.e., of the predicted outcomes
in the last $m$ trials) and of the actual outcome, minimizing the
error between the prediction and the experienced outcome. Otherwise
it can be assumed that the predicted outcome coincides with the last
experienced outcome.
\begin{enumerate}
\item Let the UCS be an active source of stimulation (e.g., a painful stimulation
or a drug administration).
\item The UCS is perceived by a given subject on successive trials, then
it exerts an active elicitation (i.e., it elicits the active response
$X$). During the first trial the response is exclusively due to the
active UCS elicitation, that is $y_{1}=X$. After the first trial
(for instance, during the UCS perception in the second trial), the
predicted (reactive) response, called $y_{predicted,1}$, is computed. 
\item In the second trial, after the UCS perception, the predicted outcome
($y_{predicted,1}$) adds up to the successive UCS active elicitation,
so that the outcome can be expressed as $y_{2}=y_{expected,1}+X$.
Furthermore, since $y_{predicted,1}$ does not coincide with the actual
experienced outcome, the new prediction $y_{predicted,2}$ is computed
after the second trial; it can be easily proved that $y_{predicted,2}>y_{predicted,1}$
(since the experienced outcome has been strengthened and the error
signal has to be minimized).
\item In the third trial the experienced outcome can be written as $y_{3}=y_{predicted,2}+X$;
since $y_{3}>y_{2}\geq y_{predicted,2}$ a new value for the predicted
response is computed, called $y_{predicted,3}$, such that $y_{predicted,3}>y_{predicted,2}$.
\item In the $n$-th trial the outcome can be expressed as $y_{n}=y_{predicted,n-1}+X$;
it is easy to prove that $y_{n}>y_{n-1}\geq y_{predicted,n-1}$. Moreover,
if the number of trials tends to infinity, the outcome grows indefinitely
(i.e., $\underset{n\rightarrow\infty}{\lim}y_{n}=\infty$).
\item The last statement is absurd, as it contradicts hypothesis H.6. 
\end{enumerate}

\subsubsection{Quantitative analysis of source devaluation and exctinction\label{sub:Quantitative-analysis-d-e}}

Let us assume now that in the $n_{0}$-th trial (corresponding to
the beginning of the extinction process), the considered source does
cease to stimulate actively the subject (e.g., an inert drug is administered,
after that the asymptotic response expressed by Eq. (\ref{eq:fixed-point-d-a})
has been reached through the administration of an effective active
drug in the previous trials). In this case, Eqs. (\ref{eq:final_d_a})-(\ref{eq:third eq})
turn into

\begin{equation}
y_{n}=\alpha\cdot y_{n-1}\label{eq:recursive-3}
\end{equation}
for $n>n_{0}$,

\begin{equation}
y_{0}=Y_{0}\label{eq:second_eq-1}
\end{equation}
and

\begin{equation}
y_{n}=y_{n-1}\label{eq:third eq-1}
\end{equation}
when the error signal $e_{n}$ is equal to zero, respectively. These
formulas show that, even if the active stimulation drops to zero,
the corresponding reactive response, which depends on the previous
source-subject interactions, does not vanish abruptly, but exhibits
a decay rate depending on the value of the parameter $\alpha$. In
particular, if $n_{0}=0$ and $Y_{0}$ denotes $y_{0}$ (note that
(\ref{eq:zero_instant}) does not hold in this case), from (\ref{eq:recursive-3})
it is easily inferred that

\begin{equation}
y_{n}=Y_{0}\cdot\alpha^{n}\label{eq:solution-d-e}
\end{equation}
This result shows that, if $\left|\alpha\right|<1$, $y_{n}$ asymptotically
tends to zero in the absence of an active elicitation.

\subsection{Emotional response dynamics in discrete time scale\label{sub:Quantitative-analysis-of}}

The mathematical results derived in Sections \ref{sub:Quantitative-analysis-d-a}
and \ref{sub:Quantitative-analysis-d-e}, taken together, can be employed
for evaluating the emotional response to an arbitrary pattern of physical
source elicitation, that is in the general case of \emph{source revaluation
}\cite{Rescorla1974,Schultz2013,Gottfried2004}. Hence, the \emph{emotional
adaptation,} on the basis of the variability of the phisical stimulation
of a given stimulus (UCS) during successive trials, can be quantitatively
described. 

Moreover, we argue that, in investigating the emotional response in
the presence of an arbitrary active stimulation, it is important to
understand under which conditions the equality (\ref{eq:third eq})
holds, i.e. the error signal is equal to zero (excluding, of course,
the trivial cases in which the source is extinguished, i.e., $y_{n}=0$,
or the response reaches its asymptotic value (\ref{eq:fixed-point-d-a})).
In fact, this should allow to acquire a deeper understanding and a
quantitative description of diverse psychophysiological phenomena,
like particular forms of placebo/nocebo effects, evaluative conditioning
phenomena, and some psychiatric and psychological disorders, like
panic attacks, \emph{post traumatic stress disorders} (PTSD) and others.
Further details about this issue are provided in the next Sections
(\ref{sub:Evaluative-conditioning} and \ref{sub:saturation}).

\subsection{On the impact of excitation decay\label{sub:Discrete-trials-with-decay}}

The proposed dynamic model can be modified to account for the\emph{
excitation decay }\cite{Bryant2003a,Hull1943,Scott2007,Zillmann1971},
because of which a certain time (denoted $\tau$) is needed to dissipate
an elicited response. In practice, this phenomenon becomes relevant
through successive trials when the inter-trial interval $T$ is relatively
small with respect to the time constant $\tau$ characterizing the
dominantly humorally controlled factor of the emotional response (and,
consequently, H.2 does not hold). In fact in \cite{Zillmann1971}
it is shown that, under some conditions, an emotional excitation can
be transferred even to a successive independent source of stimulation,
because of the residual excitation due to the incomplete decay of
the previous source stimulation. This phenomenon is called \emph{excitation
transfer} \cite{Zillmann1971,Zillmann1972}. In these conditions,
the response aroused in each trial is due to both the actual (both
physiological/active and reactive) elicitation and to the residual
excitation from the previous trial; moreover, the dissipation rate
of the emotional response (i.e., the value taken on by the parameter
$\tau$) is influenced by the intensity of the aroused response, intervening
distractions, fatigue \cite{Bryant2003a} and other factors \cite{Hull1943}.

If an exponential decay is assumed for the dominantly humorally controlled
response, the recursive equation

\begin{equation}
y_{n}=X+y_{n-1}\cdot\left(\alpha+\xi\cdot\exp\left(-T/\tau\right)\right),\label{eq:decay}
\end{equation}
can be derived for the evaluation of $y_{n}$ (following the same
line of reasoning as that adopted for the derivation of Eq. (\ref{eq:recursive-2}));
here, $\xi$ is a positive parameter representing the fraction of
the emotional response decaying according to an exponential law (consequently,
$0<\xi<1$). From the last equation the stability condition (i.e.,
the condition ensuring that the emotional response does not diverge
as the number of trials increases; see H.6 in Section \ref{sub:specific_hyp_discrete})

\begin{equation}
\alpha+\xi\cdot\exp\left(-T/\tau\right)<1\label{eq:new_stability-1-1}
\end{equation}
can be easily inferred. Moreover, on the basis of eq. (\ref{eq:decay})
it can be easily proved that the final emotional value associated
with the source of stimulation increases as $T$ gets smaller. In
some cases, however, the ITI could change over the sequence of trials.
For instance, a ``fast'' emotional acquisition, resulting from a
group of some very close trials, could be followed by another group
of trials characterized by a larger ITI. In this case, if the second
group starts only after the end of the excitation decay of first group,
then the response will naturally decrease according to Eq. (\ref{eq:final_d_a})
until the asymptote expressed by (\ref{eq:fixed-point-d-a}) is reached.
Finally, we note that, from a quantitative perspective, the effects
of an excitation decay in close trials could be illusorily perceived
as a larger value for the parametere $\alpha$ as long as the trials
are temporally close to each other.

\subsection{On the inclusion of contrast effects in the discrete-time model \label{sub:contrast-effect}}

In the literature it is well documented \cite{flaherty1982,papini1997}
that surprising reward omissions, that is, the absence or reduction
of an expected reward, are accompanied by aversive emotional reactions,
at least in mammals \cite{papini1997}. On the contrary, surprising
increases in the expected reward result in an appetitive emotional
reaction. In particular, positive and negative \emph{contrast effects},
arising from unexpected shifts in the obtained reward (whose value
is greater or smaller than that previously experienced), depend on
the comparison of the sensory property of the present stimulus with
information stored in memory \cite{Genn2004} and lead to an emotional
response overshoot or undershoot, which is independent from the absolute
value of the real reward. For instance, in \cite{Genn2004} it is
shown that rats, in the presence of a shift from 32\% to a 4\% of
the administred sucrose solution, displayed a successive negative
contrast (i.e., a \emph{depression effect} \emph{\cite{flaherty1982}})
by initiating significantly fewer bouts of licking than control rats
maintained on 4\% sucrose. Furthermore, no significant increase in
the dopamine efflux in the NAcc was observed during the consumption
of 4\% sucrose by rats that experienced the shift from 32\%; on the
contrary, the consumption of 4\% sucrose by control rats was accompained
by a significant increase in the DA efflux in the NAcc.

Generally speaking, the emotional ``overshoot'' experienced during
positive contrast is termed \emph{elation effect; }instead, the ``undershoot''
experienced after a negative contrast is termed \emph{depression effect}
\cite{flaherty1982}\emph{. }

Contrast effects can be interpreted in terms of emotional responses,
as indirectly suggested by the effects of drugs on contrast \cite{flaherty1982}.
In fact, experimental data reveal that drugs having anxiolytic effects
on humans (e.g., amobarbital, ethanol, and benzodiazepines) tend to
reduce negative contrasts. Interestingly, the barbiturate drug reduces
negative contrast, but does not have any effect on positive contrast
\cite{flaherty1982}. The hypothesis according to which emotional
responses are involved in contrasts is also supported by the experiments
reviewed in \cite{flaherty1982} and showing that an increased release
of adrenocorticosteroid hormones is detected in the presence of negative
contrasts; this proves that a negative contrast is able to activate
a component of the sympathetic response to stress. In addition, the
responses evoked by negative contrasts are often characterized by
a long duration and sometimes do not dissipate by the end of the experiment
\cite{flaherty1982}. 

Experimental evidence also shows that contrast exhibits an inverse
dependence on the \emph{inter trial interval} $T$ and a direct dependence
with the magnitude difference between the preshift and the postshift
values (in other words, it is proportional to the error signal, defined
as the difference between the expected outcome and the perceived outcome).
For this reason, prior experience (e.g., prior trials) with the source
of stimulation determining the expected value (or outcome) plays a
fundamental role in determining contrast effects. 

Given the empirical results illustrated above, we argue that contrast
effects can be included in the proposed model for implicit emotional
learning by adding a new function, called \emph{contrast function}
and denoted \emph{$C(e_{A};T)$}; this function exhibits a nonlinear
dependence on $T$ and on the \emph{actual error-signal}, defined
as

\emph{
\begin{equation}
e_{A,n}\triangleq\left(x_{n}+\alpha\cdot y_{n-1}\right)-y_{n-1}
\end{equation}
}for the $n$-th trial; note that this definition is motivated by
the fact that the error signal refers to the present trial (instead
of the previous one), since contrast effects occur in parallel with
the actual outcome. Consequently, the emotional response during the
$n$-th trial can be evaluated as (see Eq. (\ref{eq:final_d_a}))

\begin{equation}
y_{n}=x_{n}+\alpha\cdot y_{n-1}+C\left(e_{A},T\right)\cdot e_{A,n}\label{eq:acq-with-contrast}
\end{equation}
if $e_{A,n}\neq0$ and

\begin{equation}
y_{n}=y_{n-1}
\end{equation}
if $e_{n}=0$ and $e_{A,n}=0$

The following properties can be reasonably assumed for\emph{ }$C(e_{A};T)$:
a) \emph{$C(e_{A};T)\cong0$} if $0\leq e_{A}\leq T_{A}$, where $T_{A}$
is a proper threshold; b) \emph{$C(e_{A};T)\cong K$} (where $K$
is a positive constant) if $e_{A}>T_{A}$ for a fixed $T$; c) \emph{$C(e_{A};T)$}
is inversely proportional to the ITI (i.e.,$C(e_{A};T)\propto1/T$).
Property (a) derives from the fact that no contrast effect is expected
for a relatively small error signal; (b) is based on a first order
approximation and (c) comes from empirical observations \cite{flaherty1982}. 

The quantities $T_{A}$ and $K$ have to be experimentally estimated.
Moreover, it is reasonable to assume that they could depend on the
specific emotional component elicited during the stimulation (e.g.,
the dopaminergic neuronal population in the NAcc).

It is not difficult to show that a simple continuos function approximately
satisfying the above mentioned conditions is 

\begin{equation}
C(e_{A};T)=\frac{K}{1+e^{-(e_{A}-T_{A})}}\cdot\frac{1}{T}\label{eq:contrast-function}
\end{equation}

This can be approximated by the linearized model

\begin{equation}
C(e_{A})\simeq K\cdot e_{A}\label{eq:approximated-contrast-d}
\end{equation}

If Eq. (\ref{eq:approximated-contrast-d}) is adopted to model the
contrast effect, an unexpected UCS elicitation (i.e., an active UCS
stimulation which is not signalled by a CS nor by the UCS perception,
such as, for instance, a permanently-connected electric shock device
elicitation) determines the response

\begin{equation}
y_{UCS}=X+K\cdot X,\label{eq:UCS-unexpected}
\end{equation}
and is attributed to the UCS. Furthermore, depending on the value
expected for the UCS before the unexpected elicitation, the error
signal is computed and the reactive response associated with the UCS
is updated accordingly; in particular, if the expected response before
the unexpected elicitation is equal to $X+\alpha X$, the error signal
becomes $e=X\cdot(K-\alpha)$. Moreover, if another unexpected UCS
elicitation occurs, the resulting error signal is equal to zero since
the expected outcome (which coincides with the last outcome) is now
equal to the actually experienced outcome, which is given by $X+K\cdot X$
(i.e., the active elicitation and the contrast contribution due to
the unexpected elicitation). This mathematical result is important
because shows that a series of trials of unexpected UCS elicitations
lead to an error signal different from zero only during the first
unexpected elicitation, in fact, in the successive unexpected trials
this leads to a static situation in which the error signal remains
equal to zero and a constant reactive contribution (i.e., $K\cdot X$)
due to the contrast effect is elicited. 

Moreover, if the Eq. (\ref{eq:approximated-contrast-d}) is adopted
to model the contrast effect and if it is assumed that $0<K<1$, it
is easy to demonstrate that Eq. (\ref{eq:acq-with-contrast}) becomes

\begin{equation}
y_{n}=\left(1+K\right)\cdot x_{n}+\left(\alpha+K\alpha-K\right)\cdot y_{n-1}.\label{eq:acq-contrast-d-approx}
\end{equation}

Furthermore, if it is assumed that $x_{n}=X$ for every trial (see
Eq.(\ref{eq:recursive-1})), the asymptotic solution for the Eq. (\ref{eq:acq-contrast-d-approx})
coincides with the one obtained in the absence of the contrast effect
(see Eq. (\ref{eq:fixed-point-d-a})). This last result is motivated
by the fact that during successive trials the contrast effect decreases,
since the expected outcome (which is signalled by the UCS perception
before the UCS active elicitation) approaches the experienced outcome
(i.e., both the error signal and the actual error signal tend to zero
over successive trials). Note also that the contrast effect could
be negative, for instance, if the experienced outcome is smaller than
its expected counterpart a response inhibition occurs. For this reason,
if the active elicitation $x_{n}$ drops to zero, the response decreases
faster than in the case in which no contrast is considered, since
the term $\left(\alpha+K\alpha-K\right)$ in the second side of Eq.
(\ref{eq:acq-contrast-d-approx}) is smaller than $\alpha$ (see Eq.
(\ref{eq:final_d_a})).

Finally, we argue that, if the \emph{contrast effect} and the effect
of the \emph{excitation decay} are included in our discrete-time model
(see also the classical conditioning model in Section \ref{sub:Derivation-of-the}),
such a model should also be able to justify the \emph{spontaneous
recovery effect }\cite{Miller1995}, which could occur after a classical
conditioning extinction. This phenomenon, which consists in the possibility
of experiencing a conditioned response some time after a complete
conditioned extinction, cannot be described in terms of the Rescorla-Wagner
model \cite{Miller1995}. Moreover, we argue that the passive residual
response due to a contrast effect (e.g., an inhibitory response due
to the lack of an expected UCS elicitation) is able to counteract
the effect of a residual conditioned Pavlovian response, which, in
turn, it can not be detected. For this reason, after the dissipation
of a contrast passive response, the conditioned reflex can be observed
again. Our viewpoint is partially supported by the fact that spontaneus
recovery is stronger when conditioning extinction occurs through massed
trials (i.e., in the presence of a small ITI, which are known to enhance
contrast effects) and weaker when extinction occurs through widely
spaced trials (i.e., in the presence of a larger ITI which determines
a smaller contrast effect) \cite{Urcelay2009}.

\section{Misattribution of a source of stimulation and evaluative conditioning\label{sec:Misattribution-of-a}}

When a source of emotional stimulation elicits a subject, the process
of encoding emotional memory starts; this involves various interactions
between the amygdala and the hippocampus \cite{Richardson2004}. This
encoding results in the processing and storage of different pieces
of information, such as contextual information, the elements determining
internal states and the elicited response. Generally speaking, the
hippocampus (and, in particular, the \emph{dentate gyrus}, DG) encodes
contextual information, whereas the BLA encodes emotional valence
and unconditioned stimulus representation \cite{Gore2015,Redondo2014}.
The encoding of emotional memory requires that the source of stimulation
is correctly detected and attributed. When an emotional response due
to a source of stimulation is attributed to a wrong source, an event
of \emph{source misattribution }occurs (e.g., see \cite{Bryant2003a,Cotton1981,Jones2009},
and references therein). Note that misattributions may result either
from conscious, accessible and measurable controlled processes, or
from spontaneous, inaccesible, automatic processes \cite{Anderson1989,Uleman1987}.
In the last case this phenomenon is called \emph{implicit misattribution}
(e.g., see\emph{ }\cite{Anderson1989,Hutter2013,Uleman1987}).

A quantitative description of the source misattribution phenomena
can be developed on the basis of our dynamic model; moreover, the
new results about this topic shed new light on the problem of \emph{evaluative
conditioning} (EC). Both these issues are illustrated in the remaining
part of this Section.

\subsection{Source misattribution: a quantitative analysis\label{sub:Source-Misattribution:-quantitat}}

Let us focus again, like in the previous Section, on multiple trials
of the interaction between a given aource and a subject and assume
a discrete-time scale in our analysis. In a generic trial one of the
following mutually exclusive events might happen: 1) the elicited
response is correctly attributed to the source of stimulation, so
that the emotional reactive response is computed and coded according
to Eq. (\ref{eq:recursive-2}); 2) no source of stimulation is identified
and, consequently, no reactive response is encoded and associated
with the source; 3) the elicited response is misattributed to another
(others) source(s) of stimulation. In the following we focus on the
last event and assume, without any loss of generality, that the misattributed
source of stimulation is initially \emph{neutral} (i.e., it does not
elicit an active or a reactive emotional response). Actually, this
event encompasses the following three mutually exclusive cases: 

a) The misattribution occurs in the presence of an active response;
for instance, this occurs when an hidden active source of stimulation
(e.g., the hidden administration of a given drug able to elicit an
emotional component) is misattributed to another insignificant source
of stimulation. 

b) The misattribution occurs in the presence of a residual (i.e.,
passive) response decay only (in other words, no active or reactive
responses are elicited); in this case the misattribution trial follows
the elicitation trial and occurs during the excitation decay.

c) The misattribution occurs when a purely reactive source of stimulation
is eliciting the subject, so that the associated response is purely
reactive.

If the misattributed source of stimulation is not neutral but elicits
a response, the response elicited during the misattribution process
will result from the superposition of the actual source response with
the previous non-attributed emotional state \cite{Zillmann1971}.
For this reason, in this case the previous non-attributed emotional
state ``energizes'' the actual source. 

In the following the above mentioned three cases are analysed in the
framework of the dynamical model developed in the previous Section;
moreover, it is assumed that misattribution always occurs in the first
trial and that the previous reactive response is equal to zero (i.e.,
$y_{0}=0$).

\subsubsection*{Case a): Misattribution occurring in the presence of an active response.}

In this case, the response in the first trial is only due to the active
component (i.e., to $X$; see Eq. (\ref{eq:recursive-2})) elicited
by a non-attributed source of stimulation (e.g., an hidden administered
drug). Consequently, the error signal in the first trial, denoted
$e_{1}$, is equal to $X$ and the corresponding reactive component
(expressed by $\alpha\cdot X$) is coded for the new misattributed
source of stimulation. If the original (i.e., true) source of stimulation
ceases to actively elicitate the subject in the following trials (i.e.,
the physical active component becomes zero), a negative error signal
is computed. In this case, after repetitive expousures of the misattributed
source without active elicitation, the emotional response asymptotically
tends to zero (according to a mathematical law similar to that expressed
by Eq. (\ref{eq:solution-d-e}).

\subsubsection*{Case b): Misattribution occuring in the presence of a decaying residual
response only.}

This case is known in literature as \emph{transfer paradigm} (described
in the Hullian \emph{drive theory} \cite{Hull1943}) or \emph{excitation
transfer} \cite{Bunce1993,Zillmann1971,Zillmann1972}, and refers
to the influence of a prior episode of arousal on subsequent emotional
responses. In this case, since environmental cues about the actual
source of a residual arousal are missing, such an arousal is misattributed
to a subsequent stimulus; this may result in an intensification of
the subject's emotional response to the new stimulus (see \cite{Bunce1993}
and references therein). In this case, in the first trial the registered
response is due to a passive (residual) response only, which is denoted
$\xi$ in the following. Consequently, the error $e_{1}$ in the first
trial is equal to $\xi$ and the reactive response stored for the
misattributed source is equal to $\alpha\cdot\xi$. Moreover, if the
subject is elicited by the misattributed source in a successive trial,
the corresponding reactive response is given again by $\alpha\cdot\xi$,
which is smaller than what was expected for that source (i.e., $\xi$),
so that a negative error signal is computed. Therefore, if further
trials occur, the emotional response asymptotically tends to zero,
similarly as in the previous case.

\subsubsection*{Case c): Misattribution occurring in the presence of a purely reactive
source of stimulation eliciting the subject.}

In this case, the response in the first trial (when the misattribution
is occuring), called $i_{U}$, is due to a purely reactive response
elicited by another unrevealed source (e.g., a subliminal emotional
stimulation \cite{Esteves1994,Glascher2003,Mayer1999}). Hence, the
response attributed to the new source (because of the misattribution)
is equal to $i_{U}$, which, in this case, is exactly what the amygdala
is eliciting, so that the error $e_{1}$ in the first trial is equal
to zero (i.e., the elicited response coincides with the expected response).
Hence, during the second trial, the expousure to the new source determines
the reactive response elicited by the amygdala during the first trial
($i_{U}$); furthermore, since the contribution due to the error signal
is equal to zero, the overall response remains equal to $i_{U}$.
For the same reason, the response remains constant during the successive
trials. These results, that hold if habituation or \emph{mere exposure
}phenomena \cite{zajonc_mere_2001} are neglected, show that an inextinguishable
(i.e., not vanishing through repetitive trials of source perception)
reactive response can be obtained through a complete misattribution
of a purely reactive emotional response. Note that, generally speaking,
a source of stimulation is expected to elicit an active component
too, and when this active component is no longer present, a negative
error signal drives the response extinction through repetitive expousures,
as illustrated in the Section \ref{sub:Quantitative-analysis-d-e}.
On the contrary, our mathematical results show that, if the stored
response associated with a stimulus is purely reactive (i.e., no active
component is expected), the error signal is equal to zero in each
trial, because the reactive response corresponds to the expected response;
for this reason, no updating of the emotional reactive response occurs.
A concrete example of such effect is the EC through \emph{implicit
misattribution} \cite{Hutter2013,Jones2009} (see Section \ref{sub:Evaluative-conditioning}).

Let us define now a \emph{non-active stimulus }as a stimulus eliciting
a reactive (or null) response only; given this definition, our main
finding can be summarised in the following corollary.

\emph{\noun{Corollary 1}}

\emph{If a purely reactive emotional response is attributed to a non-active
stimulus, this stimulus becomes a resistant-to-extinction source of
reactive emotional stimulation.}

\medskip{}

\emph{Corollary} 1 allows us to justify the inextinguishability of
EC due to\emph{ }implicit misattribution (see \cite{Hutter2013,Jones2009}
and references therein), and could suggest a mechanism for other forms
of resistant-to-extinction responses. In practice, through the repeated
application of Corollary 1, in each misattribution stage the reactive
response is given by the \emph{sum} of the previous reactive responses
evoked by the considered stimulus (because of the resistant-to-extinction
nature of the reactive stimulus itself) with the actual misattributed
reaction (in other words, the reactive response is \emph{cumulative}).
This could explain why an emotional additive increase can be obtained
naturally in every day life through the so called \emph{incubation
effect} \cite{book_traumatic2012,Davey1989} (note that even a summation
of different reactive sources could be misattributed to one single
target stimulus). More specifically, some neurotic individuals can
rehearse a trauma in their minds, or people can misattribute some
background emotional states (such as mood and others irrilevant and
disconnected emotional events) forward a target stimulus, which, in
turn, becomes able to elicit a stronger arousal at every misattribution
or rumination stage \cite{Davey1989}. 

Furthermore, Corollary 1 provides theoretical basis for the develpment
of purely reactive emotional stimuli: in principle, an additive and
a resistant-to-extinction emotional reaction can be obtained in a
controlled environment, through repetitive interactions between a
subject, a non-detectable reactive emotional source (the misattributed
source) and a target stimulus (to be attributed). In practice, a confounding
source (or a compound of confounding sources) of reactive stimulation
could elicit a subject in the presence of a target stimulus, which
has to become the attributed source of stimulation. Moreover, through
repetitive elicitations, the target stimulus is expected to produce
an inextinguishable and increasing reactive response. One method to
obtain a compound of confounding reactive sources is to develop a
subliminal masked expousure of an emotional and sufficently strong
aroused stimulus \cite{Flykt2007,Morris1998,Morris1999,Ohman1993,Ohman1994},
while a target stimulus has to be clearly perceived by the subject.
To this aim the confounding reactive stimulus has to elicit the same
emotional components as the target, in order to obtain the superposition
(i.e., the algebraic sum) of multiple contributions, i.e. the process
we call \emph{implicit accumulation effect}. We believe that further
research activities are needed to understand how to construct and
optimize a suitable target stimulus (which must be easily attribuitable
by a subject in relation to the emotional induced response) and which
sensory elements (e.g., tactile, visual or auditory elements) should
be included, in order to optimize the effect. Furthermore, it is important
to consider that a misattribution mechanism requires that the response
evoked by the UCS could feasibly have arisen from the target CS, so
some minimal degree of feature matching is a necessary condition;
without it, source confusion (and misattribution) is unlikely to occur
\cite{Jones2009}.

Finally, it is important to point out that a practical exploitation
of our results could be represented by the development of novel methods
to mitigate an undesired emotional reaction. In fact, in principle,
an undesired reactive response could be weakened through the misattributed
elicitation of an opposite valenced reactive response (for instance
the pain perception can be weakened through the stimulation with rewarding
reactive stimulus, or it can enhanced through the stimulation with
anxiety related or emotionally threatening reactive stimulus \cite{Roy2009,Wagener2009,Wiech2009}),
or through a reactive inhibition. This might represent a valid supporting
tool for treating certain psychopathologies for which the mere exposure
treatment, even if coupled with drug administration, could fail to
extinguish an undesired emotional reactive response (see Section \ref{sub:reactive-versus-active}).

\subsection{Evaluative conditioning through implicit misattribution\label{sub:Evaluative-conditioning}}

As already mentioned above, the EC phenomenon represents the formation
(or change of the valence) of a stimulus, called CS, originating from
a prior pairing of the CS itself with another stimulus, called UCS
\cite{Baeyens1993,Gast2012,Hofmann2010,Houwer2001,Hutter2013,Jones2009};
unlike Pavlovian conditioning, a CS response acquired through EC seems
to be resistant to extinction \cite{Baeyens2005}.

In recent years various research activities have been devoted to investigate
the role played in EC by the \emph{awareness of contingencies} between
a CS and its paired UCS \cite{Baeyens1992,Baeyens1993,Dawson2007a,Hutter2013,Ruys2009}.
In particular, experimental evidence illustrated in \cite{Hutter2013}
leads to the conclusion that the EC occurrence can be justified by
different mechanisms, like classical conditioning (and, consequently,
the awareness of CS-UCS contingencies) and implicit misattribution
(in this case, awareness is not required). In the same reference it
is also shown that sequential CS-UCS presentations are subject to
UCS revaluation, retroactive inference from subsequent learning and
contingency awareness (these phenomena are typically related to classical
conditioning); on the contrary, a simultaneous CS-UCS presentation,
which, generally speaking, prevents awareness and facilitates the
source misattribution, can produce EC effects which are robust against
all the factors mentioned above. Moreover, in \cite{Jones2009} it
is shown that, according to the implicit misattribution model, responses
to UCSs can be misattributed without awareness to the CS, and that
the implicit misattribution depends on \emph{source confusability},
in other words the subject has to confuse which multiple coocuring
stimuli in her environment is evoking the evaluative response. Furthermore,
manipulations of the variables related to the potential for the misattribution
of an evaluation, (i.e., the source confusability) show that greater
EC occurs with an higher degree of confusability \cite{Jones2009}.
Hence, as already discussed in the previous Section, the inextinguishability
of the CS valence in EC phenomena due to implicit misattribution can
be explained in terms of \emph{Corollary }1; in other words, it is
motivated by the fact that the UCS (e.g., a reactive stimulus such
as an emotional picture) behaves like a purely reactive source of
stimulation, and the reactive outcome is attributed (at least partially)
to the CS, so that the response stored and expected for that stimulus
is purely reactive (see also the\emph{ Case c} in the Section \ref{sub:Source-Misattribution:-quantitat}
for the quantitative analysis). This result is also supported by Baeyens
and colleagues \cite{Baeyens1993}, who found that EC was not sensitive
to the degree of statistical contingency between the CS and US, but
EC should increased with the absolute number of pairings because each
provides an opportunity for misattribution, and such misattributions
could act cumulatively \cite{Jones2009}. It is worth mentioning that
EC could also be determined by other mechanisms, or by the combinations
of different mechanisms \cite{Jones2009}, such as, classical conditioning,
the formation of particular beliefs about the CS and\emph{ conceptual
recategorization} of the CS \cite{Jones2009}.

\section{Differences and relations between the implicit emotional model and
the predictive coding model\label{sub:diff_interoception}}

The predictive coding (or active inference) version of interoception
\cite{Barrelt2015,Pezzulo2015,Seth2013}, described in Section \ref{sub:On-error-based-emotional},
cannot be applied to implicit emotional learning because the theorem
derived in Section \ref{sub:Theorem:-On-the}. In fact, since the
prediction signal (which coincides with the reactive response in predictive
coding) and the active (bottom-up sensory) elicitation sum up, the
prediction response would increase indefinitely over successive trials.
One might argue that the precision associated with the error signal
originating from the reactive (self-made) emotional response should
be reduced by attentional mechanisms, like in active inference in
primary motor cortex (M1), where the gain (or precision) associated
with the sensations originating from the self-made action are reduced
\cite{Barrelt2015,Shipp2013} (see also the \emph{corollary discharge};
e.g. \cite{Crapse2008}). Nevertheless, this cannot occur within the
emotional system, since, the reactive and the active emotional responses
are indistinguishable and, for this reason, add up in algebraic sense,
as shown by the experimental evidence described in Section \ref{sub:On-the-specificity-of-error}. 

Despite this, we argue that the predictive coding principle could
be applied under the assumption that different hierarchical neural
networks operate in succession, and others in parallel, during implicit
emotional learning. More specifically, if an active stimulus elicitates
a given subject, a first neural network, describing perception, determines
which is the most probable stimulus responsible for the elicitation
(i.e., the source attribution process) through the maximization of
the Bayesian posterior probability distribution (or the free energy
minimization, which leads to the ``surprise'' or entropy minimization
\cite{friston2009,Friston2008}), then the network for the computation
of the expected outcome (managed by the OFC) interacts with the network
which computes the updating of the reactive responses (within the
amygdala). However, our model provides a system level representation
of implicit emotional learning and, for this reason, the details about
the iterative computations (and messages passing) between neurons
or between different neuronal hierarchical levels of the involved
signals (such as error signals, expected outcomes or reactive responses,
and the neural computations leading to source attribution) are behind
the scope of our analysis. For instance, our system-level model assumes
that a distributed network computes error signals without entering
into the details of such a computation; hence, for example, our model
cannot be account for phenomena or physiopathologies originating from
a defect in the error computation processes (such as a pathological
dopamine system unable to properly compute the precisions associated
with the error signals). Nevertheless, our model can specifically
predict phenomena and pathologies originating from system-level problems
(in other words, it provides a macro-level perspective). Furthermore,
our model does not describe the computations leading to source attribution,
instead it describes the effects originating from the cases in which,
during a stimulation, a stimulus is correctly attributed, misattributed
or even not attributed. Finally, it is worth mentioning that an important
similarity between the proposed implicit emotional learning model
and the active inference model is represented by the fact that the
brain can experience a reactive stimulation (i.e., a prior bayesian
belief, in the active inference terminology) shaped by prior experiences
and learning, even if the stimulus does not actively elicitate the
subject. However, it is important to stress out that in our model
the reactive response is differentiated from the expected outcome,
which represents a value (more precisely a function over internal
physiological states) of the elicitation that the brain (OFC) expects
to experience from a given UCS. In other words, expecting a given
response does not coincide with the ``self-induction'' of that response
(see the Theorem in Section \ref{sub:Theorem:-On-the}). Furthermore,
a variation of the active (sensory or bottom-up) stimulation leads
to an update or modulation of the expected outcome and of the reactive
emotional response associated to the given stimulus.

\section{On the emotional learning in classical conditioning and UCS evaluation}

In this Section the intrinsic differences and the relations between
the mechanism of learning in classical conditioning and that previously
described for UCS revaluation learning are illustrated. Then, it is
shown that the Rescorla-Wagner equation for classical conditioning
represents an approximation of a specific case of emotional learning,
since it can be derived from our model, exploiting the stochastic
Hebbian plasticity rule \cite{Amit1994,Fusi2002,Fusi2007,Soltani2006,Soltani2010}
and taking into account specific approximations.

\subsection{Introduction\label{sub:Introduction}}

In classical conditioning a CS, which is usually a neutral/innocuous
stimulus like a sound or a neutral visual cue, is paired with a source
of stimulation (e.g., an electric shock or food), which is called
UCS. Through repeated CS-UCS pairings the CS can elicit a CR, called
\emph{unconditioned response} (UCR), which is often similar to the
response aroused by the paired UCS \cite{Fanselow2005,Kim2006,Pavlov1927}.
When the UCS represents an aversive stimulus, the CS-UCS pairing phenomenon
is called \emph{Pavlovian fear conditioning} \cite{Fanselow2005,Kim2006}.
In fear conditioning with humans, an indirect estimate of the autonomic
CR can be obtained measuring the changes in \emph{skin conductance
level} (SCL), i.e. the so called \emph{skin conductance response}
(SCR). Note, however, that in the analysis illustrated in the previous
sections, the UCR (i.e., the response elicited by a source of stimulation)
is not simply represented by indirect measurements, such as SCR or
the degree of salivation \cite{Pavlov1927}. In fact, it represents
the response associated with the brain neural populations elicited
by the UCS and, consequently, includes both non-emotional and emotional
components. More specifically, the UCR consists of a reactive response
component ($\mathbf{y}_{rem}$), a component due to the active (physical)
stimulation of emotional brain areas ($\mathbf{y}_{aem}$; see Eq.
(\ref{eq:2})) and a contribution due to non-emotional stimulation
($\mathbf{y}_{ne}$). For this reason, the SCR acquired during a specific
UCS stimulation, like an electric shock delivery, represents only
an indirect and correlated measure of the full response $\mathbf{y}$
expressed by Eqs. (\ref{eq:rappresentaz_base},\ref{eq:em-ne-components}).

\subsection{Different emotional learning mechanisms: for a primary stimulus (UCS)
and for a stimuli association (CS-UCS)\label{sub:On-the-learning}}

Even if both a CS and the paired UCS could be able to eliciting the
same emotional reaction under some circumstances, we argue they are
qualitatively different entities and, most important, they might be
learned through independent mechanisms. In fact, on the one hand,
through classical conditioning, which represents the learning of stimulus
contingencies, a CS acquires the capacity to elicit a response because
it is able to signal a likely occurrence of the UCS elicitation. On
the other hand, during \emph{implicit emotional learning }(or\emph{
}UCS evaluation)\emph{,} an element acquires the capacity to elicit
a response because a \emph{direct attribution} of the outcome forward
that element (i.e., the UCS). Furthermore, the UCS outcome revaluation
(i.e., inflation and devaluation) is also driven by implicit emotional
learning.

Our claims are supported by various results available in the literature
\cite{Davey1989,Gottfried2004,Hosoba2001,Rescorla1974,Schultz2013}.
In particular, \cite{Rescorla1974}, on the basis of the results of
various experiments in which inflation (UCR increase) is performed
after conditioning, came to the conclusion that organisms form memories
of a given UCS independently of associative connections with a CS.
Furthermore, in \cite{Schultz2013} it is shown that the automatic
response associated with a UCS (SCR) changes through the revaluation
of the UCS itself in the absence of a variation in the probability
of the CS-UCS contingency. In \cite{Gottfried2004} it is shown that
a UCS inflation, even during trials of conditioning extinction, results
in a larger CR. In \cite{Davey1989}, starting from contemporary models
of Pavlovian conditioning in humans, it is inferred that the processes
of CS-UCS association and UCS revaluation may be largely independent.
In particular, UCS revaluation can be obtained in different ways (including
verbal suggestions and cognitive processes) and this modifies the
strength of a CR in the absence of any change in stimulus contingencies.
Moreover, in the experiment reported in \cite{Hosoba2001} thirty
subjects were randomly assigned to the inflation (UCR increase) or
the deflation (UCR decrease) group, after a common classical conditioning
acquisition procedure (i.e., after experiencing the same UCR strength).
During the test session the indicators of the CR intensity were SCRs
and subjective aversion to the conditioned stimulus (CS). The main
results obtained in that case can be summarized as follows: a) the
CR strength measured by SCR was increased by the UCS inflation and
decreased by the UCS deflation; b) the subjective aversiveness to
the CS was not sensitive to both manipulations of the UCS intensity.

All the above mentioned results lead us to the following conclusions:
a) UCS revaluation occurs out of classical conditioning and, implicitly,
is able to modify the CR; b) the strength of an autonomic CR might
be influenced by the subjective revaluation of a UCS, even when the
CS-UCS contingency remains the same; c) the probability of a CS-UCS
contingency (i.e., the CS prediction of the occurrence of a UCS) through
Pavlovian conditioning is independent of the UCS revaluation. For
these reasons, the \emph{UCS evaluation} (and revaluation) cannot
be described by classical conditioning, so that a specific model (i.e.,
a model for implicit emotional learning or UCS revaluation, like that
illustrated in Section \ref{sec:Quantitative-Analysis-of}) is necessary.
These concepts are summarized in Fig. 4, where the CS representation
is related to the UCS representation through a connection whose strength
is proportional to the probability (or the \emph{belief}) of stimulus
contingency, i.e. to the probability $\Pr\left\{ UCS/CS\right\} $
of the UCS stimulation conditioned on the CS perception (briefly,
the CS-UCS contingency probability updated through the classical conditioning
learning). It is worth noting that the probability $\Pr\left\{ UCS/CS\right\} $
can be increased through repetitive CS-UCS pairings or reduced through
a CS exposition without UCS elicitation. These probability/belief
updatings reflects the CR during testing, according to the classical
conditioning learning (e.g., through the results given by the Rescorla-Wagner
equation \cite{Rescorla1972}). Note also that, in Fig. 4 the connection
between the UCS and the corresponding reactive response ($i_{R}$,
see Eqs. (\ref{eq:function_amygdala}-\ref{eq:Ii1})) is highlighted. 

The schematic representation illustrated in Fig. 4 is in agreement
with various recent results available in \cite{Redondo2014} and \cite{Gore2015}.
In particular, the experimental results shown in \cite{Redondo2014}
have evidenced that the hippocampal engram memory (which codes the
contextual CS) is neutral and could freely associate with either positive
or negative emotions (through the UCS representation) coded within
the BLA. The connection between the contextual CS coded in the DG
in the hyppocampus and the UCS coded within the BLA could also be
switched by optogenetic technology manipulation. In practice, a CS-UCS
connection (denoted CS-UCS \#1) can be switched to another CS-UCS
connection (denoted CS-UCS \#2) of opposite valence with respect to
the first one (e.g., from a fear emotional response to a reward emotional
response). Indeed, the optogenetic reactivation of the DG engram-bearing
cells during the presentation of a UCS having valence opposite to
the original one strengthens the connectivity of these DG cells with
a new subset of BLA neurons, while weakening the connections established
during the original learning \cite{Redondo2014}. From these results
it can be inferred that the CS memory engram is neutral and that a
CS can elicit an emotional reaction through the connection to an UCS
in the BLA. Additional results about the effects of optogenetic manipulation
are illustrated in \cite{Gore2015}, where it is shown that the projection
of a CS representation onto a UCS ensemble in the BLA is required
for the expression of a learned behavior, and that a CS (e.g., auditory
or olfactory) activates an UCS representation in the BLA to generate
a learned behavior. 

\begin{figure}
\noindent \begin{centering}
\includegraphics{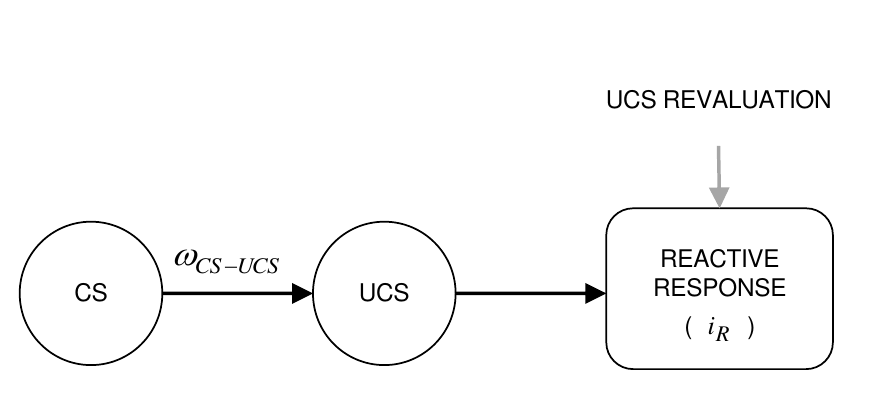}
\par\end{centering}

\caption{{\footnotesize{}Schematic representation of the connections between
a CS, the paired UCS and the reactive response ($i_{R}$) associated
with the representation of the UCS itself. The reactive response $i_{R}$
is determined by the UCS revaluation learning, instead the connection
strength between CS and UCS ($\omega_{CS-UCS}$) is determined by
classical conditioning learning. Note that this schematic representation
is in agreement with the experimental results shown in \cite{Gore2015,Redondo2014}. }}
\end{figure}

On the basis of our previous analysis we can state that a CS elicitation,
in turn, arouses a UCS reactive response ($i_{R}$) through a probabilistic
(or belief) connection; furthermore, the value of the associated probability
(or \emph{belief}) $\Pr\left\{ UCS/CS\right\} $ is determined by
classical conditioning (or even by optogenetic manipulations \cite{Redondo2014}).
For instance, at the end of a conditioning acquisition, the probability
$\Pr\left\{ UCS/CS\right\} $ is close to unity, since a CS predicts
almost certainly the UCS ``imminent'' elicitation. In this case
a trial test produces a CR expressed by $\Pr\left\{ UCS/CS\right\} \cdot i_{R}\simeq i_{R}$
and, consequently, given by $i_{R}$, which represents the reactive
response, determined by the amygdala reaction computed and stored
within the amygdala itself for the considered UCS (see Eqs. (\ref{eq:function_amygdala}-\ref{eq:Ii1})).
When the value of $\Pr\left\{ UCS/CS\right\} $ is between zero and
unity, the CR can be computed, for instance, through the Rescorla-Wagner
equation \cite{Rescorla1972} and has to equal the product $\Pr\left\{ UCS/CS\right\} \cdot i_{R}$.
Therefore, at the end of a conditioning acquisition process, there
is no real difference, in terms of emotional reactive response, if
the subject perceives a CS alone or the paired UCS alone (note that
the term ``response'', and not ``behavior'', is adopted here,
since different intensities of the components of the same emotional
response may lead to distinct observable behaviors; this issue is
discussed in more detail below). This is exemplified by the case of
the dog observed at the end of a conditioned acquisition in the conditioning
experiments performed by \cite{Pavlov1927}; in fact, as far as the
degree of salivation (i.e., the behavioral outcome due to the reactive
response $i_{R}$) was concerned, perceiving the food (UCS) or perceiving
the bell (CS) which signalled the incoming food, did not really make
any difference (at least during the first trial, before the eventual
extinction process). Note that this does not mean that the CS becomes
a substitute of the UCS, as initially supposed by Pavlov in the so
called ``\emph{Stimulus-Substitution Theory}''\cite{Chance2008},
but that the CS triggers the reactive response ($i_{R}$) through
the elicitation of the representation of the paired UCS within the
BLA (such an UCS elicitation through the CS perception could be ``partial''
or ``total'' depending of the CS-UCS connection strength; see Fig.
4 and \cite{Redondo2014}). In fact, as already stated in Section
\ref{sec:Response-representation}, a CR and the corresponding UCR
are not identical, and only in specific conditions they could be similar
(in addition, a UCR, unlike a CR, may involve an active component,
$X$). In practice, the CR mimics the emotional components elicited
by the corresponding UCR (see Section \ref{sub:On-the-specificity-of-error}
and Remarks \ref{rem:Ii+Ie} and \ref{rem:emotional_mimicking}) but,
generally speaking, with a lower intensity, since the emotional system
has to be stable (i.e., $\left|\alpha\right|<1$ is required; see
H.6 in Section \ref{sec:Quantitative-Analysis-of}) and because the
active component ($X$) is no longer present in a purely reactive
stimulation. These considerations could be useful to explain the fact
that, in specific circumstances, the CR might represent an opposite
behavioral response with respect to the original UCR; for instance,
one unconditioned response to morphine is represented by the reduced
sensitivity to painful stimuli; however, the conditioned response
to stimuli associated with morphine is represented by an increased
sensitivity to painful stimuli \cite{Siegel1975}. Phenomena like
this could be explained considering that, generally speaking, the
degree of elicitation of a neuronal population (or the quantity of
a specific type of released neurotransmitter), due to a CR is smaller
than the original one generated by the corresponding UCR (e.g., an
active morphine administration) because of the above mentioned reasons.
In particular, in neuropharmacology \cite{Goodman2006} it is well
known that an \emph{excitatory} effect is commonly observed with low
concentrations of certain \emph{depressant} drugs (e.g., alcohol,
morphine, general anesthetics) because of either the depression of
inhibitory systems or a transient increase in the release of excitatory
transmitters (note that an excitatory state occurs only with low concentrations
of the depressant). Consequently, within the complex CNS structure,
different \emph{doses} of a specific receptor agonist or of a specific
neurotransmitter (i.e., different intensities of a specific component
of a response) could lead to very different (even opposite) observable
behavioral responses (see \cite{Goodman2006} for an exahustive review
about this topic). 

Generally speaking, the typical scenario of a laboratory, where the
UCS is represented by an electric shock device, which is permanently
connected to the subject who does not know when the electric shock
will be effectively delivered (unless signalled by the paired CS),
leads to the impossibility of observing the emotional reactive response
$i_{R}$ when the source of stimulation (i.e., the UCS electric shock
device in the above mentioned scenario) is perceived by the subject.
On the contrary, we argue that, if the electric device is repeatedly
connected to the subject and disconnected from him/her in each trial
of electric stimulation, the subject perception of the connected UCS
device will elicit $i_{R}$ exactly as the CS perception does (after
a conditioned acquisition for the CS). One may argue that, in this
case, the CS is represented by the electric shock device and that
the electric shock delivery represents the UCS. Note, however, that
in the considered scenario the device represents a primary source
of direct stimulation (i.e., the UCS), whereas the shock delivery
represents the physical elicitation of that source, which supports
the active elicited response (i.e., the term $X$, see Eq. (\ref{eq:recursive-2}),
or the term $\mathbf{y}_{aem}$, see Eq. (\ref{eq:2})). Note also
that the same considerations can be expressed when the primary stimulus
is represented by food \cite{Pavlov1927}, but do not hold for a neutral
CS (e.g., a neutral sound) signalling the UCS, since, in the last
case, the CS cannot substain a direct physical stimulation.

\subsection{Conceptual difference between encoding a stimulus as a primary stimulus
(UCS) or as a conditioned stimulus (CS)\label{sub:Conceptual-difference-between} }

Another important issue related to the differences in the learning
mechanisms analyzed in the previous Section concerns the regions of
the brain in which a CS and a UCS are memorised. If a contextual CS
is considered, the CS engram and the the associated UCS are stored
in the DG and in the BLA, respectively \cite{Redondo2014}; however,
it is still unclear if the same rule applies to any non-contextual
CS. Note also that phylogenetic fear-relevant stimuli, which are sources
of stimulation according to the definition provided in Section \ref{sub:Active-and-reactive},
are natively stored in the mammial brain \cite{Ohman1993,Ohman1994}. 

In the following we summarise the most important experimental results
which concern the \emph{subliminal elicitation of} UCSs and the \emph{failure
in subliminal elicitation of ``previously-neutral'' conditioned
}CSs. Then, some important conclusions about CS/UCS memorisation in
specific regions of the brain are inferred from them.

\emph{Subliminal elicitation of UCSs} - In various experiments accomplished
by Öhman \emph{et al.} the awareness of visual stimuli was blocked
by means of \emph{backward masking} \cite{Esteves1994,Ohman1998}.
In this case a target picture, representing a CS, is presented in
an short interval (lasting less than 50 ms) and is followed by the
presentation of a masking picture having similar luminosity and colour
features, and shown in the same area of the visual field \cite{Enns2000}.
The presentation of this masking stimulus interrupts the cortical
processing of the target stimulus \cite{Noguchi2005,Rolls1999}; hence,
the target is invisible to the conscious and awareness of the subject
\cite{Kim2005}. Moreover, in these experiments snake- or spider-
fearful subjects have been exposed to phylogenetic fear-relevant masked
stimuli (snakes or spiders) and neutral masked stimuli (flowers and
mushrooms). The acquired experimental results have evidenced that
only the phylogenetic stimuli were able to elicit an automatic SCR
response in phobic patients \cite{Ohman1994}. These findings agree
with the hypothesis about amygdala functionality proposed by LeDoux
\cite{LeDoux1996,LeDoux2000}. In fact, LeDoux has hypothized the
existence of a \emph{thalamic pathway} to the amygdala; such a pathway
would allow to automatically detect evolutionary prepared visual stimuli
(like emotional faces, spiders, snakes, injuries). Note that this
model is also supported by other results acquired by different researchers
that have employed masking in normal participants \cite{Liddell2004,Morris1999}
or have observed brain activity in patients affected by cortical blindness
\cite{Gelder2006,Morris2001}. According to this model about amygdala
functionality, the superior colliculus stimulates the pulvinar nucleus
of the thalamus, which then arouses the amygdala \cite{LeDoux2000,Ohman2005,Ohman2007}.
This mechanism is also supported by various brain imaging studies,
which show that masked facial stimuli activate the amygdala exactly
as masked pictures of threatening animals (such as snakes and spiders)
do \cite{Ohman2005}. This suggests that the salient features representing
an aversive source of stimulation could be stored in the amygdala. 

\emph{Failure in subliminal elicitation of ``previously-neutral''
conditioned} CSs - Other experimental results available in the literature
help us to understand what happens when a previous neutral stimulus
(i.e., a CS) conditioned on an active source of stimulation (i.e.,
an UCS, like an electric shock) elicits a subject, and in particular,
if the CS representation is stored in the same region as a phylogenetic
aversive stimulus. Note that, from a supraliminar perspective, both
a purely reactive UCS (e.g., a snake picture) and a previously conditioned
CS (e.g., a neutral picture conditioned to an electric shock) are
able to elicit a similar autonomic response, at least in subjects
suffering from a phobia towards the phylogenetic stimulus represented
on the exposed pictures \cite{Ohman1994}. A more interesting case
is that of subliminal perception, where only the sub-cortical thalamic-amygdala
pathway is elicited because of the backward masking procedure described
above. Various results, acquired by Ohman and Soares in experiments
of \emph{differential conditioning}, are available about this case
\cite{Ohman1993}. In these experiments, a neutral stimulus, denoted
CS+ (e.g., flowers or mushrooms) was paired with an electric shock
UCS (i.e., an active source of stimulation) during the acquisition
phase, and a different neutral stimulus, denoted CS-, was presented
in the absence of the UCS. The results acquired during the extinction
phase have evidenced that the conditioned CS+ was able to elicit a
differential response (in particular, an SCR response different from
the baseline level of the CS- response) only in the presence of supraliminar
perception; on the contrary, no differential SCR response has been
observed in the case of masked perception. This means that, unlike
the case of a reactive UCS (such as spiders, snakes or angry faces),
which is able to elicit an emotional reaction both through supraliminarly
or subliminally perception \cite{Ohman1994}, a previously neutral
conditioned CS elicits an emotional response through supraliminar
perception only. Therefore, the sub-cortical thalamic-amygdala pathway
(i.e., the so called \emph{high road} \cite{LeDoux2000}) is unable
to elicit a representation of the given CS; this representation, instead,
can be elicited only through the thalamus-cortex-amygdala pathway.
These considerarions lead us to the conclusion that a CS representation
might not be directly stored within the amygdala (or, in any case,
not in the same region as an UCS representation) even if it predicts
an aversive-related stimulus. 

Finally, it is worth mentioning that the same conditioning paradigms
have been used when employing biologically fear-related (phylogenetic)
stimuli \cite{Ohman1993}, angry faces \cite{Esteves1994,Parra1997},
and even ontogenetic pictures (i.e., pictures of cultural threats
such as guns directed toward a subject) for CS+ and CS-. In all the
seen experiments, even if the employed CSs were conditioned stimuli,
they were not neutral, since they represented reactive aversive emotional
sources (i.e., reactive UCSs). This qualitative difference with respect
to a neutral CS (i.e., to a CS not representing a reactive emotional
source) reflects the different behavioral results provided by the
considered experiments and, in particular explains, the fact that,
even during the masked extinction procedure, the CSs+ have been able
to elicit an emotional response through the sub-cortical thalamic-amygdala
pathway. 

The experimental results illustrated above have lead us to the following
conclusions. It is likely that any representation of an emotional
source of stimulation, such as an innate biological fear-related threat
(in other words, a phylogenetic source) or a cultural threat (i.e.,
an ontogenetic source), is able to elicit an emotional response through
the sub-cortical thalamic-amygdala pathway; hence, this source is
expected to be stored in a \emph{rapid access site} of the amygdala.
On the contrary, in the case of classical conditioning, a previously-neutral
conditioned CS is unable to elicit a reactive emotional response through
the thalamic-amygdala pathway; consequently, it should be stored in
other regions of the brain or in a region of the amygdala different
from that employed for primary stimuli. Therefore, which is the qualitative
difference between a reactive UCS (e.g., an emotional picture) and
a previous neutral CS conditioned through classical conditioning?
Both UCS and CS are able to elicit similar responses supraliminarly,
but such stimuli are stored in different brain regions. We argue that
the main qualitative difference between these two cases is represented
by how initially the stimulus has acquired the capacity to elicit
an aversive emotional response, that is by how the stimulus has been
encoded (as a primary stimulus or as a conditioned stimulus). In fact,
a phylogenetic source might be innate and natively stored within the
amygdala; moreover, an ontogenetic emotional source, such as a gun
\cite{Flykt2007}, might be acquired through learning mechanisms different
from those of classical conditioning. It is reasonable to suppose
that a generic UCS, like an electric shock device or a weapon, is
acquired through an active and direct stimulation (or even cognitively,
exploiting aversive experiences of other individuals through social
fear learning \cite{Olsson2007,OlssonA2007}); consequently, the corresponding
outcome will be directly \emph{attributed} to that active source.
In brief, a UCS represents a stimulus able to directly elicit a response
(i.e., it is encoded as a primary stimulus); on the contrary, a CS
does not have such a capability and its role is limited to statistical
signalling of an incoming UCS. We also argue that the experiments
described above should also lead to similar conclusions if a new active
source of stimulation (i.e., an UCS which does not represent a phylogenetic
threatening stimulus nor an ontogenetic stimulus), experimentally
generated, is adopted. In particular, the correctness of the last
claim could be assessed by modifying the experimental procedure described
in \cite{Flykt2007}; in particular, this would require the following
two steps: a) introducing an additional phase in the experimental
procedure described in \cite{Flykt2007}, in order to generate (starting
from a neutral stimulus) a newly ontogenetic source of stimulation
for a group (through UCS evaluation), and a new conditioned stimulus
for another group (through classical conditioning learning); b) analyzing
the behaviors of the two different groups at the end of the learning
procedure. We expect to find out that both groups will be able to
show an emotional reaction (e.g., an SCR) through a supraliminar stimulus
perception, whereas only the group who has encoded the stimulus as
a primary stimulus (i.e., as UCS) will be able to show an emotional
reaction through subliminal perception (i.e., through the thalamo-amygdala
pathway). If such a result will be obtained, it will also definitively
prove that the mechanism through which the aversive object has been
encoded (i.e. through classical conditioning acquisition or by direct
or implicit response attribution) makes the difference and determines
the brain region in which the stimulus representation will be stored.

In the absence of further experimental results, we believe, on the
basis of the existing literature, that the only reasonable claim that
can be made about the nature of a coded stimulus, is given by the
following remark.
\begin{rem}
\label{rem:UCS-in-amygdala}
\end{rem}
\emph{If a subliminally perceived cue elicits an emotional reaction,
then this means the cue representation is stored within the amygdala
and it represents a source of emotional stimulation (i.e., an UCS).}

Note that this remark is useful when analysing experimental results
in which a subliminally perceived stimulus elicits a measurable emotional
reaction.

\subsection{Derivation of the Rescorla-Wagner Equation for Classical Conditioning
and a corrected formulation\label{sub:Derivation-of-the}}

In this section it is shown how the Rescorla-Wagner equation for CS-UCS
Pavlovian conditioning can be derived starting from our model about
the implicit learning of an UCS outcome.

Our derivation relies on the assumption that the CS-UCS synaptic connections
are governed by the mechanisms of \emph{stochastic Hebbian plasticity}
\cite{Amit1994,Fusi2002,Fusi2007,Hebb1949,Soltani2006,Soltani2010}.
This hypothesis is supported by both some experimental results shown
in \cite{Redondo2014} and \cite{Gore2015}, and other models relying
on the fact that a CS-UCS pairing entails the Hebbian potentiation
of the CS inputs onto the UCS representations in the BLA \cite{Pape2010,Pickens2004,Sah2003}.
In practice, Hebbian learning is based on the idea that synapses between
neurons being simultaneously active become stronger. Consequently,
``neurons that fire together could wire together'' through an increase
in synaptic efficacy mediated by \emph{long term potentiation} (LTP,
see \cite{Bliss1973}); on the contrary, a decrease in synaptic efficacy
is mediated by \emph{long term depression} (LTD) \cite{Artola1990}.
In particular, in \cite{Redondo2014} it is shown that the optogenetic
reactivation of the DG engram-bearing cells coding a contextual CS,
during the presentation of a new UCS having valence opposite to the
original UCS (which was previously paired with the CS itself), strengthens
the connectivity of these cells with a new subset of the BLA neurons,
while weakening the connections established during the original learning
process. In other words, the simultaneous activation of a CS neural
representation and of a new UCS strengthens a CS-UCS connection (through
LTP) and, at the same time, weakens the connection (through LTD) between
the CS and the previously associated UCS, which is not simultaneously
active. 

As illustrated in \cite{Destexhe2004}, the main task of the sensory
system is to detect (and model) correlations \cite{Barlow1985} through
neuron firing, in order to exploit \textquoteleft suspicious coincidences\textquoteright{}
in complex incoming information. These correlations may form the \textquoteleft objects\textquoteright{}
or \textquoteleft features\textquoteright{} of the representations
of any stimulus. After being detected by primary sensory areas, such
correlations can be used for binding elementary features into more
elaborate percepts. These binding phenomena \cite{Destexhe2004,Roskies1999}
are based on the concept of neuronal assemblies, which are usually
defined as a group of neurons that transiently undergo synchronous
firing \cite{Destexhe2004,Hebb1949}. This transient synchrony could
form the basis of a common input to later stages of integration, and
so promote responses that are specific to a given ensemble of features
\cite{Engel2001}. These mechanisms rely on the fact that cortical
neurons are very efficient at detecting correlations \cite{Abeles1991},
as evidenced by computational models \cite{Rudolph2001}. On the basis
of these results it can be assumed that that the representation of
a CS and that of its associated UCS are formed by a \emph{neuronal
assembly}, which, in turn, is composed by cells assemblies representing
specific features, and that connections between the CS and the UCS
neuronal assemblies are established through Hebbian plasticity. The
implications of this assumption are analysed in the following part
of this Section, where a new sequence of trials, involving CS-UCS-subject
interactions, is taken into consideration.

In these trials we assume that: a) the source of stimulation (UCS)
has been acquired (i.e., properly coded as a source stimulus), and
the emotional learning for the UCS source has encoded the reactive
response $i_{R}$ (which can be assumed to be equal to $\alpha\cdot X$,
with $X$ representing the active induced response); b) a new cue
(denoted CS) becomes paired with the source UCS in the first trial
and, hence, an UCR associated with this UCS is elicited; c) in the
first trial (i.e., for $n=1$) the strength $\omega_{CS-UCS}^{(1)}$
(ranging from 0 to 1, as evidenced below) of the CS-UCS connection
(which, in turn, is related to $i_{R}$; see Fig. 4) is equal to zero
(in other words, no connection has been established before the start
of the pairing process); d) the CS remains the same during all the
considered trial; e) the UCS elicitation is signalled by the CS presentation
only and not by the UCS presentation (e.g., the electric shock device
is permanently attached to the subject and the shock delivery is signalled
exclusively by the CS presentation). In the following analysis the
response elicited by the CS, previously named CR, is denoted $y_{CS}$.
Then, in the first trial, an unexpected active UCS elicitation ($X$)
generates the emotional response (see Eq. (\ref{eq:solution-decomposition-d-a}))

\begin{equation}
y_{UCS}^{(1)}=X,\label{eq:cond-y1}
\end{equation}
which is what was expected for the source UCS. However, since the
UCS occurs unexpectedly in time, a reactive contribution due to the
\emph{contrast effect} (see Section \ref{sub:contrast-effect}) should
be produced (provided that no previous conditioned cue or direct UCS
perception signals the active stimulation). Nevertheless, in Section
\ref{sub:contrast-effect} it has been shown that successive unexpected
UCS elicitations do not entail the computation of an error signal,
but certainly involve the elicitation of a reactive response due to
the contrast effect, (quantified by the product $K\cdot X$). However,
without any loss of generality, this reactive contribution can be
neglected; furthermore, it can be noted that, during conditioning
trials, this contrast contribution will vanish (since no ``unexpected
stimulation'' occurs as the CS becomes progressively able to predict
the UCS occurrence) and in place of this contribution the reactive
response associated with the UCS ($i_{R}$) will be elicited as the
CS is perceived (see Fig. 4). Even if no error signal has to be estimated
for the given UCS source during the first trial, the contemporary
presence of the CS during the UCS elicitation is sufficient to generate
some synaptic connections between the representation of the CS and
that of the UCS through the stochastic Hebbian rule; consequently,
the strength of this link is potentiated through LTP. In the following
it is also assumed that individual plastic synapses exhibit a binary
behavior, since they can be in a \emph{depressed state} or in a \emph{potentiated
state}. For this reason, the strength of a set of plastic synapses
is quantified by the fraction of synapse population in the potentiated
state \cite{Amit1994,Fusi2002,Fusi2007,Soltani2006,Soltani2010};
this fraction is called \emph{synaptic strength} and in the $n$-th
trial is denoted as $\omega_{CS-UCS}^{(n)}$ for the set of synapses
from the neurons representing the CS stimulus onto the encoding neurons
for the UCS. 

The mechanism through which plastic synapses learn cue-outcome contingencies
through stochastic reward-dependent Hebbian modifications is illustrated
in \cite{Soltani2010}. In practice, whenever the neurons encoding
a given CS are simultaneously elicited at the activation of the UCS
neurons, the plastic synapses from CS onto UCS in the depressed state
make a transition to the potentiated state with probability $\hat{\alpha}_{+}$
(this quantity is called \emph{potentiation rate}); otherwise, if
the CS is elicited without the contingent UCS elicitation, they make
a transition in the reverse direction with probability $\hat{\alpha}_{-}$
(this quantity is called \emph{depression rate}). It is worth mentioning
that the parameters $\hat{\alpha}_{+}$ and $\hat{\alpha}_{-}$ are
ofter called \emph{learning rates} \cite{Soltani2006} and that they
depend on the firing rate of the postsynaptic neurons of the CS and
on the state of the UCS (which is either active or non-active, that
is eliciting or not eliciting UCS). The firing rate is low for the
neurons which do not represent the perceived CS and, on the contrary,
is high for the neurons encoding the CS. Hence, if the CS features
are modified during the trials, the learning rates change too. In
the following, however, we assume that the CS perception does not
change during the trials (see H.4 in Section \ref{sub:specific_hyp_discrete}),
so that the depression and potentiation rates remain constant (and
different from zero). It is worth pointing out that the parameters
$\hat{\alpha}_{+}$ and $\hat{\alpha}_{-}$ are scalar quantities
(each of them would be replaced by a vector having identical components
if multiple emotional components were considered in the evaluation
of the reactive response) and they are not related to the parameter
$\alpha$ defined in Section \ref{sub:Quantitative-analysis-d-a}
(i.e., $\alpha$ becomes also a vector, but with different components,
if multiple emotional components are taken into consideration). 

On the basis of the plastic probabilistic Hebbian rule illustrated
above, in the $n$-th trial the synaptic strenght is updated as

\begin{equation}
\omega_{CS-UCS}^{(n)}=\omega_{CS-UCS}^{(n-1)}+\hat{\alpha}_{+}\cdot\left(1-\omega_{CS-UCS}^{(n-1)}\right)\label{eq:strength-increase}
\end{equation}
during a conditioned acquisition (through LTP), and as
\begin{equation}
\omega_{CS-UCS}^{(n)}=\omega_{CS-UCS}^{(n-1)}-\hat{\alpha}_{-}\cdot\omega_{CS-UCS}^{(n-1)}\label{eq:strength-decrease}
\end{equation}
during the extinction phase (through LTD). Note that the second term
in the right-hand side of (\ref{eq:strength-increase}) describes
the change related to the transition of synapses in the depressed
state, since a fraction $\left(1-\omega_{CS-UCS}^{(n-1)}\right)$
of synapses are potentiated with probability $\hat{\alpha}_{+}$.In
the following, we assume, without any loss of generality, that the
potentiation and depression rate are equal (i.e. $\hat{\alpha}_{+}=\hat{\alpha}_{-}=\hat{\alpha}$).
Then, during acquisition the synaptic potentiation of the CS-UCS connection
can be evaluated as follows:

\begin{equation}
\omega_{CS-UCS}^{(n)}=\omega_{CS-UCS}^{(n-1)}+\hat{\alpha}\cdot\left(1-\omega_{CS-UCS}^{(n-1)}\right)\label{eq:w-acquisition}
\end{equation}
where the term $\hat{\alpha}$ denotes the learning rate, it represents
a scalar value (or a vector with all equals components if more than
one emotional components are involved in the reactive response $i_{R}$).

The synaptic strenght $\omega_{CS-UCS}^{(n)}$ evaluated on the basis
of Eq. (\ref{eq:strength-increase}) can be exploited to assess the
response $y_{CS}^{(n)}$ to the presentation of the CS alone in the
$n$-trial; in fact, the intensity of the CR is determined by the
product of $\omega_{CS-UCS}^{(n)}$ with the reactive response $i_{R}$
associated with the paired UCS, i.e. by

\begin{equation}
y_{CS}^{(n)}=\omega_{CS-UCS}^{(n)}\cdot i_{R}.\label{eq:ycs-init}
\end{equation}
Then, substituting (\ref{eq:strength-increase}) in (\ref{eq:ycs-init})
yields the expression

\begin{equation}
y_{CS}^{(n)}=i_{R}\cdot\omega_{CS-UCS}^{(n-1)}+\hat{\alpha}_{+}\cdot\left(i_{R}-i_{R}\cdot\omega_{CS-UCS}^{(n-1)}\right),\label{eq:ycs-fin}
\end{equation}
which can be easily put in the form

\begin{equation}
y_{CS}^{(n)}=y_{CS}^{(n-1)}+\hat{\alpha}_{+}\cdot\left(i_{R}-y_{CS}^{(n-1)}\right).\label{eq:ycs-recursive}
\end{equation}
It is easy to prove that the last formula coincides with the well
known Rescorla-Wagner equation for Pavlovian conditioning \cite[Sec 1, p. 365, eq. (1-2)]{Miller1995},
\begin{equation}
V_{x}^{n+1}=V_{x}^{n}+\alpha_{x}\beta_{1}\left(\lambda_{1}-V_{total}^{n}\right)\label{eq:Rescorla-Wagner-Model}
\end{equation}
for the case in which a single CS is considered; in fact, Eq. (\ref{eq:Rescorla-Wagner-Model})
is obtained from (\ref{eq:ycs-recursive}) if $\hat{\alpha}_{+}$
and $i_{R}$ are replaced with $\alpha_{x}\cdot\beta_{1}$ and $\lambda$,
respectively, and $y_{CS}^{(n-1)}$ is assumed to represent the \emph{associative
strength} $V_{x}^{n}$. Note that the $V_{total}^{n}$ coincides with
the term $V_{x}^{n}$ if a single CS (denoted $x$) exists; otherwise
it represents the sum of the associative strengths of all CSs (including
$x$).

Equation (\ref{eq:ycs-recursive}) deserves the following comments:
\begin{itemize}
\item If the overall CS is composed by $N$ distinct CSs (i.e., a \emph{CSs
compound} is considered), the synaptic strength between the compound
and its paired UCS can be still computed on the basis of (\ref{eq:w-acquisition}).
However, in this case, a fraction of the overall strength should be
associated with each component of the compound CS (such a fraction
depends on the nature of the considered CS and its neural representation).
Then, the contribution to the synaptic strength originating from the
$k$-th component (with $k=1,2,...,K$) can be evaluated as (see ((\ref{eq:strength-increase}))
\begin{equation}
\omega_{CS_{k}-UCS}^{(n)}=\omega_{CS_{k}-UCS}^{(n-1)}+\hat{\alpha}_{k}\cdot\left(1-\omega_{totCS-UCS}^{(n-1)}\right),\label{eq:compound-rescorla}
\end{equation}
where $\omega_{CS_{k}-UCS}^{(n)}$ and $\omega_{totCS-UCS}^{(n-1)}$
represent the synaptic strength originating from the $k$-th CS and
the overall synaptic strength originating from the compound, rispectively.
The last formula is motivated by the fact that each component shares
the same full synaptic connection in reaching the neural representation
of the stored UCS. It is also worth mentioning that multiplying both
sides of (\ref{eq:compound-rescorla}) by $i_{R}$ produces
\begin{equation}
y_{CS}^{(n)}=y_{CS}^{(n-1)}+\hat{\alpha}_{+}\cdot\left(i_{R}-y_{total}^{(n-1)}\right),\label{eq:R-total}
\end{equation}
which represents the Rescorla-Wagner equation for the case of a CS
compound \cite[Sec 1, p. 365, eq. (1-2)]{Miller1995}. From Eq. (\ref{eq:compound-rescorla})
it can also inferred that, if a CS or a compound have been conditioned
to a UCS, so that their synaptic strength is equal to unity, when
a new CS is added to the compund and paired with the given UCS no
connection updating can occur, this phenomenon is known as \emph{blocking
effect} \cite{Miller1995}. 
\item From a mathematical perspective Eqs. (\ref{eq:strength-increase})
and (\ref{eq:compound-rescorla}) can be intepreted as simple rules
for updating a conditional probability. In fact, the synaptic strength
$\omega_{CS-UCS}^{(n)}$ can be interpreted as the probability (or
the belief) of that the event of UCS elicitation occurs conditioned
on the fact that the given subject perceives the considered CS.
\item During the extinction process, the CS is repeatedly shown to the subject
without UCS pairings; consequently, (\ref{eq:strength-decrease})
holds and the CS response $y_{CS}$ tends asymptotically to zero.
\item A straightforward consequence of Eq. (\ref{eq:R-total}) is represented
by the fact that the maximum response obtained at the end of the conditioning
process asymptotically approaches $i_{R}$, which represents exactly
the emotional reactive response associated with the given source of
stimulation (i.e., with the UCS). In turn, the reactive response $i_{R}$
can be increased (inflation) or decreased (devaluation) by UCS re-evaluation
(e.g., through a variation of the active stmulation $X$)), independently
of the Pavlovian conditioning process.
\item From the last point it can be inferred that, at the end of the acquisition
of a CS-UCS Pavlovian conditioning, there is no difference, in terms
of emotional response, if the CS only or the UCS only is perceived
in the absence of an active elicitation. This claim holds only in
a supraliminal stimulus elicitation (see Section \ref{sub:Conceptual-difference-between}). 
\item Actually, since the CS-UCS connection strength increases during conditioning
acquisition, the reactive response $i_{R}$ associated with the UCS
should also become stronger. As a matter of fact, the emotional response
$y_{UCS}^{(n)}$ in the $n$-th acquisition trial is due to the elicitation
of the CS response (i.e., to $y_{CS}^{(n)}=i_{R}\cdot\omega_{CS-UCS}^{(n)}$
due to the partial activation of the UCS representation from the CS)
and to the active UCS stimulation (i.e., to the term $X$ due, for
instance, to an electric shock stimulation; see Eq. (\ref{eq:cond-y1})),
since the overall response is \emph{attributed} to the UCS (which
represents, unlike the CS, a direct source of stimulation). This last
claim is supported by the experimental results obtained through optogenetic
manipulations \cite{Redondo2014}, which show that a memory engram
coding a CS is ``emotionally neutral'' and could freely associate
with different emotional responses through the corresponding UCS representations
coded within the BLA. Therefore, it is easy to prove that the associated
reactive component $i_{R}$ grows from the initial value $\alpha\cdot X$
to the value $\alpha\cdot(X+i_{R}^{(n-1)}\cdot\omega_{CS-UCS}^{(n)})$
because of the error signal (computed as the difference between the
expected response $y_{UCS}$ and the experienced outcome). For these
reasons, the process of implicit UCS inflation originates from an
indirect contribution of the CS; in fact, the CS, signalling the UCS,
is able to elicit the reactive response ($i_{R}$; see Fig. 4) associated
with the UCS itself. Therefore, $i_{R}$ does not remain constant
over consecutive acquisition trials, as assumed by the original Rescorla-Wagner
model, but evolves according to the recursive equation
\end{itemize}
\begin{equation}
i_{R}^{(n)}=\alpha\cdot\left(X+i_{R}^{(n-1)}\cdot\omega_{CS-UCS}^{(n)}\right)\label{eq:ir-updating}
\end{equation}

The last formula shows that the CR intensity influences the intensity
of the unconditioned response. This result is in agreement with some
experimental results \cite{Young1976} (see also \cite{Miller1995}
and articles therein), evidencing \emph{the dependence of asymptotic
responding on CS intensity and US intensity;} this represents one
of the results unpredicted by the Rescorla-Wagner model \cite{Miller1995}.
It is also important to point out, however, that the changes of $i_{R}$
over successive trials could be really small, since the term $\alpha\cdot i_{R}$
in Eq. (\ref{eq:ir-updating}), which is smaller than $\alpha^{2}\cdot X$
(since $0<\omega_{CS-UCS}<1$), is negligible with respect to the
term $\alpha\cdot X$ (since $\left|\alpha\right|<1$). Moreover,
it is easy to prove that the asymptotic value of $i_{R}$ is $\alpha X/(1-\alpha)$,
which is greater than the initial value $\alpha X$. This leads to
the conclusion that, since the value of the parameter $\alpha$ is
influenced by the selected CS (if the impact of other factors, such
as internal psysiological states and the selected UCS, is deemed constant),
different CSs may result in distinct asymptotic values of $i_{R}$
(and consequently of $y_{CS}$; see Eq. (\ref{eq:ycs-init})). 

In summary, we propose to adopt a new classical conditioning model
(valid for a discrete time scale), which encompasses the Rescorla-Wagner
model and coincides with it only under specific conditions and restrictions.
If a single CS is assumed to simplify the notation, this extended
model is described by Eq. (\ref{eq:strength-increase}) when the given
CS is paired with an UCS, Eq. (\ref{eq:strength-decrease}) when the
CS is presented alone, by Eq. (\ref{eq:ir-updating}), which has to
be updated only in the CS-UCS pairing trials (since $i_{R}$ does
not vary if the CS is presented alone), and, finally, by the formula

\begin{equation}
y_{CS}^{(n)}=\omega_{CS-UCS}^{(n)}\cdot i_{R}^{(n-1)},\label{eq:ycs-updatings}
\end{equation}
which expresses the $y_{CS}$ updating. 

Note that Eqs. (\ref{eq:ir-updating}) and (\ref{eq:ycs-updatings})
hold for $n\geq2$ and that the initial conditions

\begin{equation}
\omega^{(1)}=0\label{eq:ci1}
\end{equation}

\begin{equation}
i_{R}^{(1)}=\alpha\cdot X\label{eq:ci2}
\end{equation}

\begin{equation}
y_{CS}^{(1)}=0\label{eq:ci3}
\end{equation}
and

\begin{equation}
y_{UCS}^{(1)}=X\label{eq:ci4}
\end{equation}
should be adopted when employing them. Our extended model provides
a more general and accurate description of the emotional response
during conditioning than the original Rescorla-Wagner model for at
least three different reasons. First of all, it includes the contributions
of both the active response ($X$) due to the elicitation of the physical
UCS and the reactive response associated with the UCS representation
within the BLA ($i_{R}$). Moreover, the recursive equations describing
it are \emph{causal} unlike those representing the Rescorla-Wagner
model. Note that causality ensures that the currently computed response
depends only on the past and present values of the stimulus and the
response itself, but not on their future values; unluckily, this does
not occur for the Rescorla-Wagner model since the evaluation of the
current response requires the knowledge of the final asymptotic response
(i.e., the term $\lambda$ in Eq. (\ref{eq:Rescorla-Wagner-Model})),
which is actually unknown to the brain. Finally, in our model the
CS-UCS synaptic strength (associative learning) and the consequent
UCS inflation are jointly considered: the model shows that classical
conditioning learning influences the reactive response associated
with the paired UCS; this is due to the misattribution of the CR contribution
forward the UCS response. Hence, generally speaking, classical conditioning
(i.e., the CS-UCS contingency or associative learning) and implicit
emotional learning (i.e., UCS revaluation) are different learning
mechanisms, but are closely related; in particular, classical conditioning,
is able to indirectly determine the inflation of the UCS. It is also
interesting to note that the asymptotic value 

\begin{equation}
i_{R}^{(\infty)}=\alpha\cdot y^{*}=\frac{\alpha\cdot X}{1-\alpha}\label{eq:ir-asymptotic}
\end{equation}
of $i_{R}$ provided by Eq. (\ref{eq:ir-updating}) is identical to
the asymptotic reactive response resulting from implicit learning
(see Eq. (\ref{eq:solution-decomposition-d-a})). This means that
perceiving the source of stimulation (i.e., the UCS) or perceiving
a signalling conditioned stimulus before the administration of the
active stimulation ($X$) leads to the same asymptotic reactive response
to the given UCS. However, the CS is only a means to strenghten the
reactive response associated with its paired UCS, and classical conditioning
represents a secondary (indirect) mechanism of emotional learning.
In fact, the CS remains neutral \cite{Redondo2014} and the corresponding
CR (i.e., $y_{CS}$) is related to the synaptic connections between
the CS representation and the UCS representation in the BLA, which,
in turn, is connected to the reactive response which has been strenghtened
through UCS revaluation (because of a misattribution of the CR). It
is also worth mentioning that direct implicit emotional learning (i.e.,
perceiving the UCS at every trial in place of the CS) allows to achieve
the asymptotic response expressed by Eq. (\ref{eq:ir-asymptotic})
in fewer trials than classical conditioning, since in last case the
initial value of the synaptic strength $\omega_{CS-UCS}$ able to
elicit the UCS is smaller than unity and increases with trials. 

It is also important to note that the proposed model is able to properly
predict the fact that, if the CS is presented with a UCS of lower
intensity (i.e., with an active component $X^{'}<X$) during a generic
trial, the reactive response $i_{R}$ will be reduced by the influence
of the error signal and, at the same time, the synaptic strength $\omega_{CS-UCS}$
will be independently increased (or will remain constant if the asymptotic
value has been already reached), since the synaptic connections between
the CS and the UCS are still reinforced by the stimuli contingency
(i.e., through associative learning, rather than through an error
signal computation), even in the presence of a negative error signal.
This result is in agreement with the experimental evidencies shown
in Section \ref{sub:On-the-learning}; note, however, that classical
conditioning may determine, indirectly (through the elicitation of
the reactive response and its misattribution forward the UCS), an
error signal. It is also important to mention that classical conditioning
experiments involving the contingency of two neutral stimuli (e.g.,
the pairing of a neutral tone and a neutral light; see \cite{Young1998}
and articles therein) are in agreement with our results, since these
show that the associative learning can occur with the pairing of two
neutral (not emotional nor motivational) stimuli and, hence, with
no computation of any error signal. From these results and considerations
it can be inferred that the associative learning is not driven by
the emotional system, on the countrary the implicit UCS revaluation
learning is driven by the emotional system through the computation
of the emotional error signals. 

Finally, it is worth mentioning that the asymptotic reactive response
associated with a given UCS cannot be increased by means of the conditioning
acquisition of a new CS (called CS\#2) . In fact, if the CS\#2 becomes
paired with the UCS, in the initial trials the total outcome $y_{UCS}$
is smaller than the asymptotic value reached during the previous conditioning
due to the first CS (CS\#1); this is due to the fact that, at the
beginning, the connection strength $\omega_{CS\#2-UCS}$ is smaller
than unity. Therefore, in the initial trials a negative error signal
will lower the reactive response $i_{R}$ and, in the successive trials,
a positive error signal will increase it again untill the asymptotic
value in Eq. (\ref{eq:ir-asymptotic}) is reached.

\subsubsection{Stochastic Hebbian rule in more complex pattern}

When complicated CS patterns are conditioned to an UCS, Hebbian learning
leads to self-organizing networks of synaptic connections. Such networks
are able to represent complicated statistical regularities characterizing
the considered environment \cite{Munakata2004,Soltani2010}; for instance,
temporal CS-UCS relations can be also learned assuming the \emph{time}
variable\emph{ }(and temporal relations between stimuli) as a contextual
CS \cite{Bouton1993}. We argue that, even during the implicit emotional
learning of the UCS, the temporal trend of the stimulation represents
an important contextual information, which might be coded and stored.
In addition, we feel that more complicated Hebbian networks simultaneously
taking into account several variables could be able to predict more
sophisticated phenomena related to classical conditioning paradigms
(e.g., \emph{second order conditioning} \cite{Miller1995}).

\subsubsection{Conditioning to a reactive source stimulus \label{sub:Conditioning-to-a-reactive-source}}

We argue that the following three possible events can occur when a
CS is conditioned to a purely reactive UCS (i.e., an UCS for which
no active component elicitation is expected, e.g., an emotional picture):
1) a simple associative connection CS-UCS is generated; 2) the CS
is misattributed to be the source of the elicited response, so becoming
a new (and independent) reactive source like the original UCS, so
that a new reactive response, equal (but independent) to the original
$i_{R}$, is generated and associated to the CS; furthermore, this
misattribution process could occur even during the presentation of
the CS alone after conditioning, since the reactive elicitation equals
to $\omega_{CS-UCS}\cdot i_{R}$ (see Eq. (\ref{eq:ycs-init})) could
be misattributed forward the CS, in this case the reactive response
being associated with CS corresponds to the quantity $\omega_{CS-UCS}\cdot i_{R}$;
3) a combination of the previous two events happens. Moreover, if
one of the two last mentioned events occurs, the conditioned CS will
become an inextinguishable element of emotional reaction. In summary,
we argue that a CS conditioned to a reactive UCS can become an inextinguishable
reactive source of stimulation through different mechanisms; furthermore,
a given stimulus could act as both a CS and a primary stimulus (UCS)
at the same time, owning even opposite valenced potentials; for instance,
the considered stimulus could be conditioned to an aversive UCS and,
at the same time, could operate as an appetitive source of reactive
stimulation.

\subsection{On the conditioned inhibitor}

In this Section the quantitative description of the so called \emph{conditioned
inhibitor }is developed on the basis of our theory.

Operationally speaking, a conditioned inhibitor is a CS that passes
a negative summation and retardation tests for conditioned inhibition
(see \cite{Miller1995} and articles therein). A CS is said to pass
a negative summation test for conditioned inhibition if, when it is
presented together with a conditioned excitor (i.e., a CS wich has
been paired with an UCS), it reduces the level of conditioned responding
to that excitor that would otherwise occur. Moreover, a CS is said
to pass a retardation test for conditioned inhibition if it requires
a larger number of pairings with the UCS to become an effective conditioned
excitor than those required if the CS were novel \cite{Miller1995}
(i.e., if the CS had not undergone inhibitory training).

The most widely used method to create a conditioned inhibitor consists
of not reinforcing the intended inhibitory CS in the presence of another
cue that itself has been previously reinforced \cite{Miller1995}.
Let us analyze the conditioned inhibition on the basis of the model
we proposed for emotional learning. 

One of the main assumptions made in the development of our model is
that the functions involved in the computation of emotional reactions
are time invariant (see Section \ref{sub:General-hypothesis}). Under
this assumption, if in a sequence of trials an UCS elicits a subject
that always maintains the same internal physiological states (for
instance, this case is exemplified by a dog that undergoes some conditioninng
trials, pairing a ring and some food, in which the intensity of its
hunger remains the same), the only source of variation of the response
(and of the expected outcome) is represented by the learning process.
For this reasons, if the expected outcome for an UCS after the first
elicitation trial is equal to $X$, the reactive response associated
with the UCS ($i_{R}$) is equal to $\alpha X$ (see Eqs. (\ref{eq:Ii1},
\ref{eq:approx-famygdala} and \ref{eq:approx-fchain})), whereas
the asymptotic value is equal to $\alpha X/(1-\alpha)$ (see Eq. (\ref{eq:asymptotic reactive})).
Hence, in both cases, the reactive response $i_{R}$ is expressed
by 

\begin{equation}
i_{R}=\alpha\cdot y_{expected}.\label{eq:ir-function-of-expected-l}
\end{equation}
that is by the product of the parameter $\alpha$ depending on the
amygdala and the system chain functions (see Eqs. (\ref{eq:approx-famygdala}-\ref{eq:alfa}))
with the expected outcome. 

Generalizing the last consideration leads to the conclusion that,
if the expected outcome for a specific UCS is also influenced by internal
physiological states (and not only by the learning process), the reactive
response associated with the considered UCS varies according to Eq.
(\ref{eq:ir-function-of-expected-l}). From a neurophysiological perspective,
we argue that the last result is motivated by the fact that the expected
outcome and the biological functions determining the reactive response
are controlled by the same internal physiological states (the degree
of hunger in the example mentioned above). In practice, if food is
presented to a satiated dog, the expected outcome (i.e., the rewarding
response which is expected to occur) is very weak and, consequently,
the associated reactive response will be weak too (e.g., a low degree
of salivation will be observed). However, this phenomenon does not
lead to the extinction (or devaluation) of the UCS, since no error
signal is computed; this is due to the fact that the expected outcome
coincides with the experienced outcome (actually, in this case the
dog could even avoid to eat and hence to experience the actual UCS
elicitation). It is also reasonable to assume that, if a tone (CS)
is presented together with some food (UCS) to a satiated dog, the
CS-UCS association will be reinforced even if no emotional response
occurs, and, in any cases, no CS extinction occurs. In other words,
the perception of the UCS in a trial in which the internal physiological
states of the subject lead to a low expected response (or even to
a response equal to zero for the considered UCS) does not lead to
the extinction of the UCS itself (nor to a correction of the reactive
response through the computation of an error signal); furthermore,
the paired presentation of a previously paired CS (excitor) with the
considered UCS, during the aforementioned conditions, does not lead
to extinction of the CS. 

On the basis of the considerations illustrated above the following
quantitative analysis of a conditioned inhibitor\emph{ }can be derived.
We argue that an inhibitory CS (denoted CSi in the following) represents
a stimulus whose neural representation is directly connected to an
inhibitory reactive response, denoted $i_{Ri}$ (in other words, a
causal \emph{attribution} of the inhibition effect forward the CSi
occurs, so that the CSi is encoded as a source of stimulation by the
emotional system). For instance, if an excitor CS previously conditioned
with an electric shock delivery (characterized by the intensity $X$)
is paired with a new CSi and, simultaneously, with an intensity of
electric stimulation equal to zero, the CSi becomes a conditioned
inhibitor and, more specifically, becomes connected to an inhibitory
(e.g., negative) reactive response $i_{Ri}=\alpha(-X$). If these
considerations are generalized like in the previous case of an excitor
source of stimulation (see Eq. (\ref{eq:ir-function-of-expected-l})),
the reactive response associated with the CSi is a function of the
expected active stimulation which has to be inhibited, so that

\begin{equation}
i_{Ri}=\alpha(-y_{expected}).\label{eq:ir-inhibit1}
\end{equation}
On the basis of the last result it should be expected that, if a new
CS excitor (denoted CS2 in the following), predicting an UCS stimulation
intensity weaker than the one predicted by the conditioned excitor
which has been employed for the inhibitory conditioning (to obtain
the CSi), called $X^{'}$ (with $X^{'}<X$, where $X$ represents
the UCS intensity predicted by the conditioned excitor employed for
the inhibitory conditioning, denoted CS1 in the following), or, equivalently,
conditioned with the active stimulation $X$ at a lower reinforcement
rate (which results in the inequality $\omega_{CS2-UCS}^{'}<\omega_{CS1-UCS}$)
and such that $\omega_{CS2-UCS}^{'}\cdot X=X^{'}$, is presented in
compound with the CSi, the reactive inhibition becomes equal to $\alpha(-X^{'})$.
More specifically, we argue that the CSi is able to proportionally
inhibit only the CSs excitors signalling an expected excitatory outcome
equal or smaller than $X$; furthermore, the inhibition for expected
outcomes greater than $X$ remains equal to $-\alpha X$, since the
emotional system does not know how the inhibitor behaves for larger
excitatory intensities.

Conversely, if during inhibitory conditioning the CSi has been obtained
through a \emph{partial reduction} of the active UCS elicitation,
say from $X$ to $X-\triangle X$, the inhibitory reactive response
is able to proportionally inhibit only an expected outcome comprised
between $X-\triangle X$ and $X$, but not smaller than $X-\triangle X$;
this is due to the fact that the emotional system has learned that
an outcome smaller than $X-\triangle X$ represents a residual response
which cannot be inhibited. For these reasons, for the last mentioned
case, the reactive inhibition response can be expressed as

\begin{equation}
i_{Ri}=-\alpha\left[y_{expected}-\left(X-\triangle X\right)\right]\label{eq:ir-inhibit2a}
\end{equation}
when $X-\triangle X\leq y_{expected}\leq X$, and as

\begin{equation}
i_{Ri}=-\alpha\cdot\triangle X\label{eq:ir-inhibit2b}
\end{equation}
when $y_{expected}\geq X$, and, finally, as

\begin{equation}
i_{Ri}=0\label{eq:ir-inhibit2c}
\end{equation}
when $y_{expected}<X-\triangle X$.

From the above results, it is easily inferred that the CSi inhibitory
effect depends not only on the expected excitatory outcome (i.e.,
$y_{expected}$), but also on the inhibition intensity and on the
residual (not inhibited) response generated during the learning process
(i.e., $-\triangle X$ and $X-\triangle X$, respectively). These
concepts are summarised in Fig. 5.

It is worth noting that, if a previously conditioned CSi (which it
has to remember is coded as a source of stimulation) is presented
alone (i.e., without a paired UCS or CS), an extinction cannot occur
since the expected outcome is equal to zero; consequently, the associated
reactive response $i_{Ri}$ is equal to zero too (see Eqs. (\ref{eq:ir-inhibit1})
and (\ref{eq:ir-inhibit2c})) and no error signal is computed. Moreover,
if the conditioning inhibition has occured by pairing a CS excitor,
the CSi and an UCS stimulation intensity lower than the one employed
during CS excitatory conditioning, the presentation of the CSi alone
can even strenghten its inhibitory effect. This is due to the fact
that the synaptic connection between the CSi and the CS, which in
turn is connected to the UCS representation (which is excitatory)
is extinguished (or weakened) by repeated presentations of the CSi
alone \cite{Harris2014}. This effect, indeed, can also be obtained
by extinguishing the CS employed in the inhibitory conditioning \cite{Harris2014};
this experimental evidence supports the existence of the associative
connections CSi-CS-UCS. For these reasons, the CSi could exert even
an excitatory response when paired with a new CS2 (i.e., a different
CS than that empolyed during the inhibitory conditioning) predicting
(i.e., signalling) an excitatory outcome smaller than $X-\triangle X$
(i.e., smaller than the residual excitatory response generating during
the inhibitory conditioning) \cite{devito1987,Harris2014}. In fact,
in this case, the reactive inhibitory response $i_{Ri}$ associated
with CSi is equal to zero (see Eq.(\ref{eq:ir-inhibit2c})), and the
associative synaptic connections, generated during inhibitory conditioning,
(CSi-CS-UCS) from CSi to CS (which in turn is connected to the excitatory
UCS, which is associated with the reactive response $i_{R}$) determine
the reactive response equal to $\omega_{CSi-CS}\cdot\omega_{CS-UCS}\cdot i_{R}$
(if not extinguished through repeated presentations of the CSi alone,
or through the original CS extinction). All the results illustrated
above, namely the impossibility of extinguishing a CS inhibitor through
its repeated presentation (or even its inhibitory strengthen), and
the fact that the inhibitory effect of the conditioned inhibitor (CSi)
depends on the expected outcome, on the UCS intensity decrease and
on the residual (i.e., not inhibited) response generating during the
inhibitory conditioning process (see Eqs. (\ref{eq:ir-inhibit1}-\ref{eq:ir-inhibit2c})),
are supported by experimental results illustrated in a growing body
of literature (e.g., see \cite{Cotton1982,devito1987,Miller1995,Wheeler2008,zimmer1974})
and reviewed in \cite{Harris2014}. It is worth to mention that the
above obtained results cannot be predicted by the simple Rescorla-Wagner
model \cite{Miller1995}, nor by the classical conditioning theory.

Finally, as in the case of some phylogenetic ``excitatory'' stimuli
(i.e., prepared biological and evolutionary fear relevant stimuli,
such as angry faces, spiders, snakes and others) which are coded in
the mammalian amygdala since birth \cite{Esteves1994,Flykt2007,Ohman1993,Ohman1994},
we argue that also \emph{phylogenetic inhibitors }could exist since
birth; in particular, the \emph{food} may represent a phylogenetic
inhibitor of \emph{hunger} (which in turn depends on physiological
internal states). The last interpretation is in line with the observation
that when the degree of hunger is relatively high, the reactive response
intensity $i_{Ri}$ associated with food representations is particularly
strong (see Eq. (\ref{eq:ir-inhibit1})). It is also important to
note that the inhibitory effect originated by consuming food, could
lead to a rewarding experience also because the \emph{contrast effect}.
This last consideration is supported by the Berlyne's \emph{``arousal
jag'' theory} \cite{Berlyne1960}, which postulates that a drop from
unusually high level of arousal (which is experienced as unpleasant;
represented by an high level of hunger in the example mentioned above)
to a low level is associated with a feeling of pleasure \cite{Joseph2015}.
This effect is described in more detail (also from a quantitative
perspective) in Section \ref{sub:contrast-continuous}.

\begin{figure*}
\noindent \begin{centering}
\includegraphics{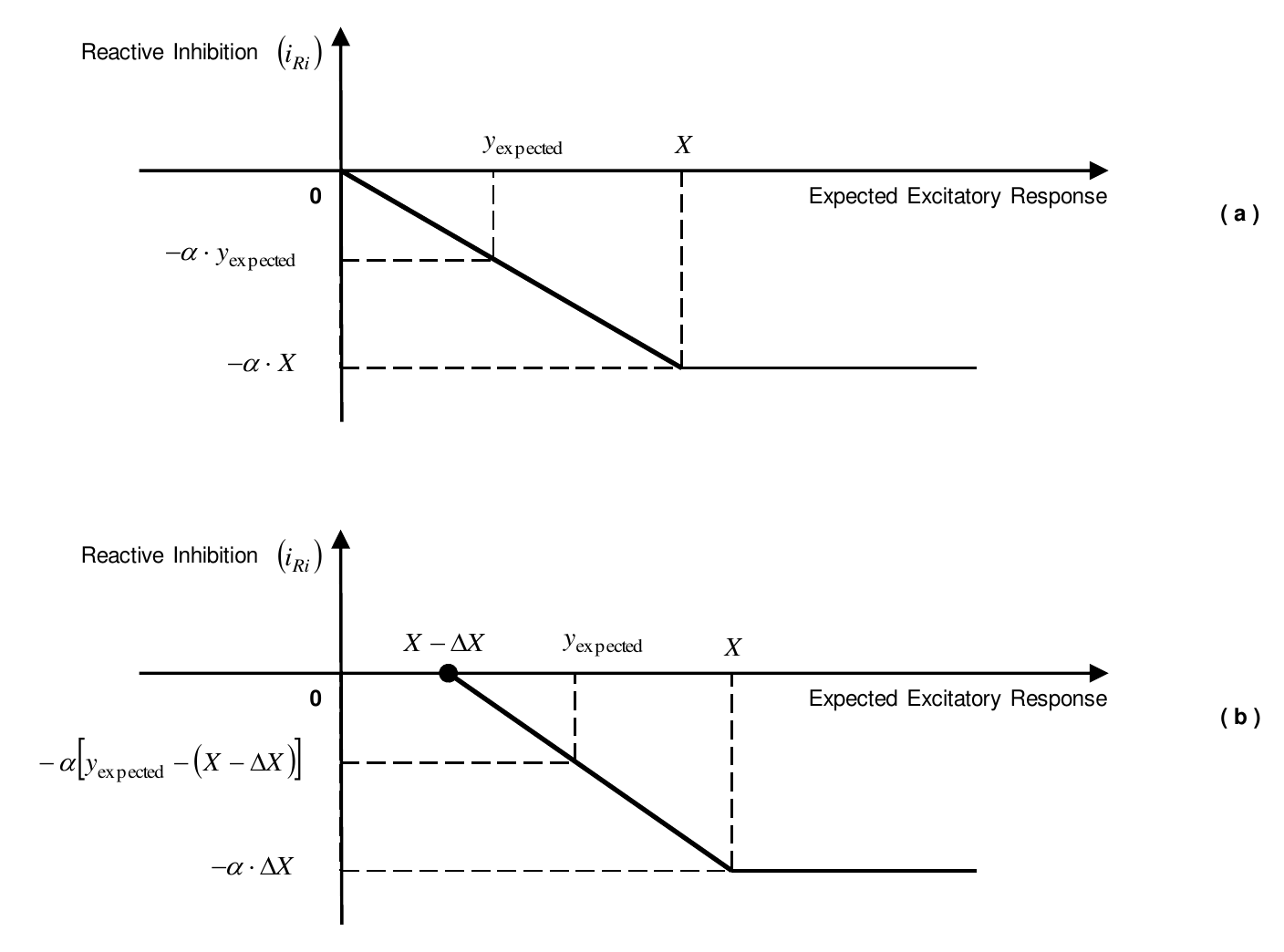}
\par\end{centering}

\caption{Representation of the reactive inhibitory response $i_{Ri}$, associated
with a conditioned inhibitor CSi (which is encoded by the emotional
system as a source of stmulation rather than as a conditioned stimulus),
over the expected excitatory outcome. In case a) the reactive inhibition
associated with CSi is effective on all the expected excitatory responses
greater than zero, since, during inhibitory conditioning, the emotional
system learned that the CSi is able to completely inhibit the excitatory
response. In case b) the reactive inhibition associated with CSi is
not effective for expected excitatory responses smaller than the value
given by $X-\triangle X$; this is due to the fact that the emotional
system has learned, during conditioning inhibition, that excitatory
intensities smaller than $X-\triangle X$ represent a residual response
which cannot be inhibited{\small{}. }In both cases, a) and b), if
the expected excitatory outcome is greater than $X$, which represents
the greatest expected excitatory intensity generated during inhibitory
conditioning, the $i_{Ri}$ coincides with the inhibitory intensity
associated with $X$; this is due to the fact that the emotional system
has not been trained for values higher than{\small{} $X$.}}
\end{figure*}

\section{Emotional Response Dynamics in Continuous Time Scale\label{sec:Emotional-Response-Dynamics}}

In the previous Sections a discrete-time model has been developed
for the evaluation of an emotional response in the presence of discrete
time trials. In real word conditions, however, an emotional source
might elicit continuously a subject in a certain time interval, so
that the inter-trial interval $T$ and the single discrete trial duration
$\triangle T$ tend to zero (in other words, a continuous elicitation
can be seen as a series of an infinite number of discrete active elicitations,
each of which has an infinitesimal time duration $\triangle T$ and
the temporal spacing between them tends to zero) . In these conditions,
the emotional response during the source-subject interaction cannot
be deemed constant and the dynamics of the response variation have
to be carefully assessed. Moreover, the reactive response is continuosly
updated driven by the continuous time counterpart of the error signal.
In the following, the problem of developing a mathematical model for
describing the continuous time evolution of an emotional response
is investigated, in order to extend our previous findings to the continuous
time scale. In principle, the theory of \emph{time scale calculus}
\cite{Bohner2001} should be applied to the considered problem in
order to devise a solution independent of the involved time scale.
In our analysis, however, a simpler approach, based on standard engineering
methods, is developed.

\subsection{A continuous time model for emotional dynamic learning \label{sub:continuous_model}}

To begin, we consider the proposed discrete-time model (see Eqs. (\ref{eq:final_d_a}-\ref{eq:third eq})
and, without any loss of generality, we focus on the model not accounting
for the response decay over time nor for the contrast effect (the
contrast effect will be discussed in Section \ref{sub:contrast-continuous};
moreover, from a quantitative perspective the effects of the exponential
decay can be accounted for increasing the value of the parameter $\alpha$).
The so-called \emph{bilinear transform} method \cite{Schafer2009}
is applied to this model in order to derive a continuous time counterpart
of it. It is not difficult to prove that this results in the differential
equation 

\begin{equation}
\begin{array}{c}
y'(t)=x(t)\cdot\frac{2}{T\cdot\left(1+\alpha\right)}+x'(t)\cdot\frac{1}{1+\alpha}-y(t)\cdot\frac{2\cdot\left(1-\alpha\right)}{T\cdot\left(1+\alpha\right)}\end{array}\label{eq:differential_eq}
\end{equation}
which describes the desired continuous dynamic model. In Eq. (\ref{eq:differential_eq})
the functions $y(t)$, $y'(t)$, $x(t)$ and $x'(t)$ represent the
elicited response, its first derivative, the active stimulation and
its first derivative over time, respectively; the parameter $T$,
which represents the ITI in the discrete time scale, in the continuous
time scale model represents the \emph{sensory} \emph{time discrimination
threshold }\cite{Luna2005} (i.e., the smallest temporal interval
for which the CNS neurons can discriminate between two distinct consecutive
stimulations).\footnote{Note that, generally speaking, the excitation decay between one stimulation
and the successive one is not negligible and the $\alpha$ parameter
increases as $T$ decreases (see Section \ref{sub:Discrete-trials-with-decay}).
Whenever the value of the parameter $T$ is experimentally estimated,
the associated value of the parameter $\alpha$ has also to be estimated
and employed in the continuous time equations (the estimation of a
vector $\mathbf{\mathbf{\alpha}}$ is required if multiple emotional
components are elicited).} Hence, the value of $T$ depends on the involved perceptive modality
(e.g., somatosensory stimulation, visual stimulation or acoustic stimulation).

If the condition $y_{n}=y_{n-1}$ (see Eq. (\ref{eq:third eq})) is
taken into account, solving this equation by standard methods produces

\begin{equation}
\begin{array}{c}
y(t)=\exp\left(\frac{2\left(\alpha-1\right)}{T(1+\alpha)}t\right)\Biggl(\frac{4\cdot\alpha}{T\left(\alpha+1\right)^{2}}\intop_{0}^{t}\exp\left(-\frac{2\left(\alpha-1\right)}{T(1+\alpha)}\tau\right)x(\tau)d\tau\\
+\frac{x(t)}{\alpha+1}\exp\left(-\frac{2\left(\alpha-1\right)}{T(1+\alpha)}t\right)\Biggl)-\exp\left(\frac{2\left(\alpha-1\right)}{T(1+\alpha)}t\right)\frac{x(0)}{\alpha+1}
\end{array}\label{eq:cont_time}
\end{equation}
with
\begin{equation}
y(t)=y\left(t^{*}\right)\:\textrm{if}\:y'(t^{*})=0\:\textrm{and}\:x(t)=x(t^{*})\:\forall t\geq t^{*}\label{eq:sec_eq}
\end{equation}
and
\begin{equation}
y(t)=y\left(t^{*}\right)\:\textrm{if}\:y'(t^{*})=0\:\textrm{and}\:x(t)=0\:\forall t\geq t^{*}.\label{eq:third_eq}
\end{equation}
Note that Eq. (\ref{eq:sec_eq}) represents the continuous time counterpart
of $y_{n}=y_{n-1}$ (see Eq. (\ref{eq:third eq})). In fact, the condition
$y_{n}-y_{n-1}=0$ turns into the differential condition $y'(t^{*})=0$
in the continuous time domain. Therefore, Eq. (\ref{eq:sec_eq}) means
that the emotional reactive response is not updated if the the first
derivative of the response itself is equal to zero at a generic time
instant $t^{*}$; consequently, the response at an instant $t\geq t^{*}$
equals the response at the instant $t^{*}$, provided that the active
response $x(t)$ remains constant in the time interval $\left(t,t^{*}\right)$.
In practice, however, it should be expected that $x(t)$ is a time
varying excitation, so that the response will experience continuous
changes. A simple interpretation can be also provided for Eq. (\ref{eq:third_eq}).
In fact, this refers to the case in which both the first derivative
of the response and the active response are equal to zero at the istant
$t=t^{*}$. In this case, the response will take on the constant value
$y(t^{*})$ for $t\geq t^{*}$ in the absence of an active elicitation
(i.e. if $x(t)=0,\forall t>t^{*}$); this means the resulting response
will be purely reactive. In practice, it should be expected that,
if $y'(t^{*})<\epsilon$ and $x(t^{*})<\kappa$ , where $\varepsilon$
$\kappa$ are small quantities, the response $y(t)$ will remain approximately
constant over a certain time interval. Moreover, if the active response
$x(t)$ becomes greater than zero at the instant $t=t^{*}$ when the
condition expressed by Eq. (\ref{eq:third_eq}) holds, it is not difficult
to prove that the response can be expressed as

\begin{equation}
y(t)=y_{1}+y_{2}(t)\label{eq:cont_time-1}
\end{equation}
for $t\geq t^{*}$, where
\begin{equation}
y_{1}(t)=y(t^{*})\label{eq:final_MISO-1}
\end{equation}
and

\begin{equation}
\begin{array}{c}
y_{2}(t)=e^{\left(\frac{2\left(\alpha-1\right)(t-t^{*})}{T(1+\alpha)}\right)}\Biggl(\frac{4\cdot\alpha\cdot}{T\left(\alpha+1\right)^{2}}\intop_{t^{*}}^{t}e^{\left(\frac{2\left(\alpha-1\right)}{T(1+\alpha)}\tau\right)}x(\tau)d\tau\\
+\frac{x(t)}{\alpha+1}e^{\left(-\frac{2\left(\alpha-1\right)}{T(1+\alpha)}(t-t^{*})\right)}\Biggl)
\end{array}\label{eq:final_MISO}
\end{equation}

The last result shows that the response $y(t)$ can considered as
the sum of two separate contributions, one representing the inextinguishable
reactive response ($y_{1}$), the other one ($y_{2}(t)$) accounting
for the possible variations of the physical (active) elicited response
$x(t)$. It is also important to point out that, if the condition
$y'(t^{+})=0\:\textrm{and }x(t^{+})=0$ occurs again at a successive
instant $t=t^{+}$, a new purely reactive response is elicited and
the model (\ref{eq:cont_time-1}) can be adopted to represent it.
Therefore, our mathematical model leads to the conclusion that an
inextinguishable\footnote{The adjective \emph{inextinguishable} is employed to indicate that
the source of stimulation cannot be extinguished through simple repetitive
expousures.} reactive response to a given source of stimulation (e.g., an electric
shock source or an acoustic noise source) can be obtained through
a suitable choice of the dynamics of an induced active response and
that, in principle, this emotional reaction could be indefinitely
strenghtened by repeating similar dynamics. 

Note also that, in order to obtain a reactive resistant-to-extinction
response, the dynamics of the active emotional response $x(t)$ have
to meet specific conditions, which, in turn, depend on the dynamics
of the physical features (e.g., electric voltage, frequency and amplitude
of a noise sound, etc.) of the adopted source. In other words, it
should be expected that a specific mapping function between the features
of a given physical source and the corresponding active emotional
response induced by it exists. If this function is known, the physical
features of the source can be controlled in a way to generate specific
dynamics in the active response; this, in turn, results in the generation
of an inextinguishable form of emotional reactive response. Therefore,
these theoretical findings suggest that, in principle, the inertial
nature of the \emph{emotional dynamic system} can be exploited to
originate a resistant-to-extinction emotional reaction.

It is worth mentioning that the asymptotic response which is obtained
from the Eq. (\ref{eq:cont_time}) assuming $T=1$ and a costant active
elicitation (i.e., $x(t)=X$) is $y=X/(1+\alpha)$. This last result
shows that if a continuous active and constant elicitation occurs
over time, the overall response reaches the same asymptotic value
which is reached in the discrete-time model during successive trials
(see Eq.(\ref{eq:fixed-point-d-a})). 

Finally, it is worth mentioning that an ``hybrid'' time scale should
be adopted, in some cases, in real world; for instance we could have
different discrete trials, each of which is sufficiently extended
over time so that the continuous time model should be used to analyse
it (note that this type of time scale is indicated by $\mathcal{\mathfrak{\mathscr{\mathbb{P}}}}$
in \cite{Bohner2001}). In this case both the discrete model and the
continuous time model have to be employed to describe the dynamics
of emotional learning.

\subsection{On the inclusion of contrast effects in the proposed continuous time
emotional model \label{sub:contrast-continuous}}

In the case of a continuous time source of stimulation (e.g., a continuous
acoustic stimulation, such as music) contrast effects can be exploited
to evoke specific emotional responses. It is well known that music
is able to evoke emotions, for instance, violating expectations or
shifting in time the rewards in a balanced mechanism based on frustration
(i.e., tension, as a state of dissonance, instability and uncertainty
\cite{Joseph2015}) and satisfaction (resolution towards consonant
and stable sounds experienced as pleasurable) \cite{Koelsch2014}.
Violation or retardation in resolution produces a tension increase
which may result in a successive stronger satisfaction during resolution
\cite{Joseph2015}. In this scenario, the contrast effect can explain,
for instance, why a slowly increasing tension could result in a successive
higher pleasure and rewarding effect after a sudden and unexpected
resolution. A qualitative description of the contrast effect on a
continuous time scale is expressed in Berlyne's \emph{``arousal jag''
theory} \cite{Berlyne1960}, which postulates that a drop from unusually
high level of arousal (which is experienced as unpleasant) to a low
level is associated with a feeling of pleasure \cite{Joseph2015}.
Apparently, the low level arousal alone is not able to elicit a pleasure
response; this means that only the \emph{contrast} between arousal
levels produces an emotional pleasure (i.e., a contrast effect). These
considerations motivate our interest in including the contrast effect
in our continuous-time dynamic model for the emotional system. Unluckily,
the discrete-time counterpart described in the Section \ref{sub:contrast-effect}
is nonlinear; consequently, the procedure adopted in the derivation
of Eq. (\ref{eq:differential_eq}) cannot be applied to this case.
Hence, the contrast effect on a continuous time scale can be analysed
by numerical simulation. A less refined approach is based on linearizing
the function $C(e_{A};T)$ (see Eq.(\ref{eq:contrast-function}))
and applying the bilinear transform method to derive a new differential
equation. In particular, if the Eq. (\ref{eq:approximated-contrast-d})
is adopted for the modeling of the contrast effect (and assuming that
$0<K<1$), the bilinear transform method applied to Eq. (\ref{eq:acq-contrast-d-approx})
leads to the following differential equation

\begin{equation}
\begin{array}{c}
y'(t)=x(t)\cdot\frac{2\left(1+K\right)}{T\cdot\left(1+\alpha+K\alpha-K\right)}+x'(t)\cdot\frac{1+K}{1+\alpha+K\alpha-K}+\\
-y(t)\cdot\frac{2\cdot\left(1-\alpha-K\alpha+K\right)}{T\cdot\left(1+\alpha+K\alpha-K\right)}
\end{array}\label{eq:eq-diff-2}
\end{equation}

The solution of the last equation is not reported here for the sake
of brevity. Finally, it is important to mention that the contrast
effect on a continuous time scale represents a further tool that can
be exploited to obtain a desired dynamics for an emotional reactive
response (e.g., for the elicitation of positive/negative emotional
responses, or to obtain a resistant-to-extinction response exploiting
the dynamics in a continuous time scale).

\section{Resistant-to-extinction emotional reaction through response saturation\label{sub:saturation}}

Generally speaking, any mathematical function $f(x)$ representing
a specific \emph{biological response} cannot take on arbitrarily large
values, because of the limited dynamics of the response itself. In
practice, as the value taken on by the argument $x$ grows, the corresponding
value $f(x)$ of the function does not steadily increase in proportion
to it and, when $x$ crosses a certain threshold, a certain saturation
level is reached; in other words, $f(x)$ exhibits a \emph{nonlinear}
behavior for sufficiently large values of $x$. In particular, these
considerations hold for the \emph{system chain function} $F_{Ch}(x)$
defined in our dynamic model (see Eq. (\ref{eq:yn-d-a})) and involved
in the computation of the emotional reaction, and for the \emph{amygdala
function} $F_{A}(x)$ (see Eq. (\ref{eq:function_amygdala})). In
the following, we focus on the second function and analyse the implications
of its nonlinear behevior; similar considerations, however, can be
expressed for the system chain function. In developing our model of
UCS revaluation (see Section \ref{sec:Quantitative-Analysis-of}),
we have assumed that $F_{A}(x)$ can be approximated as a linear function
(see Eq. (\ref{eq:approx-famygdala})) if in the $n$-th trial (or
at a specific time instant) the error signal takes on a value smaller
than a \emph{saturation threshold} $T_{S}$, which defines the linearization
range for $F_{A}(x)$ (see Fig. 6).

If, however, the error signal exceeds the saturation threshold $T_{S}$
(i.e., if the error signal becomes excessively large), a phenomenon
of \emph{emotional saturation} should be expected. This phenomenon
could occur during an extremely traumatic event, represented by a
unique event over a continuous time scale (see Section \ref{sec:Emotional-Response-Dynamics}),
or by an event involving successive acquisition trials and stimulations
on a discrete time scale (see Section \ref{sec:Quantitative-Analysis-of}).
It is worth noting, for instance, that if during a continuous-time
active elicitation the reactivity of the amygdala is relatively high
(which could be increased by the stress hormones and other factors
related to stress \cite{Kim2002}), the parameter $\alpha$ increases
accordingly (see Eqs. (\ref{eq:approx-famygdala}-\ref{eq:alfa}),
so that the reached asymptotic response, which can be expressed as
$y=X/(1-\alpha)$ (see Eq. (\ref{eq:fixed-point-d-a}) and Section
\ref{sub:continuous_model}) could grow until several times the intensity
of the active elicitation $X$ (for instance if $\alpha=0.9$ then
$y=10\cdot X$). It is also worth to mention that a purely reactive
response could grow until a saturation level in the absence of any
active stimulations, for instance through the so called ``social
learning of fear'' \cite{OlssonA2007,Olsson2007}. When emotional
saturation occurs, the source of stimulation (or any associated cue)
generating it could produce inextinguishable effects, even in the
absence of an active component (which corresponds to the term $X$
in eq. (\ref{eq:yn-d-a})). In fact, in the linear case, a purely
reactive elicitation in the absence of an active component (i.e.,
with $X=0$) leads to a negative error signal, since an active component
is expected. In turn this error signal weakens the amygdala response
(i.e., more precisely it produces a reduction by the amount $-\gamma\cdot X$;
see Eqs. (\ref{eq:function_amygdala}, \ref{eq:approx-famygdala})),
and hence leads to the extinction through successive trials (see Eq.
(\ref{eq:solution-d-e})). Conversely, if the amygdala function has
previously reached its saturation level, in the first trial of reactive
elicitation in the absence of the active component (e.g., a purely
reactive visual or auditory cue triggers the reactive emotional response
associated with the considered traumatic event) the computed negative
error signal is unable to reduce the amygdala response. Hence, in
the successive trials the error signal becomes equal to zero and no
response updating occurs. More specifically, this is true if the error
signal computed in the first trial, whose amplitude is equal to that
of the expected active component (i.e., $e_{1}=-X$), is smaller than
the saturation level reached by the amygdala function (see Fig. 6).
These considerations lead us to the conclusion that any stimulus perception,
even in the absence of an active elicitation, can trigger a reactive
emotional response. This holds both for the perception of the source
of stimulation and for any related cue; in fact, a cue (CS) conditioned
to the source (UCS) can trigger the reactive response associated with
the UCS (i.e., $i_{R}$), and since such a response is evident and
constant at every trial (i.e., the error signal is equal to zero)
the contingency between CS and the UCR ($i_{R}$) is reinforced at
every trial. Furthermore, the considerations made in the Section \ref{sub:Conditioning-to-a-reactive-source}
about a cue conditioned to a purely reactive source of stimulation
apply to this case too. We argue that this phenomenon could happen
in panic disorders and PTSD \cite{book_traumatic2012,Parsons2013,Perusini2016}.
As a matter of fact, in some forms of PTSD and panic disorders the
mere repetitive exposure to cues related to a traumatic event does
not lead to an extinction of emotional responses or results in a very
slow extinction \cite{Paunovic1999,Perusini2016,vanRooij2015}. It
is worth noting that during the reactive stimulation (e.g., a panic
attack) other previously neutral cues could be associated (i.e., conditioned
or misattributed) to the occuring reactive response, and successively
these cues might trigger the reactive response; hence, in turn, this
phenomenon can lead to a generalization of triggered panic attacks
(i.e., panic disorders). Note that the standard classical conditioning
model is unable able to explain these psychopathologies. Nevertheless,
one might argue that in PTSD patients the traumatic stimulus is generally
perceived in a context (and also in the presence of boundary conditions)
that differ from the one which originally caused the emotional response
saturation. Hence, the different perceived context should lead to
an (at least partial) inhibition of the elicited reactive response.
Nonetheless, it should be kept into account that: a) the inhibitory
strength originating from the discrimination of the contextual information
could be smaller than the degree of saturation reached by the amygdala
response during the considered traumatic event (see Fig. 6); b) contextual
information are primarily coded and stored in the hippocampus (conversely,
the representation of an aversive stimulus is coded within the BLA),
so that, if during the traumatic event the hippocampus does not properly
code contextual information, then no effective contextual discrimination
can be obtained during further expousures of the stimulus. As far
as this last point is concerned, it is worth mentioning that hippocampus
functioning and its ability to encode information (especially contextual
information) are impaired by uncontrollable stress (see \cite{Kim2002}
for a review on this topic). In particular, it is well known that
the hippocampus of a mammalian brain is a target of stress hormones
since it has one of the highest concentrations of receptors for \emph{corticosteroids}.
Consequently, certain hippocampal functions, such as learning and
memory, are susceptible to disruption by stress partly mediated by
corticosteroid receptors (mainly by the \emph{glucocorticoid receptors},
GRs). Interestingly, the model developed in \cite{Kim2002} shows
that alterations in hippocampal functioning require both stress hormones
and the active output from the amygdala, which in turn projects both
directly and indirectly to the hippocampus. For this reason, the amygdala
output is a crucial component on the stress-induced modulation of
hippocampal plasticity; in fact, when there is an experimentally induced
reduction in the amygdalar input to the hippocampus (as a result of
an inactivation or damage of the amygdala), plasticity in the hippocampus
remains intact under stress conditions \cite{Kim2002}. These results
support our hypothesis according to which the saturation of the amygdala
response during an extreme traumatic event, together with the release
of the stress hormones, could lead to an impairment of the hippocampus
functioning and contextual information encoding. In parallel, the
saturation of an emotional response component could cancel the effect
of the error signal, which, on the contrary, in physiological conditions
should produce an extinguishment of the emotional response in few
trials. 

It is also worth mentioning that cronic stress (e.g., PTSD) may induces
hippocampus atrophy \cite{Bremner1999} and impair neurogenesis, specifically
in the DG \cite{Kim2002}. This, in turn, may lead to a greater generalization
of anxiety disorders, in the impairment of emotional reactions inhibition
and of contextual cue processing \cite{vanRooij2015}.

\begin{figure*}
\noindent \begin{centering}
\includegraphics{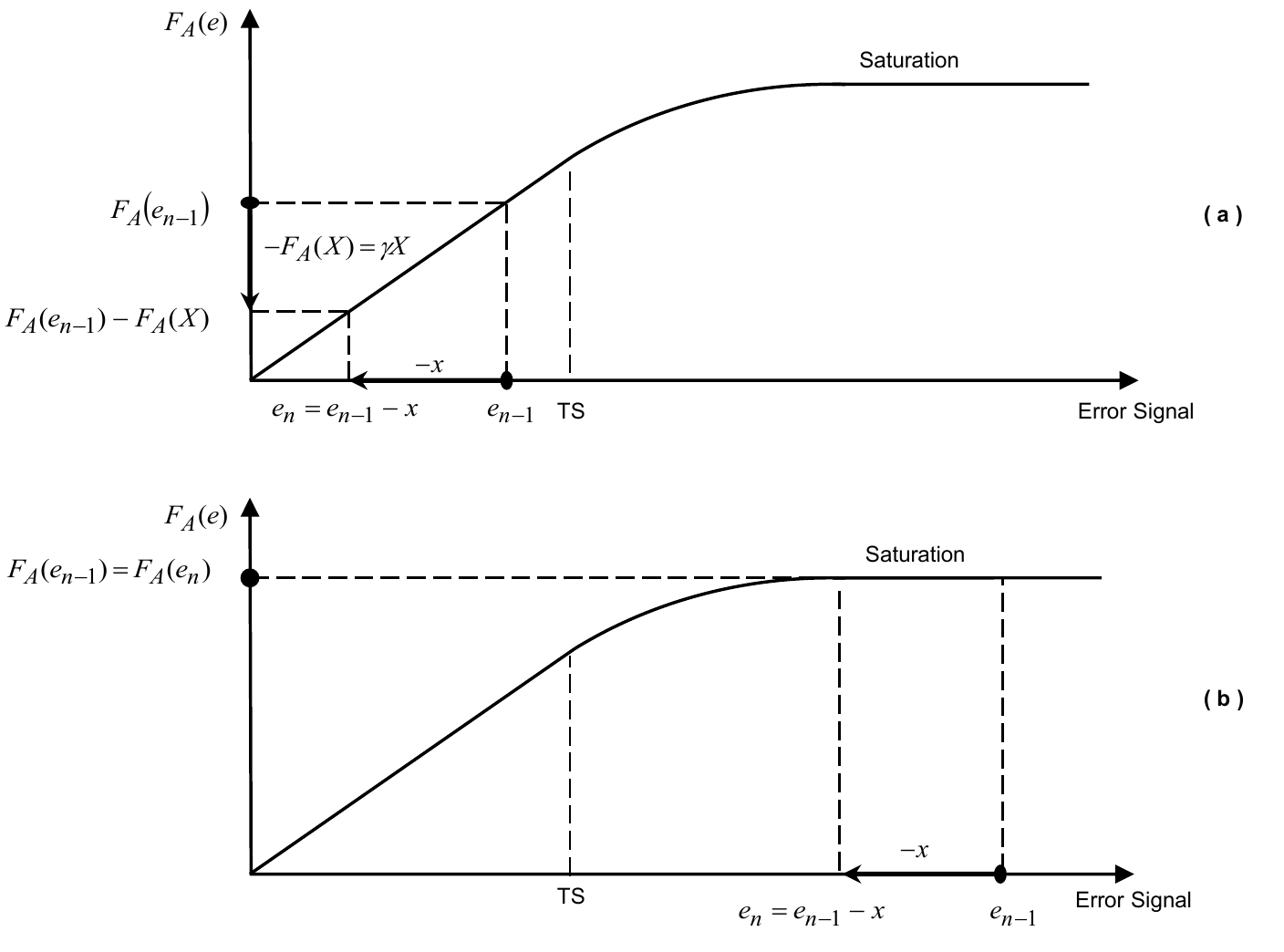}
\par\end{centering}

\caption{Schematic representation of the biological behavior of the amygdala
function $F_{A}(x)$ in its linear zone a) and in its saturation zone
b). In case a) the error signal (equal to $-x$) is able to reduce
the elicited emotional response. On the contrary, in case b) a negative
error signal is unable to produce a similar effect, so that the emotional
response remains at its saturation level. In particular, the case
b) occurs if the negative error signal (due, for instance, to the
fact that the active response $x$ is no more elicited during the
stimulation) is smaller than the degree of saturation reached by the
amygdala in the previous stimulation(s).}
\end{figure*}

\subsection{On the reactive response to a stimulus under drug administration\label{sub:reactive-versus-active}}

The proposed dynamic model can be also exploited to analyse the emotional
response over multiple trials when a reactive (and resistant-to-extinction)
response is contrasted by an active drug treatment (e.g., an anxiolytic
drug is employed to lower an anxiety reaction or panic attack elicited
by a given stimulus). Our interest in this analysis is motivated by
the possibility of quantitatively assessing the long term mitigation
of an undesired pathological emotional reaction provided that it is
administered an active drug during the elicitation of a reactive response;
in particular, we are interested in analysing the effects of the drug
withdrawal.

One might argue that, in the considered scenario, the mere exposition
to the source stimulus during the administration of a suitable drug
could be effective in reducing or extinguishing an undesired pathological
reactive response, such as those occurring in panic attacks, PTSD
or phobias. Indeed, as shown in Section \ref{sec:Quantitative-Analysis-of},
if a negative error signal (due to a response reduction) is computed
during a trial expousure, the reactive response is modified accordingly,
and this results in a less intense emotional reaction. Nonetheless,
it can be shown that, after the trial in which the drug administration
is suspendend, the reactive response asymptotically increases to its
original value, that is to the value of the response occuring before
the trial in which drug was administred. In fact, this result is due
to the computed error signal and holds in the absence of any cognitive
modulation of the reactive response (i.e., in the absence of emotional
\emph{reappraisal} \cite{Banks2007}). 

In our study of a reactive response contrasted by an active drug treatment
it is assumed that a given stimulus produces an inextinguishable reactive
response $i_{R}=Y_{0}$. Such a response could originate from different
events, like \emph{response saturation} during a traumatic event (see
Section \ref{sub:saturation}) or repetitive energizing trials of
a reactive response through an \emph{accumulation effect }(see Section
\ref{sub:Source-Misattribution:-quantitat}), or it can be natively
stored within the amygdala \cite{Ohman1994,Ohman1993} (e.g., arachnophobia).
For this reason, the initial condition

\begin{equation}
y_{0}=Y_{0}\label{eq:init_cond}
\end{equation}
is given. In the sequence of trials the effect of the administered
drug is always represented by an active elicitation, denoted $A=-\triangle Y_{0}$,
opposite to the reactive response (i.e., the active drug effect consists
of an inhibition of the reactive response). Then, from Eq. (\ref{eq:2})
it is easily inferred that

\begin{equation}
y_{1}=i_{R}+A_{1}=Y_{0}-\triangle Y_{0};\label{eq:drug-treatment}
\end{equation}
so that the error signal computed in the first trial is

\begin{equation}
e_{1}\triangleq y_{1}-y_{0}=-\triangle Y_{0}.
\end{equation}
Note that the active drug (represented by the term$-\triangle Y_{0}$
in the last equation) should inhibit the reactive response $y$, which
is expected to became weaker (i.e, should be obtained $y_{1}\simeq0$).
Furthermore, if the drug withdrawal starts in the second trial, the
resulting response $y_{2}$ can be expressed as (see Eq. \ref{eq:final_d_a})

\begin{equation}
\begin{array}{c}
y_{2}=Y_{0}+\mathrm{\alpha}\cdot e_{1}\\
=Y_{0}-\alpha\cdot\triangle Y_{0},
\end{array}
\end{equation}
so that the resulting error signal is 

\begin{equation}
\begin{array}{c}
e_{2}\triangleq y_{2}-y_{1}\\
=\left(1-\alpha\right)\cdot\triangle Y_{0}.
\end{array}
\end{equation}
Consequently, the response in the third trial is

\begin{equation}
\begin{array}{c}
y_{3}=i_{R}+\alpha\cdot e_{1}+\alpha\cdot e_{2}\\
=Y_{0}-\alpha^{2}\cdot\triangle Y_{0}.
\end{array}
\end{equation}
Following this line of reasoning, the response elicited in $n$-th
trial (with $n\geq2$) can be expressed as

\begin{equation}
y_{n}=Y_{0}-\alpha^{n-1}\cdot\triangle Y_{0}\label{eq:final-drug}
\end{equation}
The last equation can be also derived from the mathematical model
developed for the case of a phylogenetic source of stimulation, that
is from Eq. (\ref{eq:phylogenetic-d-a}) (a phylogenetic stimulus
is conceptually different from an undesired reactive response acquired
through emotional learning, but the emotional responses appearing
in these two cases are functionally identical). Moreover, from Eq.
(\ref{eq:final-drug}) it is easily inferred that, after drug withdrawal,
the emotional response asymptotically tends to its initial value ($Y_{0}$)
as the number of trials increases (since the inequality $\left|\alpha\right|<1$
holds). This result confirms that an active response (e.g., the effect
of an active drug) can certainly mitigate a reactive one; however,
it also shows that, if the active response vanishes (i.e., if the
drug is withdrawn in this case), the error signal becomes positive,
since the actual outcome is greater than the expected counterpart,
and the reactive response progressively returns to its initial value.
Note also that an active drug can exert its inhibition effect also
through a reduction of the amygdala reactivity (i.e., a reduction
of the parameter $\gamma$ derived from the amygdala function $F_{A}(x)$;
see Eqs. (\ref{eq:function_amygdala}, \ref{eq:approx-famygdala}))
and, consequently, of the $\alpha$ parameter. However, it is easy
to show that the mathematical result derived above also applies to
this case if it is assumed that the amygdala reactivity (i.e., the
behavior of $F_{A}(x)$) returns to its original conditions after
the drug withdrawal (actually, a time invariant behavior of the amygdala
function has been previously assumed to simplify the development of
our model). 

Unlike the case of an active drug treatment, if an undesired reactive
response (target response) is counteracted through an \emph{opposite}
\emph{reactive} elicitation (e.g., through a reactive inhibition),
then the effect of the counteraction will be permanent, since no active
component is expected. Hence, ideally, a reactive counteraction will
be learned and permanently sustained by the emotional learning system;
on the contrary, an active counteraction needs the presence of an
active response at every trial. For instance, psychotherapy and behavioral
treatments could be effective \emph{inhibitors of a} \emph{reactive
response} since they are able to modulate (e.g., reduce) a reactive
response within the amygdala through cognitive manipulations. This
is due to the fact that the human \emph{dorsolateral prefrontal cortex}
(DLPC) is able to direct higher level cognitive information, such
as ``self regulation thoughts'', to the OFC, which, in turn, can
integrate these information and modulate the original emotional reaction
stored within the amygdala (so that a \emph{reappraisal} effect is
obtained) \cite{Banks2007,Delgado2008a,Li2011,Plassmann2008}. In
summary, a \emph{conscious down-regulation of emotions} \cite{Banks2007}
allows to generate an error signal able to mitigate a reactive response
without requiring an external active source, like a drug. 

From a mathematical viewpoint, our last considerations can be motivated
as follows. The reappraisal (e.g., reduction) of the reactive response
of the amygdala by a quantity $-\bigtriangleup Y_{0}^{'}/\alpha$
through OFC modulation results in the response 

\begin{equation}
y_{1}=Y_{0}-\bigtriangleup Y_{0}^{'},
\end{equation}
so that the error signal computed in the first trial is

\begin{equation}
e_{1}\triangleq y_{1}-y_{0}=-\triangle Y_{0}^{'}.
\end{equation}
Hence, if it is assumed that no further session of cognitive reappraisal
occurs, the response in the second trial can be expressed as

\begin{equation}
\begin{array}{c}
y_{2}=Y_{0}-\bigtriangleup Y_{0}^{'}+\mathrm{\alpha}\cdot e_{1}\\
=Y_{0}-\bigtriangleup Y_{0}^{'}-\alpha\cdot\triangle Y_{0}^{'},
\end{array}
\end{equation}
Note the last equation still contains the term $-\bigtriangleup Y_{0}^{'}$,
since it originates directly from a reactive emotional downregulation
(and it is not due to the computation of an error signal after that
an active elicitation has been occured, like in the previous case;
see Eq. (\ref{eq:drug-treatment})). Such a reduction can be obtained,
for instance, by lowering the amygdala reactivity during the response
(e.g., by reducing the release of stress hormones, since these are
known to affect the amygdala output; \cite{Kim2002}); moreover, multiple
sessions of reactive modulation could lead to the extinction of the
considered reactive response. Note also that, in the absence of additional
sessions of emotional modulation, the asymptotic response can be expressed
as 

\begin{equation}
y_{\infty}=Y_{0}-\frac{\triangle Y_{0}^{'}}{1-\alpha}
\end{equation}
and this shows that the effect of the reactive inhibition persists
over successive trials.

\subsection{Drug treatments, reactive methods and emotional responses}

Since pathologies involving emotional disorders, such as panic disorders
or PTSD, usually lead to a generalization of their triggered responses
(i.e., an increased number of new stimuli becomes able to trigger
a pathological reactive response) and to the atrophy or impaired neurogenesis
of the hippocampus (in particular, of the DG) \cite{Kim2002}, specific
pharmacological treatments could be effective in blocking and reversing
these biological conditions. Nevertheless, as shown in the previous
Section, a treatment based on an active drug alone could be unable
to counteract the pathological reactive response on a long term (i.e.,
after the drug withdrawal). This means that active drugs, even if
are able to mitigate the symptoms due to strong inextinguishable responses,
cannot durably extinguish them. For this reason, the adoption of \emph{reactive}
\emph{methods} (e.g., psychotherapy treatments \cite{Lipka2014,Paunovic1999})
able to extinguish (or mitigate) a pathological reactive response
is required. These methods include all\emph{ }the techniques\emph{
}able to counteract a pathological emotional response through an opposite
reactive response or a reactive inhibition (consequently, the use
of any active component is avoided). We argue that different technologies,
like optogenetic manipulation of memory engrams and memory photostimulation,
could be adopted to achieve this target. In fact, on the one hand,
the optogenetic manipulation of the memory engrams within the DG or
the BLA \cite{Gore2015,Ramirez2015,Redondo2014} allows to switch
the valence of a CS in the DG from a positive (negative) UCS to an
opposite one \cite{Redondo2014}. Some results have evidenced that
the chronic reactivation of hippocampal cells associated with a positive
memory is able to suppress depression-like behavior in mice \cite{Ramirez2015}.
On the other hand, the memory photostimulation of UCS-responsive cells
on the BLA has been shown to produce valence-specific behaviors \cite{Gore2015}.
On the basis of the aforementioned results we argue that, if during
trials in which a subject is exposed to the cues related to a traumatic
event, positive valenced UCS-responsive cells were photostimulated,
then the outcome triggered by such cues (i.e., the target response)
could be weakened. In fact, the photostimulation-induced response
should be misattributed to the target cue and, consequently, the final
cue-related outcome should be due to the superposition of the original
target response with the optically elicited response; this should
result in a net decrease of the target response. Unluckily, optogenetic
neuronal manipulations and memory photostimulation cannot be used
on humans today because of their invasiveness. We argue, however,
that similar reactive counteractions could be obtained through a subliminal
expousure, exploiting the \emph{implicit accumulation effect }(see
Section \ref{sub:Source-Misattribution:-quantitat}). Finally, it
is useful to mention that techniques exploiting the misattribution
of reactive responses (both conditioned responses and purely reactive
responses) could be also employed to strengthen a desired response
(e.g., an unconscious placebo response).

\part*{Discussion}

In this manuscript a novel theory of emotional learning has been illustrated.
The theory shows the differentiation (and the relations) between classical
conditioning and the UCS revaluation, and provides various new insights
on well known psychophysiological phenomena and psychiatric diseases
(e.g., panic disorders and PTSD), and a number of new ideas for further
research. One of its most interesting implications is represented
by the identification of well defined mathematical and neurophysiological
conditions ensuring the inextinguishability of specific emotional
reactions. In particular, it allows us to establish the following
four different mechanisms through which a stimulus can produce a resistant-to-extinction
emotional reaction: 1)~misattribution of a reactive response (see
Corollary 1); 2) classical conditioning of a stimulus to a purely
reactive source of stimulation; 3) saturation of emotional response
(e.g., of the amygdala reactive response); 4) the exploitation of
the inertial dynamics of emotional learning system on a continuous
time scale. Further relevant contributions are represented by the
proof that the Rescorla-Wagner model for classical conditioning can
be obtained as a special case of the proposed model; the derivation
of a new model for conditioning, which accounts for the implicit UCS
revaluation and that is able to quantitatively describe important
experimental results (such as the impossibility to extinguish a conditioned
inhibitor through its repeated presentations), which are unpredictable
by existing classical conditioning models. Indeed, classical conditioning
has been deeply studied in the last century, while UCS revaluation
did not obtain the due attention, and, for this reason, actually,
appropriate analytical and experimental models for resistant-to-extinction
responses do not exist.

Our result paves the way for various new research activities. First
of all, various potential applications of our theory can be envisaged
in the hot research area concerning the manipulation of human behaviours
and emotional reactions. Note that our theory unveils the real differences
between an UCS and a CS from the perspective of the emotional system
and, in particular, shows that removing a CS does not ensure the extinction
of an undesired emotional reaction, since, as long as a UCS reaction
remain stored (or relatively strength), it is able to form other CS-UCS
connections, or to duplicate its associated reaction through a misattribution
effect or even to strenghten itself through incubation mechanisms.
Furthermore, if an UCS reaction is extinguished or weakened, all the
associated CSs are weakened. Thus, establishing a well defined differentiation
between a cue (i.e., a CS) and a source of stimulation (i.e., UCS)
is fundamental in order to eradicate an emotional undesired reaction.
This fundamental differentiation could be effectively assessed exploiting
\emph{Remark} (\ref{rem:UCS-in-amygdala}), i.e. testing the subliminal
emotional reactive response through subliminal exposure. Indeed, as
shown in our analysis, no difference between CS and UCS might be sensed
in the course of a supraliminar perception. In addition, from a conscious
perspective, a source or a cue-stimulus could be hardly differentiated,
because of misattribution effects or response duplications (e.g.,
an initial CS could be misattributed and become an UCS); note also
these and other related attribution phenomena often occur automatically
and unconsciously \cite{Anderson1989,Uleman1987}, and are not explained
by classical conditioning theory.

Another interesting research topic is represented by the potential
applications of Corollary 1, i.e. by the possibility of devising UCSs
able to elicitate reactive responses. Such responses could be strenghtened
through an \emph{accumulation effect} and potentially exploited to
modify the emotional reaction to a given stimulus (e.g., to weaken
an opposite valenced emotional reaction or to enhance the unconscious
placebo effect of a drug \cite{Jensen2012,Jensen2014}). The idea
of generating an accumulated reactive response could be really useful
from a therapeutic viewpoint since, in practice, the active modulation
of a given reactive response (accomplished, for instance, through
the action of a pharmacological drug) vanishes over time if the associated
active component decays too (in the considered example this occurs
if the administered drug is withdrawn). Furthermore, other ``reactive
methods'' could be exploited in order to modulate (i.e., to increase
or decrease) a target response; these methods include but are not
limited to: the misattribution of a CR forward the target response,
the cognitive reappraisal of an emotional response \cite{Banks2007}
(e.g., psychotherapy), optogenetic manipulations \cite{Gore2015,Ramirez2015,Redondo2014}. 

Once again, it is worth stressing that these results rely on the differentation
between the types of emotional response (such as \emph{active}, \emph{reactive}
and \emph{passive residual} responses) that can come into play during
a stimulation (and, more precisely, during UCS revaluation) and the
concepts of \emph{source attribution} and \emph{misattribution}, and
that these relevant factors are not taken into consideration in classical
conditioning models.

A further relevant research topic concerns the applications of our
model of emotional learning on a continuous time scale. Generally
speaking, this model could be exploited to analyse the emotional reaction
generated by any physical stimulation which varies continuosly over
time (e.g., a time-varying acoustic source of stimulation, such as
music \cite{Koelsch2014}). We feel, however, that our model is preliminary
and that it should be improved by including some features of hippocampus
filtering (for instance, the temporal pattern recognition, that is
the detection of regular patterns within the time-varying elicited
response, should be included). 

Finally, it is important to mention that our theoretical framework
can be exploited for the development of psychophysiological experimental
models for both animals and humans; these, in turn, can potentially
provide new insights into emotion-related phenomena and pathologies.

\newpage{}

\section*{Appendix}

In this Appendix we describe an experiment useful to prove that the
encoding of an aversive object as a conditioned stimulus (CS) to a
primary source of \emph{punishment} (i.e., CS-UCS encoding through
aversive classical conditioning acquisition) or as a primary source
of stimulation (UCS) involves different regions of the mammalian brain. 

The procedure adopted in this experiment is obtained by including
a new part in the procedure followed in \cite{Flykt2007} and is sketched
below. The proposed experiment is based on the use of a specific device,
never observed by the involved subjects and having a neutral shape;
this device, however, is able to produce an aversive stimulus, like
an electric shock or strong noise. Morever, it involves two distinct
groups of subjects, one denoted group+, the other one group-. In practice,
the experiment consists of three different phases, each involving
multiple trials. In the first phase the selected device has to directly
elicit an aversive response for some trials in the members of group+,
so that their emotional learning systems attribute the elicited responses
to that source of stimulation, learn the associated reactive responses,
and finally code and store them within their amygdalas. In the group-,
the same device has to be conditioned to an identical aversive response
elicited in the group+; however, all the subjects of group- must be
aware of the fact that the aversive stimulus originates from another
source of stimulation (denoted UCSa) and must learn the stimulus contingencies
(in other words, classical conditioning learning occurs for group-,
instead of the direct response attribution characterizing group+).
In particular, the picture of the new device pointed toward the subject
can be easily conditioned instead of the device itself (since, otherwise,
it would be quite difficult to perform a conditioning procedure by
which the physical object is paired with the UCSa elicitation). In
the second phase, a \emph{differential conditioning} paradigm is adopted.
For this reason, each subject of group+ has to be conditioned with
the pairing of the picture of this device, pointed toward the subject
and denoted CS+, with a new UCS (e.g. and electric shock); the last
UCS, called UCSb, must be different from UCSa, that is employed for
the conditioning of group-. Moreover, another threatening element
(e.g., a gun), called CS-, is presented to the subjects of group+
in the absence of an UCS. The same procedure is followed for the subjects
of group-; these have previously acquired the given device as a conditiond
cue and not as a direct source of stimulation. Finally, in the third
phase, a masked (unmasked) extinction phase for half subjects of group+
(group-) will be able to reveal which CSs+ are able to elicit a differential
autonomic response (i.e. a SCR) even under a subliminally (masked)
perception. If group+ only will exhibit a differential autonomic response
in the masked extinction (with respect to the SCR elicited by the
masked CS- perception), it will be inferred that only the representation
of the CS+ within group+ can be activated through the sub-cortical
thalamus-amygdala pathway; therefore, we will come to the conclusion
that the the representation of CS+ has been stored within the amygdala
in the same site as other UCSs (e.g., phylogenetic or ontogenetic
threatening stimuli \cite{Esteves1994,Flykt2007,Ohman1993,Ohman1993b,Ohman1994})
have been stored (i.e., the BLA). Conversely, if the subjects of group-
are considered, the CS+ (which is the same picture as that employed
for group+), represents an aversive object which has not been stored
within the BLA, or, in any case, not in a brain region activated by
the sub-cortical thalamo-amygdala pathway (like the phylogenetic or
ontogenetic stimuli).

We believe that, if the envisaged results will be obtained, they will
also definitively prove that the mechanism through which the selected
aversive object has been encoded (i.e., through classical conditioning
acquisition, or by direct or implicit response attribution) makes
the difference.

\newpage{}

\section*{References }

\noindent \bibliographystyle{vancouver/vancouver}
\bibliography{references_29-11sidita_APA}

\begin{thebibliography}{100}

\bibitem{Flykt2007}
Flykt A, Esteves F, Ohman A.
\newblock Skin conductance responses to masked conditioned stimuli:
  phylogenetic/ontogenetic factors versus direction of threat?
\newblock Biological Psychology. 2007;74(3):328--336.

\bibitem{Ohman1993}
Ohman A, Soares JJ.
\newblock On the automatic nature of phobic fear: conditioned electrodermal
  responses to masked fear-relevant stimuli.
\newblock Journal of Abnormal Psychology. 1993;102(1):121--132.

\bibitem{Ohman1993b}
Ohman A.
\newblock Fear and anxiety as emotional phenomena.
\newblock In: Arne~Lewis M, Haviland JM, editors. Handbook of emotions. New
  York: Guilford Press; 1993. p. 511--536.

\bibitem{Damasio1996}
Damasio AR.
\newblock The somatic marker hypothesis and the possible functions of the
  prefrontal cortex.
\newblock Philosophical Transactions of the Royal Society of London Series B,
  Biological Sciences. 1996;351(1346):1413--1420.

\bibitem{Benedetti2008}
Benedetti F.
\newblock Mechanisms of placebo and placebo-related effects across diseases and
  treatments.
\newblock Annual Review of Pharmacology and Toxicology. 2008;48:33--60.

\bibitem{Enck2008}
Enck P, Benedetti F, Schedlowski M.
\newblock New insights into the placebo and nocebo responses.
\newblock Neuron. 2008;59(2):195--206.

\bibitem{Petrovic2002}
Petrovic P, Kalso E, Petersson KM, Ingvar M.
\newblock Placebo and opioid analgesia-- imaging a shared neuronal network.
\newblock Science. 2002;295(5560):1737--1740.

\bibitem{Watson2009}
Watson A, El-Deredy W, Iannetti GD, Lloyd D, Tracey I, Vogt BA, et~al.
\newblock Placebo conditioning and placebo analgesia modulate a common brain
  network during pain anticipation and perception.
\newblock Pain. 2009;145(1-2):24--30.

\bibitem{Weimer2015}
Weimer K, Colloca L, Paul E.
\newblock Placebo effects in psychiatry: mediators and moderators.
\newblock Lancet Psychiatry. 2015;2(3):246--257.

\bibitem{Blanchfield2014}
Blanchfield AW, Hardy J, Marcora SM.
\newblock Non-conscious visual cues related to affect and action alter
  perception of effort and endurance performance.
\newblock Frontiers in Human Neuroscience. 2014;8(967).

\bibitem{Radel2009}
Radel R, Sarrazin P, Pelletier L.
\newblock Evidence of subliminally primed motivational orientations: the
  effects of unconscious motivational processes on the performance of a new
  motor task.
\newblock Journal of Sport and Exercise Psychology. 2009;31(5):657--674.

\bibitem{Roy2009}
Roy M, Piche M, Chen JI, Peretz I, Rainville P.
\newblock Cerebral and spinal modulation of pain by emotions.
\newblock Proceedings of the National Academy of Sciences of the United States
  of America. 2009;106(49):20900--20905.

\bibitem{Wagener2009}
Wagner G, Koschke M, Leuf T, Schlosser R, Bar KJ.
\newblock Reduced heat pain thresholds after sad-mood induction are associated
  with changes in thalamic activity.
\newblock Neuropsychologia. 2009;47(4):980--987.

\bibitem{Wiech2009}
Wiech K, Tracey I.
\newblock The influence of negative emotions on pain: behavioral effects and
  neural mechanisms.
\newblock Neuroimage. 2009;47(3):987--994.

\bibitem{Pavlov1927}
Pavlov IP.
\newblock Conditioned Reflexes: An Investigation of the Physiological Activity
  of the Cerebral Cortex.
\newblock Oxford University Press. 1927;.

\bibitem{Fanselow2005}
Fanselow M, Poulos A.
\newblock The neuroscience of mammalian associative learning.
\newblock Annual Review of Psychology. 2005;p. 56:207--234.

\bibitem{Kim2006}
Kim JJ, Jung MW.
\newblock Neural circuits and mechanisms involved in Pavlovian fear
  conditioning: a critical review.
\newblock Neuroscience and Biobehavioral Reviews. 2006;30(2):188--202.

\bibitem{Rescorla1974}
Rescorla RA.
\newblock Effect of inflation of the unconditioned stimulus value following
  conditioning.
\newblock Journal of Comparative and Physiological Psychology.
  1974;86(1):101--106.

\bibitem{Davey1989}
Davey GC.
\newblock UCS revaluation and conditioning models of acquired fears.
\newblock Behaviour Research and Therapy. 1989;27(5):521--528.

\bibitem{Gottfried2004}
Gottfried JA, Dolan RJ.
\newblock Human orbitofrontal cortex mediates extinction learning while
  accessing conditioned representations of value.
\newblock Nature Neuroscience. 2004;7(10):1144--1152.

\bibitem{Hosoba2001}
Hosoba T, Iwanaga M, Seiwa H.
\newblock The effect of UCS inflation and deflation procedures on 'fear'
  conditioning.
\newblock Behaviour Research and Therapy. 2001;39(4):465--475.

\bibitem{Schultz2013}
Schultz DH, Balderston NL, Geiger JA, Helmstetter FJ.
\newblock Dissociation between implicit and explicit responses in
  postconditioning UCS revaluation after fear conditioning in humans.
\newblock Behavioral Neuroscience. 2013;127(3):357--368.

\bibitem{Schultz2000}
Schultz W.
\newblock Multiple reward signals in the brain.
\newblock Nature Reviews Neuroscience. 2000;1(3):199--207.

\bibitem{Schultz2006}
Schultz W.
\newblock Behavioral theories and the neurophysiology of reward.
\newblock Annual Review of Psychology. 2006;57:87--115.

\bibitem{Amanzio1999}
Amanzio M, Benedetti F.
\newblock Neuropharmacological Dissection of Placebo Analgesia:
  Expectation-Activated Opioid Systems versus Conditioning-Activated Specific
  Subsystems.
\newblock Journal of Neuroscience. 1999;19(1):484--94.

\bibitem{Guo2010}
Guo JY, Wang JY, Luo F.
\newblock Dissection of placebo analgesia in mice: the conditions for
  activation of opioid and non-opioid systems.
\newblock J Psychopharmacol (Oxford). 2010;24(10):1561--1567.

\bibitem{Haour2005}
Haour F.
\newblock Mechanisms of the placebo effect and of conditioning.
\newblock Neuroimmunomodulation. 2005;12(4):195--200.

\bibitem{Zillmann1971}
Zillmann D.
\newblock Excitation transfer in communication-mediated aggressive behavior.
\newblock Journal of Experimental Social Psychology. 1971;7(4):419 -- 434.

\bibitem{Zillmann1972}
Zillmann D, Katcher AH, Milavsky B.
\newblock Excitation transfer from physical exercise to subsequent aggressive.
\newblock Journal of Experimental Social Psychology. 1972;8(3):247--259.

\bibitem{Anderson1989}
Anderson CA.
\newblock Temperature and aggression: ubiquitous effects of heat on occurrence
  of human violence.
\newblock Psychological Bulletin. 1989;106(1):74--96.

\bibitem{Bryant2003a}
Bryant J.
\newblock 2.
\newblock In: Bryant J, Roskos-Ewoldsen DR, Cantor J, editors. Communication
  and Emotion: Essays in Honor of Dolf Zillmann. Routledge Communication
  Series. Routledge; New Ed edition; 2003. p. 39--40.

\bibitem{Cotton1981}
Cotton JL.
\newblock A review of research on Schachter's theory of emotion and the
  misattribution of Arousal.
\newblock European Journal of Social Psychology. 1981;11(4):365--397.

\bibitem{Jones2009}
Jones CR, Fazio RH, Olson MA.
\newblock Implicit misattribution as a mechanism underlying evaluative
  conditioning.
\newblock Journal of Personality and Social Psychology. 2009;96(5):933--948.

\bibitem{Uleman1987}
Uleman JS.
\newblock Consciousness and Control The Case of Spontaneous Trait Inferences.
\newblock Peronality and Social Psychology Bulletin. 1987;13(3):337--354.

\bibitem{flaherty1982}
Flaherty CF.
\newblock Incentive contrast: A review of behavioral changes following shifts
  in reward.
\newblock Animal Learning and Behavior. 1982;10(4):409--440.

\bibitem{Miller1995}
Miller RR, Barnet RC, Grahame NJ.
\newblock Assessment of the Rescorla-Wagner model.
\newblock Psychological Bulletin. 1995;117(3):363--386.

\bibitem{Rescorla1972}
Rescorla RA, Wagener AR.
\newblock 3.
\newblock In: Black AH, Prokasy WF, editors. A theory of Pavlovian
  conditioning: Variations in the effectiveness of reinforcement and
  nonreinforcement. Appleton-Century-Crofts, New York; 1972. p. 64--99.

\bibitem{Doherty2003}
O'Doherty JP, Dayan P, Friston K, Critchley H, Dolan RJ.
\newblock Temporal difference models and reward-related learning in the human
  brain.
\newblock Neuron. 2003;38(2):329--337.

\bibitem{Schultz1997}
Schultz W, Dayan P, Montague PR.
\newblock A neural substrate of prediction and reward.
\newblock Science. 1997;275(5306):1593--1599.

\bibitem{Sutton1988}
Sutton RS.
\newblock Learning to predict by the methods of temporal differences.
\newblock Machine Learning. 1988;p. 3, 9--44.

\bibitem{Sutton1990}
Sutton RS, Barto AG.
\newblock Time-derivative models of pavlovian reinforcement.
\newblock In: Gabriel M, J~Moore E, editors. Learning and Computational
  Neuroscience: Foundations of Adaptive Networks. MIT Press.; 1990. p. 59,
  229--243. 497--537.

\bibitem{Friston2008}
Friston K.
\newblock Hierarchical models in the brain.
\newblock PLoS Computational Biology. 2008;4(11):e1000211.

\bibitem{friston2009}
Friston KJ, Daunizeau J, Kiebel SJ.
\newblock Reinforcement learning or active inference?
\newblock PLoS ONE. 2009;4(7):e6421.

\bibitem{friston2010}
Friston KJ, Daunizeau J, Kilner J, Kiebel SJ.
\newblock Action and behavior: a free-energy formulation.
\newblock Biological Cybernetics. 2010;102(3):227--260.

\bibitem{Berns2001}
Berns GS, McClure SM, Pagnoni G, Montague PR.
\newblock Predictability modulates human brain response to reward.
\newblock Journal of Neuroscience. 2001;21(8):2793--2798.

\bibitem{Garrison2013}
Garrison J, Erdeniz B, Done J.
\newblock Prediction error in reinforcement learning: a meta-analysis of
  neuroimaging studies.
\newblock Neuroscience and Biobehavioral Reviews. 2013;37(7):1297--1310.

\bibitem{Bray2007}
Bray S, O'Doherty J.
\newblock Neural coding of reward-prediction error signals during classical
  conditioning with attractive faces.
\newblock Journal of Neurophysiology. 2007;97(4):3036--3045.

\bibitem{Delgado2008b}
Delgado MR, Li J, Schiller D, Phelps EA.
\newblock The role of the striatum in aversive learning and aversive prediction
  errors.
\newblock Philosophical Transactions of the Royal Society of London Series B,
  Biological Sciences. 2008;363(1511):3787--3800.

\bibitem{Li2014}
Li SS, McNally GP.
\newblock The conditions that promote fear learning: prediction error and
  Pavlovian fear conditioning.
\newblock Neurobiology of Learning and Memory. 2014;108:14--21.

\bibitem{McNally2011}
McNally GP, Johansen JP, Blair HT.
\newblock Placing prediction into the fear circuit.
\newblock Trends in Neurosciences. 2011;34(6):283--292.

\bibitem{schultz2000b}
Schultz W, Dickinson A.
\newblock Neuronal coding of prediction errors.
\newblock Annual Review of Neuroscience. 2000;23(1):473--500.

\bibitem{Steinberg2013}
Steinberg EE, Keiflin R, Boivin JR, Witten IB, Deisseroth K, Janak PH.
\newblock A causal link between prediction errors, dopamine neurons and
  learning.
\newblock Nature Neuroscience. 2013;16(7):966--973.

\bibitem{Waelti2001}
Waelti P, Dickinson A, Schultz W.
\newblock Dopamine responses comply with basic assumptions of formal learning
  theory.
\newblock Nature. 2001;412(6842):43--48.

\bibitem{Gore2015}
Gore F, Schwartz EC, Brangers BC, Aladi S, Stujenske JM, Likhtik E, et~al.
\newblock Neural Representations of Unconditioned Stimuli in Basolateral
  Amygdala Mediate Innate and Learned Responses.
\newblock Cell. 2015;162(1):134--145.

\bibitem{Redondo2014}
Redondo RL, Kim J, Arons AL, Ramirez S, Liu X, Tonegawa S.
\newblock Bidirectional switch of the valence associated with a hippocampal
  contextual memory engram.
\newblock Nature. 2014;513(7518):426--430.

\bibitem{Amit1994}
Amit DJ, Fusi S.
\newblock Dynamic learning in neural networks with material synapses.
\newblock Neural Computation. 1994;p. 6, 957--982.

\bibitem{Fusi2002}
Fusi S.
\newblock Hebbian spike-driven synaptic plasticity for learning patterns of
  mean firing rates.
\newblock Biological Cybernetics. 2002;87(5-6):459--470.

\bibitem{Fusi2007}
Fusi S, Abbott LF.
\newblock Limits on the memory storage capacity of bounded synapses.
\newblock Nature Neuroscience. 2007;10(4):485--493.

\bibitem{Hebb1949}
Hebb DO.
\newblock The organization of behavior.
\newblock New York: Wiley; 1949.

\bibitem{Soltani2006}
Soltani A, Wang XJ.
\newblock A biophysically based neural model of matching law behavior:
  melioration by stochastic synapses.
\newblock Journal of Neuroscience. 2006;26(14):3731--3744.

\bibitem{Soltani2010}
Soltani A, Wang XJ.
\newblock Synaptic computation underlying probabilistic inference.
\newblock Nature Neuroscience. 2010;13(1):112--119.

\bibitem{Young1976}
Young RA, Cegavske CF, Thompson RF.
\newblock Tone-induced changes in excitability of abducens motoneurons and of
  the reflex path of nictitating membrane response in rabbit (Oryctolagus
  cuniculus).
\newblock Journal of Comparative and Physiological Psychology.
  1976;90(5):424--434.

\bibitem{Rescorla1969}
Rescorla RA.
\newblock Pavlovian conditioned inhibition.
\newblock Psychological Bulletin. 1969;72(2):77--94.

\bibitem{devito1987}
DeVito PL, Fowler H.
\newblock Enhancement of conditioned inhibition via an extinction treatment.
\newblock Animal Learning and Behavior. 1987;15(4):448--454.

\bibitem{Harris2014}
Harris JA, Kwok DW, Andrew BJ.
\newblock Conditioned inhibition and reinforcement rate.
\newblock Journal of Experimental Psychology: Animal Learning and Cognition.
  2014 Jul;40(3):335--354.

\bibitem{Koelsch2014}
Koelsch S.
\newblock Brain correlates of music-evoked emotions.
\newblock Nature Reviews Neuroscience. 2014;15(3):170--180.

\bibitem{Meuret2006}
Meuret AE, White KS, Ritz T, Roth WT, Hofmann SG, Brown TA.
\newblock Panic attack symptom dimensions and their relationship to illness
  characteristics in panic disorder.
\newblock Journal of Psychiatric Research. 2006;40(6):520--527.

\bibitem{book_traumatic2012}
Beck JG, Sloan DM.
\newblock In: The Oxford Handbook of Traumatic Stress Disorders. Oxford
  University Press; 2012. p. 176--177.

\bibitem{Parsons2013}
Parsons RG, Ressler KJ.
\newblock Implications of memory modulation for post-traumatic stress and fear
  disorders.
\newblock Nature Neuroscience. 2013;16(2):146--153.

\bibitem{Perusini2016}
Perusini JN, Meyer EM, Long VA, Rau V, Nocera N, Avershal J, et~al.
\newblock Induction and Expression of Fear Sensitization Caused by Acute
  Traumatic Stress.
\newblock Neuropsychopharmacology. 2016;41(1):45--57.

\bibitem{Baeyens1992}
Baeyens F, Eelen P, Crombez G, Van~den Bergh O.
\newblock Human evaluative conditioning: acquisition trials, presentation
  schedule, evaluative style and contingency awareness.
\newblock Behaviour Research and Therapy. 1992;30(2):133--142.

\bibitem{Baeyens2005}
Baeyens F, Diaz E, Ruiz G.
\newblock Resistance to extinction of human evaluative conditioning using a
  between-subjects design.
\newblock Cognition and Emotion. 2005;19(2):245--268.

\bibitem{Dawson2007a}
Dawson ME, Rissling AJ, Schell AM, Wilcox R.
\newblock Under what conditions can human affective conditioning occur without
  contingency awareness? Test of the evaluative conditioning paradigm.
\newblock Emotion. 2007;7(4):755--766.

\bibitem{DeHouwer2001}
De~Houwer J.
\newblock Contingency awareness and evaluative conditioning: when will it be
  enough?
\newblock Consciousness and Cognition. 2001;10(4):550--558.

\bibitem{Gast2012}
Gast A, Gawronski B, De~Houwer J.
\newblock Evaluative conditioning: recent developments and future directions.
\newblock Learning and Motivation. 2012;43(3):79--88.

\bibitem{Gawronski2014}
Gawronski B, Gast A, De~Houwer J.
\newblock Is evaluative conditioning really resistant to extinction? Evidence
  for changes in evaluative judgements without changes in evaluative
  representations.
\newblock Cognition and Emotion. 2014;p. 1--15.

\bibitem{Hardwick2000}
Hardwick SA, Lipp OV.
\newblock Modulation of Affective Learning: An Occasion for Evaluative
  Conditioning?
\newblock Learning and Motivation. 2000;31(3):251 -- 271.

\bibitem{Hofmann2010}
Hofmann W, De~Houwer J, Perugini M, Baeyens F, Crombez G.
\newblock Evaluative conditioning in humans: a meta-analysis.
\newblock Psychological Bulletin. 2010;136(3):390--421.

\bibitem{Hutter2013}
Hutter M, Sweldens S.
\newblock Implicit misattribution of evaluative responses: Contingency-unaware
  evaluative conditioning requires simultaneous stimulus presentations.
\newblock Journal of Experimental Psychology: General. 2013;142(3):638--643.

\bibitem{Sweldens2010}
Sweldens S, Van~Osselaer SMJ, Janiszewski CA.
\newblock Evaluative conditioning procedures and the resilience of conditioned
  brand attitudes.
\newblock Journal of Consumer Research. 2010;37(3):473--489.

\bibitem{Sweldens2014}
Sweldens S, Corneille O, Yzerbyt V.
\newblock The role of awareness in attitude formation through evaluative
  conditioning.
\newblock Personality and Social Psychology Review. 2014;18(2):187--209.

\bibitem{Wise2009}
Wise RA.
\newblock Roles for nigrostriatal - not just mesocorticolimbic - dopamine in
  reward and addiction.
\newblock Trends in Neurosciences. 2009;p. 32:517--524.

\bibitem{Paladini2004}
Paladini CA, Williams JT.
\newblock Noradrenergic inhibition of midbrain dopamine neurons.
\newblock Journal of Neuroscience. 2004;24(19):4568--4575.

\bibitem{Goodman2006}
Brunton LL, Lazo JS, Parker KL.
\newblock Goodman \& Gilman's The Pharmacological Basis of Therapeutics.
\newblock 11th ed. McGraw-Hill; 2006.

\bibitem{Parsons2015}
Parsons LH, Hurd YL.
\newblock Endocannabinoid signalling in reward and addiction.
\newblock Nature Reviews Neuroscience. 2015;16(10):579--594.

\bibitem{Barak2006}
Barak Y.
\newblock The immune system and happiness.
\newblock Autoimmunity Reviews. 2006;5(8):523--527.

\bibitem{cacioppo2007}
Cacioppo JT, Tassinary LG, Berntson G.
\newblock Handbook of psychophysiology.
\newblock Cambridge University Press; 2007.

\bibitem{Leary1990}
O'Leary A.
\newblock Stress, emotion, and human immune function.
\newblock Psychological Bulletin. 1990;108(3):363--382.

\bibitem{Steinman2004}
Steinman L.
\newblock Elaborate interactions between the immune and nervous systems.
\newblock Nature Immunology. 2004;5(6):575--581.

\bibitem{Ohman1994}
Ohman A, Soares JJ.
\newblock ''Unconscious anxiety": phobic responses to masked stimuli.
\newblock Journal of Abnormal Psychology. 1994;103(2):231--4.

\bibitem{Bechara1995}
Bechara A, Tranel D, Damasio H, Adolphs R, Rockland C, Damasio AR.
\newblock Double dissociation of conditioning and declarative knowledge
  relative to the amygdala and hippocampus in humans.
\newblock Science. 1995;269(5227):1115--1118.

\bibitem{Hull1943}
Hull CL.
\newblock Principles of Behavior: An Introduction to Behavior Theory.
\newblock The Century psychology series. D. Appleton-Century Company,
  Incorporated; 1943.

\bibitem{Scott2007}
Scott DJ, Stohler CS, Egnatuk CM, Wang H, Koeppe RA, Zubieta JK.
\newblock Individual differences in reward responding explain placebo-induced
  expectations and effects.
\newblock Neuron. 2007;55(2):325--336.

\bibitem{Colloca2008a}
Colloca L, Sigaudo M, Benedetti F.
\newblock The role of learning in nocebo and placebo effects.
\newblock Pain. 2008;136(1-2):211--218.

\bibitem{Colloca2008}
Colloca L, Tinazzi M, Recchia S, Le~Pera D, Fiaschi A, Benedetti F, et~al.
\newblock Learning potentiates neurophysiological and behavioral placebo
  analgesic responses.
\newblock Pain. 2008;139(2):306--314.

\bibitem{Colloca2011}
Colloca L, Miller FG.
\newblock How placebo responses are formed: a learning perspective.
\newblock Philosophical Transactions of the Royal Society of London Series B,
  Biological Sciences. 2011;366(1572):1859--1869.

\bibitem{Colloca2010}
Colloca L, Petrovic P, Wager TD, Ingvar M, Benedetti F.
\newblock How the number of learning trials affects placebo and nocebo
  responses.
\newblock Pain. 2010;151(2):430--439.

\bibitem{Jensen2012}
Jensen KB, Kaptchuk TJ, Kirsch I, Raicek J, Lindstrom KM, Berna C, et~al.
\newblock Nonconscious activation of placebo and nocebo pain responses.
\newblock Proceedings of the National Academy of Sciences of the United States
  of America. 2012;109(39):15959--15964.

\bibitem{Jensen2014}
Jensen KB, Kaptchuk TJ, Chen X, Kirsch I, Ingvar M, Gollub RL, et~al.
\newblock A Neural Mechanism for Nonconscious Activation of Conditioned Placebo
  and Nocebo Responses.
\newblock Cerebral Cortex. 2014;.

\bibitem{Lui2010}
Lui F, Colloca L, Duzzi D, Anchisi D, Benedetti F, Porro CA.
\newblock Neural bases of conditioned placebo analgesia.
\newblock Pain. 2010;151(3):816--824.

\bibitem{Montgomery1997}
Montgomery GH, Kirsch I.
\newblock Classical conditioning and the placebo effect.
\newblock Pain. 1997;72(1-2):107--113.

\bibitem{Nolan2012}
Nolan TA, Price DD, Caudle RM, Murphy NP, Neubert JK.
\newblock Placebo-induced analgesia in an operant pain model in rats.
\newblock Pain. 2012;153(10):2009--2016.

\bibitem{Williams2004}
Stewart-Williams S, Podd J.
\newblock The placebo effect: dissolving the expectancy versus conditioning
  debate.
\newblock Psychological Bulletin. 2004;130(2):324--340.

\bibitem{Ito2000}
Ito R, Dalley JW, Howes SR, Robbins TW, Everitt BJ.
\newblock Dissociation in conditioned dopamine release in the nucleus accumbens
  core and shell in response to cocaine cues and during cocaine-seeking
  behavior in rats.
\newblock Journal of Neuroscience. 2000;20(19):7489--7495.

\bibitem{Singer2004}
Singer T.
\newblock Empathy for Pain Involves the Affective but not Sensory Components of
  Pain.
\newblock Science. 2004;303(5661):1157--1162.

\bibitem{DeLaFuente2001}
De~la Fuente-Fernandez R, Ruth TJ, Sossi V, Schulzer M, Calne DB, Stoessl AJ.
\newblock Expectation and dopamine release: mechanism of the placebo effect in
  Parkinson's disease.
\newblock Science. 2001;293(5532):1164--1166.

\bibitem{DelaFuente2002}
De~la Fuente-Fernandez R, Stoessl AJ.
\newblock The placebo effect in Parkinson's disease.
\newblock Trends in Neurosciences. 2002;25(6):302--6.

\bibitem{Vits2011}
Vits S, Cesko E, Enck P, Hillen U, Schadendorf D, Schedlowski M.
\newblock Behavioural conditioning as the mediator of placebo responses in the
  immune system.
\newblock Philosophical Transactions of the Royal Society of London Series B,
  Biological Sciences. 2011;366(1572):1799--1807.

\bibitem{Chance2008}
Chance P.
\newblock 3.
\newblock In: Learning and Behavior. Belmont/CA: Wadsworth; 2008. p. p. 84.

\bibitem{Cardinal2002}
Cardinal RN, Parkinson JA, Hall J, Everitt BJ.
\newblock Emotion and motivation: the role of the amygdala, ventral striatum,
  and prefrontal cortex.
\newblock Neuroscience and Biobehavioral Reviews. 2002;26(3):321--352.

\bibitem{Davis1992}
Davis M.
\newblock The role of the amygdala in fear and anxiety.
\newblock Annual Review of Neuroscience. 1992;15:353--375.

\bibitem{Namburi2015}
Namburi P, Beyeler A, Yorozu S, Calhoon GG, Halbert SA, Wichmann R, et~al.
\newblock A circuit mechanism for differentiating positive and negative
  associations.
\newblock Nature. 2015;520(7549):675--678.

\bibitem{Richardson2004}
Richardson MP, Strange BA, Dolan RJ.
\newblock Encoding of emotional memories depends on amygdala and hippocampus
  and their interactions.
\newblock Nature Neuroscience. 2004;7(3):278--285.

\bibitem{Uwano1995}
Uwano T, Nishijo H, Ono T, Tamura R.
\newblock Neuronal responsiveness to various sensory stimuli, and associative
  learning in the rat amygdala.
\newblock Neuroscience. 1995;68(2):339--361.

\bibitem{Maren2005}
Maren S.
\newblock Synaptic Mechanisms of Associative Memory in the Amygdala.
\newblock Neuron. 2005;47(6):783 -- 786.

\bibitem{Sigurdsson2007}
Sigurdsson T, Doyere V, Cain CK, LeDoux JE.
\newblock Long-term potentiation in the amygdala: A cellular mechanism of fear
  learning and memory.
\newblock Neuropharmacology. 2007;52(1):215 -- 227.

\bibitem{Amano2011}
Amano T, Duvarci S, Popa D, Pare D.
\newblock The fear circuit revisited: contributions of the basal amygdala
  nuclei to conditioned fear.
\newblock Journal of Neuroscience. 2011;31(43):15481--15489.

\bibitem{Choi2010}
Choi JS, Jeansok JK.
\newblock Amygdala regulates risk of predation in rats foraging in a dynamic
  fear environment.
\newblock Proceedings of the National Academy of Sciences of the United States
  of America. 2010;107(50):21773--21777.

\bibitem{Glascher2003}
Glascher J, Adolphs R.
\newblock Processing of the arousal of subliminal and supraliminal emotional
  stimuli by the human amygdala.
\newblock Journal of Neuroscience. 2003;23(32):10274--10282.

\bibitem{paton2006}
Paton JJ, Belova MA, Morrison SE, Salzman CD.
\newblock The primate amygdala represents the positive and negative value of
  visual stimuli during learning.
\newblock Nature. 2006;439(7078):865--870.

\bibitem{Sangha2013}
Sangha S, Chadick JZ, Janak PH.
\newblock Safety encoding in the basal amygdala.
\newblock Journal of Neuroscience. 2013;33(9):3744--3751.

\bibitem{Schoenbaum1999}
Schoenbaum G, Chiba AA, Gallagher M.
\newblock Neural encoding in orbitofrontal cortex and basolateral amygdala
  during olfactory discrimination learning.
\newblock Journal of Neuroscience. 1999;19(5):1876--1884.

\bibitem{Tovote2015}
Tovote P, Fadok JP, Luthi A.
\newblock Neuronal circuits for fear and anxiety.
\newblock Nature Reviews Neuroscience. 2015;16(6):317--331.

\bibitem{Janak2015}
Janak PH, Tye KM.
\newblock From circuits to behaviour in the amygdala.
\newblock Nature. 2015;517(7534):284--292.

\bibitem{Sah2003}
Sah P, Faber ES, Lopez De~Armentia M, Power J.
\newblock The amygdaloid complex: anatomy and physiology.
\newblock Physiological Reviews. 2003;83(3):803--834.

\bibitem{LeDoux2000}
LeDoux JE.
\newblock Emotion circuits in the brain.
\newblock Annual Review of Neuroscience. 2000;23:155--184.

\bibitem{Killcross1997}
Killcross S, Robbins TW, Everitt BJ.
\newblock Different types of fear-conditioned behaviour mediated by separate
  nuclei within amygdala.
\newblock Nature. 1997;388(6640):377--380.

\bibitem{Balleine2006}
Balleine BW, Killcross S.
\newblock Parallel incentive processing: an integrated view of amygdala
  function.
\newblock Trends in Neurosciences. 2006;p. 29, 272--279.

\bibitem{Blundell2001}
Blundell P, Hall G, Killcross S.
\newblock Lesions of the basolateral amygdala disrupt selective aspects of
  reinforcer representation in rats.
\newblock Journal of Neuroscience. 2001;p. 21, 9018--9026.

\bibitem{Corbit2005}
Corbit LH, Balleine BW.
\newblock Double dissociation of basolateral and central amygdala lesions on
  the general and outcome-specific forms of pavlovian-instrumental transfer.
\newblock Journal of Neuroscience. 2005;p. 25, 962--970.

\bibitem{Muramoto1993}
Muramoto K, Ono T, Nishijo H, Fukuda M.
\newblock Rat amygdaloid neuron responses during auditory discrimination.
\newblock Neuroscience. 1993;52(3):621--636.

\bibitem{Shabel2009}
Shabel SJ, Janak PH.
\newblock Substantial similarity in amygdala neuronal activity during
  conditioned appetitive and aversive emotional arousal.
\newblock Proceedings of the National Academy of Sciences of the United States
  of America. 2009;106(35):15031--15036.

\bibitem{Bourdy2012}
Bourdy R, Barrot M.
\newblock A new control center for dopaminergic systems: pulling the VTA by the
  tail.
\newblock Trends in Neurosciences. 2012;35(11):681--690.

\bibitem{Schultz1998}
Schultz W.
\newblock Predictive reward signal of dopamine neurons.
\newblock Journal of Neurophysiology. 1998;p. 80: 1--27.

\bibitem{Doherty2007}
O'Doherty JP.
\newblock Lights, camembert, action! The role of human orbitofrontal cortex in
  encoding stimuli, rewards, and choices.
\newblock Annals of the New York Academy of Sciences. 2007;p. 1121:254--272.

\bibitem{Doherty2004}
O'Doherty JP.
\newblock Reward representations and reward-related learning in the human
  brain: insights from neuroimaging.
\newblock Current Opinion in Neurobiology. 2004;14(6):769--776.

\bibitem{Takahashi2009}
Takahashi YK, Roesch MR, Stalnaker TA, Haney RZ, Calu DJ, Taylor AR, et~al.
\newblock The orbitofrontal cortex and ventral tegmental area are necessary for
  learning from unexpected outcomes.
\newblock Neuron. 2009;62(2):269--280.

\bibitem{stalnaker2015}
Stalnaker TA, Cooch NK, Schoenbaum G.
\newblock What the orbitofrontal cortex does not do.
\newblock Nature Neuroscience. 2015;18(5):620--627.

\bibitem{Schoenbaum2009}
Schoenbaum G, Roesch MR, Stalnaker TA, Takahashi YK.
\newblock A new perspective on the role of the orbitofrontal cortex in adaptive
  behaviour.
\newblock Nature Reviews Neuroscience. 2009;10(12):885--892.

\bibitem{Takahashi2011}
Takahashi YK, Roesch MR, Wilson RC, Toreson K, O'Donnell P, Niv Y, et~al.
\newblock Expectancy-related changes in firing of dopamine neurons depend on
  orbitofrontal cortex.
\newblock Nature Neuroscience. 2011;14(12):1590--1597.

\bibitem{Morrison2011}
Morrison SE, Salzman CD.
\newblock Representations of appetitive and aversive information in the primate
  orbitofrontal cortex.
\newblock Annals of the New York Academy of Sciences. 2011;1239:59--70.

\bibitem{EdmundT.Rolls2008}
Rolls ET, Grabenhorst F.
\newblock The orbitofrontal cortex and beyond : From affect to decision-making.
\newblock Progress in Neurobiology. 2008;86:216--244.

\bibitem{Kennerley2011}
Kennerley SW, Walton ME.
\newblock Decision making and reward in frontal cortex: complementary evidence
  from neurophysiological and neuropsychological studies.
\newblock Behavioral Neuroscience. 2011;125(3):297--317.

\bibitem{wallis2007}
Wallis JD.
\newblock Orbitofrontal cortex and its contribution to decision-making.
\newblock Annual Review of Neuroscience. 2007;30:31--56.

\bibitem{Plassmann2008}
Plassmann H, O'Doherty J, Shiv B, Rangel A.
\newblock Marketing actions can modulate neural representations of experienced
  pleasantness.
\newblock Proceedings of the National Academy of Sciences of the United States
  of America. 2008;105(3):1050--1054.

\bibitem{Li2011}
Li J, Delgado MR, Phelps EA.
\newblock How instructed knowledge modulates the neural systems of reward
  learning.
\newblock Proceedings of the National Academy of Sciences of the United States
  of America. 2011;108(1):55--60.

\bibitem{Gallagher1999}
Gallagher M, McMahan RW, Schoenbaum G.
\newblock Orbitofrontal cortex and representation of incentive value in
  associative learning.
\newblock Journal of Neuroscience. 1999;19(15):6610--6614.

\bibitem{bishop2006}
Bishop CM.
\newblock Pattern recognition and machine learning.
\newblock springer; 2006.

\bibitem{Friston2015}
Friston KJ, Frith CD.
\newblock Active inference, communication and hermeneutics.
\newblock Cortex. 2015;68:129--143.

\bibitem{Barrelt2015}
Barrett LF, Simmons WK.
\newblock Interoceptive predictions in the brain.
\newblock Nature Reviews Neuroscience. 2015;16(7):419--429.

\bibitem{Pezzulo2015}
Pezzulo G, Rigoli F, Friston K.
\newblock Active Inference, homeostatic regulation and adaptive behavioural
  control.
\newblock Progress in Neurobiology. 2015;134:17--35.

\bibitem{Seth2013}
Seth AK.
\newblock Interoceptive inference, emotion, and the embodied self.
\newblock Trends in Cognitive Sciences. 2013;17(11):565--573.

\bibitem{Toates1986}
Toates F.
\newblock Motivational systems.
\newblock Cambridge, England: Cambridge University Press.; 1986.

\bibitem{Cabanac1992}
Cabanac M.
\newblock Pleasure: the common currency.
\newblock Journal of Theoretical Biology. 1992;155(2):173--200.

\bibitem{Dickinson1994}
Dickinson A, Balleine B.
\newblock Motivational control of goal-directed action.
\newblock Animal Learning and Behavior. 1994;22(1):1--18.

\bibitem{Balleine1998}
Balleine BW, Dickinson A.
\newblock Goal-directed instrumental action: contingency and incentive learning
  and their cortical substrates.
\newblock Neuropharmacology. 1998;37(4-5):407--419.

\bibitem{Bohner2001}
Bohner M, Peterson A.
\newblock Dynamic Equations On Time Scales: an introduction with applications.
\newblock Birkhauser, editor; 2001.

\bibitem{Benedetti2003}
Benedetti F, Pollo A, Lopiano L, Lanotte M, Vighetti S, Rainero I.
\newblock Conscious expectation and unconscious conditioning in analgesic,
  motor, and hormonal placebo/nocebo responses.
\newblock Journal of Neuroscience. 2003;23(10):4315--4323.

\bibitem{Siegel1975}
Siegel S.
\newblock Evidence from rats that morphine tolerance is a learned response.
\newblock Journal of Comparative and Physiological Psychology.
  1975;89(5):498--506.

\bibitem{May1976}
MAY RM.
\newblock Simple mathematical models with very complicated dynamics.
\newblock Nature. 1976;261:459--467.

\bibitem{Kim2002}
Kim JJ, Diamond DM.
\newblock The stressed hippocampus, synaptic plasticity and lost memories.
\newblock Nature Reviews Neuroscience. 2002;3(6):453--462.

\bibitem{Esteves1994}
Esteves F, Parra C, Dimberg U, Ohman A.
\newblock Nonconscious associative learning: Pavlovian conditioning of skin
  conductance responses to masked fear-relevant facial stimuli.
\newblock Psychophysiology. 1994;p. 31:375--385.

\bibitem{papini1997}
Papini MR, Dudley RT.
\newblock Consequences of surprising reward omissions.
\newblock Review of General Psycholology. 1997;1(2):175.

\bibitem{Genn2004}
Genn RF, Ahn S, Phillips AG.
\newblock Attenuated dopamine efflux in the rat nucleus accumbens during
  successive negative contrast.
\newblock Behavioral Neuroscience. 2004;118(4):869--873.

\bibitem{Urcelay2009}
Urcelay GP, Wheeler DS, Miller RR.
\newblock Spacing extinction trials alleviates renewal and spontaneous
  recovery.
\newblock Learning and Behavior. 2009;37(1):60--73.

\bibitem{Bunce1993}
Bunce SC, Larsen RJ, Cruz M.
\newblock Individual differences in the excitation transfer effect.
\newblock Personality and Individual Differences. 1993;15(5):507 -- 514.

\bibitem{Mayer1999}
Mayer B, Merckelbach H.
\newblock Unconscious processes, subliminal stimulation, and anxiety.
\newblock Clinical Psychology Review. 1999;19(5):571--590.

\bibitem{zajonc_mere_2001}
Zajonc RB.
\newblock Mere exposure: A gateway to the subliminal.
\newblock Current Directions in Psychological Science. 2001;10(6):224--228.

\bibitem{Morris1998}
Morris JS, Ohman A, Dolan RJ.
\newblock Conscious and unconscious emotional learning in the human amygdala.
\newblock Nature. 1998;393(6684):467--470.

\bibitem{Morris1999}
Morris JS, Ohman A, Dolan RJ.
\newblock A subcortical pathway to the right amygdala mediating "unseen" fear.
\newblock Proceedings of the National Academy of Sciences of the United States
  of America. 1999;96(4):1680--1685.

\bibitem{Baeyens1993}
Baeyens F, Hermans D, Eelen P.
\newblock The role of CS-US contingency in human evaluative conditioning.
\newblock Behaviour Research and Therapy. 1993;31(8):731--737.

\bibitem{Houwer2001}
De~Houwer J, Thomas S, Baeyens F.
\newblock Associative learning of likes and dislikes: a review of 25 years of
  research on human evaluative conditioning.
\newblock Psychological Bulletin. 2001;127(6):853--869.

\bibitem{Ruys2009}
Ruys KI, Diederik AS.
\newblock Learning to like or dislike by association: No need for contingency
  awareness.
\newblock Journal of Experimental Social Psychology. 2009;45(6):1277 -- 1280.

\bibitem{Shipp2013}
Shipp S, Adams RA, Friston KJ.
\newblock Reflections on agranular architecture: predictive coding in the motor
  cortex.
\newblock Trends in Neurosciences. 2013;36(12):706--716.

\bibitem{Crapse2008}
Crapse TB, Sommer MA.
\newblock Corollary discharge across the animal kingdom.
\newblock Nature Reviews Neuroscience. 2008;9(8):587--600.

\bibitem{Ohman1998}
Ohman A, Soares JJ.
\newblock Emotional conditioning to masked stimuli: expectancies for aversive
  outcomes following non-recognized fear-relevant stimuli.
\newblock Journal of Experimental Psychology. 1998;p. General, 127, 69--82.

\bibitem{Enns2000}
Enns JT, Di~Lollo V.
\newblock What's new in visual masking?
\newblock Trends in Cognitive Sciences. 2000;p. 4, 345--352.

\bibitem{Noguchi2005}
Noguchi Y, Kakigi R.
\newblock Neural mechanisms of visual backward masking revealed by high
  temporal resolution imaging of human brain.
\newblock Neuroimage. 2005;p. 27, 178--187.

\bibitem{Rolls1999}
Rolls ET, Tovee MJ, Panzeri S.
\newblock The neurophysiology of backward visual masking: Information analysis.
\newblock Journal of Cognitive Neuroscience. 1999;p. 11, 300--311.

\bibitem{Kim2005}
Kim CY, Blake R.
\newblock Psychophysical magic: Rendering the visible invisible.
\newblock Trends in Cognitive Sciences. 2005;p. 9, 381--388.

\bibitem{LeDoux1996}
LeDoux JE.
\newblock The Emotional Brain: The Mysterious Underpinnings of Emotional Life.
\newblock New York: Simon \& Schuster; 1996.

\bibitem{Liddell2004}
Liddell BJ, Williams LM, Rathjen J, Shevrin H, Gordon E.
\newblock A temporal dissociation of subliminal versus supraliminal fear
  perception: an event-related potential study.
\newblock Journal of Cognitive Neuroscience. 2004;16(3):479--486.

\bibitem{Gelder2006}
De~Gelder B, Hadjikhani N.
\newblock Non-conscious recognition of emotional body language.
\newblock Neuroreport. 2006;p. 17, 583--586.

\bibitem{Morris2001}
Morris JS, De~Gelder B, Weiskrantz L, Dolan RJ.
\newblock Differential extrageniculostriate and amygdala responses to
  presentation of emotional faces in a cortically blind field.
\newblock Brain. 2001;p. 124, 1241--1252.

\bibitem{Ohman2005}
Ohman A.
\newblock The role of the amygdala in human fear: Automatic detection of
  threat.
\newblock Psychoneuroendocrinology. 2005;p. 30, 953--958.

\bibitem{Ohman2007}
Ohman A, Carlsson K, Lundqvist D, Ingvar M.
\newblock On the unconscious subcortical origin of human fear.
\newblock Physiology and Behavior. 2007;p. 92, 180--185.

\bibitem{Parra1997}
Parra C, Esteves F, Flykt A, Ohman A.
\newblock Pavlovian conditioning to social stimuli: Backward masking and the
  dissociation of implicit and explicit cognitive processes.
\newblock European Psychologist. 1997;p. 2, 106--117.

\bibitem{Olsson2007}
Olsson A, Nearing KI, Phelps EA.
\newblock Learning fears by observing others: the neural systems of social fear
  transmission.
\newblock Social Cognitive and Affective Neuroscience. 2007;2(1):3--11.

\bibitem{OlssonA2007}
Olsson A, Phelps EA.
\newblock Social learning of fear.
\newblock Nature Neuroscience. 2007;10(9):1095--102.

\bibitem{Pape2010}
Pape HC, Pare D.
\newblock Plastic synaptic networks of the amygdala for the acquisition,
  expression, and extinction of conditioned fear.
\newblock Physiological Reviews. 2010;p. 90, 419--463.

\bibitem{Pickens2004}
Pickens CL, Holland PC.
\newblock Conditioning and cognition.
\newblock Neuroscience and Biobehavioral Reviews. 2004;p. 28, 651--661.

\bibitem{Bliss1973}
Bliss TVP, Lomo T.
\newblock Long-lasting potentiation of synaptic transmission in the dentate
  area of the anaesthetized rabbit following stimulation of the perforant path.
\newblock Journal of Physiology (London). 1973;p. 232 , 331--356.

\bibitem{Artola1990}
Artola A, Brocher S, Singer W.
\newblock Different voltagedependent thresholds for inducing long-term
  depression and long-term potentiation in slices of rat visual cortex.
\newblock Nature. 1990;p. 347, 69--72.

\bibitem{Destexhe2004}
Destexhe A, Marder E.
\newblock Plasticity in single neuron and circuit computations.
\newblock Nature. 2004;431(7010):789--795.

\bibitem{Barlow1985}
Barlow H.
\newblock In: (eds Rose V D \&~Dobson, editor. Models of the Visual Cortex.
  Wiley, Chichester; 1985. p. 37--46.

\bibitem{Roskies1999}
Roskies A.
\newblock The binding problem: special issue.
\newblock Neuron. 1999;p. 24, 7--125.

\bibitem{Engel2001}
Engel AK, Fries P, Singer W.
\newblock Dynamic predictions: oscillations and synchrony in top-down
  processing.
\newblock Nature Reviews Neuroscience. 2001;2(10):704--716.

\bibitem{Abeles1991}
Abeles M.
\newblock Corticonics: Neuronal Circuits of the Cerebral Cortex.
\newblock Cambridge, editor. Cambridge University Press; 1991.

\bibitem{Rudolph2001}
Rudolph M, Destexhe A.
\newblock Correlation detection and resonance in neural systems with
  distributed noise sources.
\newblock Physical Review Letters. 2001;p. 86, 3662--3665.

\bibitem{Young1998}
Young AMJ, Ahier RG, Upton RL, Joseph MH, Gray JA.
\newblock Increased extracellular dopamine in the nucleus accumbens of the rat
  during associative learning of neutral stimuli.
\newblock Neuroscience. 1998;83(4):1175 -- 1183.

\bibitem{Munakata2004}
Munakata Y, Pfaffly J.
\newblock Hebbian learning and development.
\newblock Developmental Science. 2004;7(2):141--148.

\bibitem{Bouton1993}
Bouton ME.
\newblock Context, Time and Memory Retrieval in the Interference Paradigms of
  Pavlovian Learning.
\newblock Psychological Bulletin. 1993;114:1, 80--99.

\bibitem{Cotton1982}
Cotton MM, Goodall G, Mackintosh NJ.
\newblock Inhibitory conditioning resulting from a reduction in the magnitude
  of reinforcement.
\newblock Quarterly Journal of Experimental Psychology B, Comparative and
  Physiological Psychology. 1982;34 (Pt 3):163--180.

\bibitem{Wheeler2008}
Wheeler DS, Sherwood A, Holland PC.
\newblock Excitatory and inhibitory learning with absent stimuli.
\newblock Journal of Experimental Psychology: Animal Behavior Processes.
  2008;34(2):247--255.

\bibitem{zimmer1974}
Zimmer-Hart CL, Rescorla RA.
\newblock Extinction of Pavlovian conditioned inhibition.
\newblock Journal of Comparative and Physiological Psychology. 1974;86(5):837.

\bibitem{Berlyne1960}
Berlyne DE.
\newblock Conflict, arousal, and curiosity.
\newblock New York, NY: McGraw-Hill; 1960.

\bibitem{Joseph2015}
Huston JP, Nadal M, Mora F, Agnati LF, Cela~Conde CJ.
\newblock Art, Aesthetics, and the Brain.
\newblock Oxford University Press; 2015.

\bibitem{Schafer2009}
Schafer RW, Oppenheim AV.
\newblock Discrete-Time Signal Processing.
\newblock 3rd ed. Prentice Hall; 2009.

\bibitem{Luna2005}
Luna R, Hernandez A, Brody CD, Romo R.
\newblock Neural codes for perceptual discrimination in primary somatosensory
  cortex.
\newblock Nature Neuroscience. 2005;8(9):1210--1219.

\bibitem{Paunovic1999}
Paunovic N.
\newblock Exposure counterconditioning (EC) as a treatment for severe PTSD and
  depression with an illustrative case.
\newblock Journal of Behavior Therapy and Experimental Psychiatry.
  1999;30(2):105--117.

\bibitem{vanRooij2015}
Van~Rooij SJ, Geuze E, Kennis M, Rademaker AR, Vink M.
\newblock Neural correlates of inhibition and contextual cue processing related
  to treatment response in PTSD.
\newblock Neuropsychopharmacology. 2015;40(3):667--675.

\bibitem{Bremner1999}
Bremner JD.
\newblock Alterations in brain structure and function associated with
  post-traumatic stress disorder.
\newblock Seminars in Clinical Neuropsychiatry. 1999;4(4):249--255.

\bibitem{Banks2007}
Banks SJ, Eddy KT, Angstadt M, Nathan PJ, Phan KL.
\newblock Amygdala - frontal connectivity during emotion regulation.
\newblock Social Cognitive and Affective Neuroscience. 2007;2(4):303--312.

\bibitem{Delgado2008a}
Delgado MR, Nearing KI, Ledoux JE, Phelps EA.
\newblock Neural circuitry underlying the regulation of conditioned fear and
  its relation to extinction.
\newblock Neuron. 2008;59(5):829--838.

\bibitem{Lipka2014}
Lipka J, Hoffmann M, Miltner WH, Straube T.
\newblock Effects of cognitive-behavioral therapy on brain responses to
  subliminal and supraliminal threat and their functional significance in
  specific phobia.
\newblock Biological Psychiatry. 2014;76(11):869--877.

\bibitem{Ramirez2015}
Ramirez S, Liu X, MacDonald CJ, Moffa A, Zhou J, Redondo RL, et~al.
\newblock Activating positive memory engrams suppresses depression-like
  behaviour.
\newblock Nature. 2015;522(7556):335--339.

\end{thebibliography}

\end{document}